%% file: main.tex
\documentclass[preprintnumbers,aps,prd,floatfix,nofootinbib,twocolumn]{revtex4}
\usepackage{qcd}
\usepackage{epsfig}
\usepackage{graphicx}
\usepackage{url,hyperref}
\usepackage{bm}

\begin{document}

\title{Combining nonperturbative transverse momentum dependence 
with TMD evolution}

\preprint{JLAB-THY-22-3615}

\author{J.~O.~Gonzalez-Hernandez}
\email{joseosvaldo.gonzalezhernandez@unito.it}
\affiliation{Dipartimento di Fisica, Universit\`a degli Studi di Torino, Via P. Giuria 1, 1-10125, Torino, Italy}
\author{T.~C.~Rogers}
\email{tedconantrogers@gmail.com}
\affiliation{Department of Physics, Old Dominion University, Norfolk, VA 23529, USA}
\affiliation{Jefferson Lab, 12000 Jefferson Avenue, Newport News, VA 23606, USA}
\author{N.~Sato}
\email{nsato@jlab.org}
\affiliation{Jefferson Lab, 12000 Jefferson Avenue, Newport News, VA 23606, USA}

\begin{abstract}
 Central to understanding the nonpertubative, intrinsic partonic nature of hadron structure are the concepts of transverse momentum dependent (TMD) parton distribution and fragmentation functions. 
 A TMD factorization approach to the phenomenology of semi-inclusive processes that includes evolution, higher orders, and matching to larger transverse momentum, is ultimately necessary for reliably connecting with phenomenologically extracted nonperturbative structures, especially when widely different scales are involved. 
 In this paper, we will address some of the difficulties that arise when phenomenological techniques that were originally designed for very high energy applications are extended to studies of hadron structures, and we will solidify the connection between standard high energy TMD implementations and the more intuitive, parton model based approaches to phenomenology that emphasize nonperturbative hadron structure. In the process, we will elaborate on differences between forward and backward TMD evolution, which in the context of this paper we call ``bottom-up'' and ``top-down'' approaches, and we will explain the advantages of a bottom-up strategy. We will also emphasize and clarify the role of the integral relations that connect TMD and collinear correlation functions. We will show explicitly how they constrain the nonperturbative ``$g$-functions'' of standard Collins-Soper-Sterman (CSS) implementations of TMD factorization.
 This paper is especially targeted toward phenomenologists and model builders who are interested in merging specific nonperturbative models and calculations (including lattice QCD) with TMD factorization at large $Q$. 
 Our main result is a recipe for incorporating 
 nonperturbative models into TMD factorization, and for constraining their parameters in a way that matches to perturbative QCD and evolution.
\end{abstract}

\date{\today}

\maketitle

\section{Introduction}
\label{s.intro}

The techniques of TMD factorization and evolution that are rooted in the CSS~\cite{Collins:1981uk,Collins:1984kg,Collins:2011qcdbook} approach are by now a very widespread, but applications tend to take on superficially different forms depending on their use and the specific phenomenological context. This fact will be very important for understanding the goals of the present paper. To unify different TMD applications, and especially to compare the phenomenologically extracted nonperturbative transverse momentum dependence across many different processes, some translation is still necessary. For our purposes, we will need to consider at least the following two phenomenological settings where TMD factorization has often been applied in the past:
\begin{enumerate}
\item \underline{Very large $Q$ (Type II)}: With treatments of the transverse momentum 
spectra of processes at very large hard scales (say $Q \gtrsim 10$~GeV), it is common to assume (sometimes implicitly) that transverse momentum dependence is describable mostly as a form of perturbative radiation in collinear factorization. An example is weak boson production in hadron-hadron collisions, where the hard scale is fixed by a heavy boson mass, larger than $80$~GeV. When the transverse momentum $\Tsc{q}{}$ of the boson is of order the hard scale, calculations can follow a fixed order collinear factorization treatment. The challenge is then to describe the behavior as $\Tsc{q}{}$ decreases relative to $Q$. Term-by-term in fixed order calculations there are logarithms like
\begin{equation} 
\sim \alpha_s(Q)^n \ln^m \parz{\frac{Q}{\Tsc{q}{}}}  \, , \label{e.logsa}
\end{equation}
with integer $m,n > 0$, that grow until they 
spoil truncated perturbation theory. With this as the starting point, the natural strategy is to try to resum as many such logarithms as possible. Most traditional TMD factorization techniques~\cite{Collins:1981uk,Collins:1984kg,Collins:2011qcdbook}, along with soft-collinear effective theory (SCET)-based approaches~\cite{Becher:2006mr,Becher:2010tm,GarciaEchevarria:2011rb}, as well as approaches that directly resum transverse momentum logarithms~\cite{Ellis:1997sc,deFlorian:2000pr,Bozzi:2005wk,Catani:2012qa,Catani:2013tia}, effectively account for these types of logarithms while also allowing for at least some contribution from nonperturbative transverse momentum in the $\Tsc{q}{} \approx \Lambda_\text{QCD}$ region. 
\item \underline{Hadron structure and moderate $Q$ (Type I)}: 
TMD parton distribution functions (pdfs) and fragmentation functions (ffs) also feature prominently in studies whose focus is more directly on the nonperturbative structure of hadrons. In these types of applications, the relevant hard scales tend to be at much lower $Q$ than in Type II situations, such as the $Q \approx$ few GeVs common in many
semi-inclusive deep inelastic scattering (SIDIS) measurements. It is possible to trace the origin of many of the hadron-structure oriented approaches in Type I applications to intuitive pictures of colliding hadrons in a parton model. The hadrons, in this view, are composed entirely of nonperturbative quark and gluon constituents~\cite{Gardiner:1970wy,Tangerman:1994eh,Mulders:1996dh,Anselmino:1994tv}, and the earliest versions  of phenomenological Type I applications usually adopted the approximation that all transverse momentum dependence is nonperturbative in origin. As such, they could mostly ignore the role of perturbative tails at large $\Tsc{q}{}$ and of  evolution~\cite{Bacchetta:2006tn,Anselmino:2007fs,Schweitzer:2010tt,Anselmino:1999pw}.
\end{enumerate}
The Type I and Type II classification roughly follows that at Feynman, Field and Fox~\cite[Fig.~6]{Feynman:1978dt}.
At a formal level, it is now very well understood that approaches to Type I and type II observables can be made equivalent. And, of course, there is no sharp distinction between what constitutes a Type I or a type II scenario. It is possible to merge treatments of hadron structure with the evolution formulas that were traditionally applied at much larger $Q$~\cite{Ji:2004wu,Ji:2004xq,Collins:2011qcdbook,Aybat:2011zv,Anselmino:2012aa}, and indeed much activity over the past decade was devoted to implementing TMD evolution, in the context of hadron structure studies, in ways that include nonperturbative parts~\cite{Aybat:2011ge,Godbole:2013bca,Echevarria:2014xaa,Hautmann:2014kza,Bacchetta:2015ora,Scarpa:2019fol,Boglione:2017jlh,Bacchetta:2017gcc,Anselmino:2018psi,Bacchetta:2019sam,Scimemi:2019cmh,Bastami:2020asv,Boglione:2020auc}. (The versions of TMD factorization and evolution that we will focus on in this paper are those rooted in, or very similar to, the CSS formalism as described in Ref.~\cite{Collins:2011qcdbook} -- see also Ref.~\cite{Collins:2017oxh} for translations to other approaches.)

There are, nevertheless, some remaining open issues related to the interpretation of intrinsic  nonperturbative transverse momentum dependence that is extracted pheomenologically and its role in cross section calculations, and these can have practical consequences. We will explain what we mean here in much more detail in the main body of this paper. For now, we will prepare the reader by noting how some of the remaining complications originate in a clash between the natural phenomenological strategies that are often implicit in Type I and type II situations. 

Starting from a typical Type II perspective, the main issue is that large transverse momentum perturbation theory calculations (with very large $Q$) receive correction 
terms like \eref{logsa} that diverge as $\Tsc{q}{}$ approaches zero. So, the natural strategy is to try to resum as many transverse momentum dependent logarithms as possible as $\Tsc{q}{}$ decreases until one is essentially forced to incorporate a nonperturbative transverse momentum dependent component. We call this a ``top-down'' view because it starts by optimizing a large $\Tsc{q}{}$ dependence at large $\Tsc{q}{} \approx Q$ in collinear perturbation theory and then it extends it via evolution and resummation downward to more moderate $Q$ and $\Tsc{q}{} \approx 0$. 

But, as an alternative strategy, one might instead start from the perspective more common in Type I scenarios. That is, one may begin by considering a moderate input scale $Q_0$, low enough so that the accessible range of $\Tsc{q}{}$ is either comparable to $Q_0$ and is perturbative in origin, or is smaller than $Q_0$ and is mostly nonperturbative.\footnote{More specifically, there exists no significant transverse momentum region where $m/\Tsc{q}{}$ and $\Tsc{q}{}/Q_0$ are both simultaneously small, though each is small in some region of $\Tsc{q}{}$. Note that it is possible for this to be the situation even if $Q_0$ is large enough that $\alpha_s(Q_0)$ is small.} 
A TMD parton model type of description is the  most natural and appropriate approach here.
There is no region within the range of $0 < \Tsc{q}{} \lesssim Q_0$ where calculations involve large logarithms analogous to~\eref{logsa}. Thus, for $Q \approx Q_0$, the need to resum them does not arise. The only task then is to match a nonperturbative parametrization of transverse momentum dependence to a fixed order $\Tsc{q}{} \approx Q_0$ calculation of transverse momentum dependence in collinear factorization. 

However, the diverging logarithms will reappear later if we evolve to large $Q$. In this case, the uncontrolled perturbation theory errors will reappear at large $\Tsc{q}{}$ once we consider $Q \gg Q_0$ where the transverse momentum region $\Tsc{q}{} \gg Q_0$ becomes accessible. The problematic logarithms are in the correlation functions that were originally defined at the input scale, and they take the form
\begin{equation} 
\sim \alpha_s(Q_0)^n \ln^m \parz{\frac{\Tsc{q}{}}{Q_0}}  \, . \label{e.logsaa}
\end{equation}
Unlike with the Type II oriented approaches discussed previously, the large logarithms from the bottom-up perspective are due to $\Tsc{q}{}$ getting \emph{large} rather than small.  
We call it a ``bottom-up'' viewpoint because it starts by optimizing $\Tsc{q}{} \sim 0$ treatments at moderate $Q_0$, and then uses evolution equations to extend to large $\Tsc{q}{}$ and large $Q \gg Q_0$ where terms like \eref{logsaa} need to be addressed. These aspects of a bottom-up style of approach will be made much clearer in our review of TMD factorization in \sref{tmd_fact}. 

In a more general sense, within typical Type II oriented approaches, transverse momentum dependence is seen as primarily perturbative and requiring corrections from non-perturbative behavior, while in typical Type I oriented approaches it is seen as a primarily non-perturbative phenomenon that needs to be improved with perturbative corrections from large transverse momentum. 
In the former, divergences are seen as a low $\Tsc{q}{}$ problem while in the latter they are a large $\Tsc{q}{}$ problem.

There is no formal difference between the bottom-up and top-down viewpoints just discussed, and they are generally described by identical sets of factorization and evolution equations. The logarithms like \eref{logsaa} are simply the mirror images of those in~\eref{logsa}. 
However, they suggest different phenomenological strategies and this can have practical consequences, as we will see.  
The focus of this paper will be to give a full account of a bottom-up approach, which to our knowledge has never been made entirely explicit. We will also formally link it back to top-down approaches. 

As we will see, adopting a bottom-up strategy brings with it some significant practical advantages in situations where extracting nonperturbative transverse momentum dependence is the primary goal. It also raises several important aspects of nonperturbative parametrizations that are often hidden to the foreground. All of this will be made clearer in the main body of the text.

One advantage is that, at the input scale $Q_0$, it will be natural from the bottom-up perspective to categorize regions as perturbative or nonperturbative in transverse momentum space rather than in transverse coordinate space. This accords well with standard approaches to interpreting experimental data. For example, Ref.~\cite{COMPASS:2017mvk} offers some interpretations of recent COMPASS measurements with their Figure 17 and the comment ``...the two exponential functions in our parameterisation $F_1$
can be attributed to two completely different underlying physics mechanisms...". Much earlier, Feynman, Field, and Fox similarly identified different physical TMD mechanisms in transverse momentum space -- see Figures 6 and 7 and the surrounding discussion from Ref.~\cite{Feynman:1978dt}. Associated with observations like these is an implicit choice to categorize physical mechanisms as perturbative or nonperturbative in momentum
space.

The most prevalent models of nonperturbative TMDs were also constructed in momentum space, so they also sort regions into perturbative and nonperturbative categories in transverse momentum space. In many models, the large transverse momentum tails are suppressed on the grounds that it is only the small transverse momentum dependence that is nonperturbative or intrinsic to the hadron structure. Examples include at least spectator models~\cite{Gamberg:2003ey,Bacchetta:2008af,Kang:2010hg,Guerrero:2020hom}, light-cone wave function descriptions~\cite{Pasquini:2008ax,Pasquini:2011tk,Bacchetta:2017vzh,Pasquini:2014ppa,Hu:2022ctr}, bag models~\cite{Sakai:1979rf,Sakai:1979ez,Yuan:2003wk,Avakian:2010br,Signal:2021aum}, the Nambu-Jones Lasinio model~\cite{Matevosyan:2011vj,Matevosyan:2012ga,Noguera:2015iia}, calculations based on the Dyson-Schwinger equations~\cite{Shi:2018zqd}, classic quark-hadron model approaches and many others~\cite{Broniowski:2017wbr,Bastami:2020asv,Bastami:2020rxn}. As will become clear, it is easier to incorporate models like these into full evolution treatments when starting from a bottom-up approach.  

It will also turn out to be easier to match these models to perturbative transverse momentum calculations at large $\Tsc{q}{}$ in transverse momentum space than in coordinate space. Generally, fixed order perturbation theory calculations in collinear factorization for large transverse momentum ($\Tsc{q}{} \approx Q$) observables are performed in momentum space.
When we categorize behavior as perturbative or nonperturbative in TMD factorization according to momentum space regions, we will find that we are better able to ensure a consistent transition to the large-$\Tsc{q}{}$, collinear description where transverse momentum is generated perturbatively. (See also our comments below regarding the ``asymptotic'' term.) The difficulty of matching regions of $\Tsc{q}{}$ with different physics has led to proposals to implement all Type II style resummation entirely in momentum space, such as~\cite{Ellis:1997ii,Kulesza:1999gm}, inspired by earlier pioneering work that resummed leading transverse momentum logarithms~\cite{Dokshitzer:1978hw}.

Another issue that will become clearer from a bottom-up perspective is the question of how to describe the transition between perturbative and nonperturbative transverse \emph{coordinate} space dependence. Working in coordinate space has significant advantages for both setting up TMD factorization theoretically and implementing it. For example, most versions of the CSS formalism~\cite{Collins:1981uw,Collins:1981uk,Collins:1981tt,Collins:1984kg} are in coordinate space. In this paper, we will continue to use coordinate space for implementing evolution.
But in most top-down approaches to phenomenology, which are almost always formulated in coordinate ($\Tsc{b}{}$) space,
it is only when $\Tsc{b}{}$ is very large, usually above some specified transition point $\bmax$, that there is an explicit allowance for a nonperturbative transverse coordinate dependence. 
The strategy is often to push reliance on perturbation theory, with its well-understood collinear pdfs, normally considered valid only at small $\Tsc{b}{}$, as far into the large $\Tsc{b}{}$ region as possible and thereby maximize the predictive power supplied by collinear factorization. 
However, unless the functions that parametrize the nonperturbative $\Tsc{b}{} > \bmax$ region are constructed very carefully, there can be undesired side effects, particularly if the nonperturbative $\Tsc{b}{}$-dependent contribution becomes non-negligible. 
In the original setup of TMD factorization, $\bmax$ marks a totally arbitrary line of demarcation between what are labeled ``large/nonperturbative'' and ``small/perturbative'' transverse coordinate space regions. Formally, $\bmax$-dependence vanishes exactly from any physical observables. (See, for example, our discussion in section 13.10.4 of \cite{Collins:2011qcdbook}.) A setup like this ensures that purely perturbative calculations are sequestered from dependence on nonperturbative parameters.
However, in practical implementations there is an unsettling tendency for $\bmax$-dependence to propagate into collinear perturbative calculations~\cite{Qiu:2000hf} and affect final results. Thus, purely perturbative calculations can
appear to get contaminated by nonperturbative modeling. 
A significant $\bmax$-dependence in a calculation is a symptom that something is likely non-optimal about the underlying approximations being used. 
Part of the difficulty is that a valid description over the whole range of $\Tsc{b}{}$ needs to interpolate between the perturbative and nonperturbative regions, and varying $\bmax$ amounts to simply reshuffling contributions between factors.
 In many practical applications, however, there is a rather abrupt switch between a purely collinear perturbation-theory  calculation and one that involves a nonperturbative ansatz.  
 We will show how switching from a top-down to a bottom-up perspective clarifies the steps that are needed to minimize $\bmax$-dependence in the parametrizations used in phenomenological applications. 
 Indeed, we will find that we are able to avoid the usual $\bstarsc$-prescription entirely, along with a $\bmax$, and rely instead only on the input hard scale $Q_0$ to demarcate any transition between large and small transverse momentum. 

Part of the work that is necessary in setting up reasonable phenomenological TMD parametrizations involves ensuring that they retain at least an approximate hadron structure interpretation. 
For example, the probability density interpretation of an unpolarized quark TMD pdf $f_{i/p}(x,\Tsc{k}{})$ gives
\begin{equation}
\label{e.intrel}
\int \diff{^2\T{k}{}} f_{i/p}(x,\T{k}{}) \approx f_{i/p}(x) \, 
\end{equation}
where $i$ is the flavor of a quark in hadron $p$ and $f_{i/p}(x)$ is the corresponding collinear pdf. We use the ``$\approx$'' symbol here because the actual transverse momentum integral in \eref{intrel} is ultraviolet (UV) divergent in QCD and needs to be regulated. After accounting for the UV divergence (by renormalizing the operator definition of the pdf, for example, and introducing regulators), the equality only holds up to order $\alpha_s$ and power-suppressed corrections. 

The role that identities like~\eref{intrel} play in constraining nonperturbative transverse momentum dependence is often unclear in implementations of the CSS and related formalisms. The main reason is that the relation seems to be satisfied automatically in the way that final CSS cross section formulas normally are usually presented. To see how this happens, first consider the TMD pdf expressed in terms of its transverse coordinate space version, $\tilde{f}_{i/p}(x,\T{b}{})$
\begin{equation}
\label{e.bspaceTMDpdf}
f_{i/p}(x,\T{k}{}) = 
\frac{1}{(2 \pi)^2} \int \diff{^2\T{b}{}}{} e^{i \T{k}{} \T{b}{}}
\tilde{f}_{i/p}(x,\T{b}{}) \, .
\end{equation}
The UV divergence in~\eref{intrel} manifests itself in \eref{bspaceTMDpdf} as a divergence in $\tilde{f}_{i/p}(x,\T{b}{})$ as $\Tsc{b}{} \to 0$, so 
it may be regulated by freezing the $\T{b}{}$ argument in $\tilde{f}_{i/p}(x,\T{b}{})$ below some very small ${\boldsymbol{b}}_\text{min}$. In that case, the integral of \eref{intrel} applied to \eref{bspaceTMDpdf} is just 
\begin{equation}
\label{e.intrel2}
\int \diff{^2\T{k}{}} f_{i/p}(x,\T{k}{}) = \tilde{f}_{i/p}(x,{\boldsymbol{b}}_\text{min}) \, .
\end{equation}
In the next step, one observes that an operator product expansion (OPE) applies to 
$\tilde{f}_{i/p}(x,{\boldsymbol{b}}_\text{min})$ in the limit of small ${\boldsymbol{b}}_\text{min}$ and relates it to the collinear pdf $f_i(x)$,
\begin{equation}
\label{e.OPEintro}
\tilde{f}_{i/p}(x,{\boldsymbol{b}}_\text{min}) = \sum_j \mathcal{C}_{i/j} \otimes f_{j/p} = f_i(x) + \order{\alpha_s} + \text{p.s.} \, .
\end{equation}
The $C_{i/j}$ is a perturbative hard coefficient,  $\mathcal{C}_{i/j} \otimes f_{j/p}$ is the usual convolution integral of collinear factorization, the sum is over parton flavors, and ``$\text{p.s.}$'' refers to power suppressed corrections. The perturbation theory expansion on the right hand side is optimized by choosing renormalization scales appropriately in relation to ${\boldsymbol{b}}_\text{min}$.
So, dropping small corrections from the right side of \eref{intrel2}/\eref{OPEintro} recovers \eref{intrel}. 

At first sight, the steps leading from \eref{intrel2} to \eref{OPEintro} can appear to constitute a \emph{derivation} of \eref{intrel}. 
The normal presentations (for example, Eq.~(22) of~\cite{Collins:2014jpa}) of the evolved CSS cross sections amplify that impression because they almost always include the transition to the OPE in \eref{OPEintro} at small $\T{b}{}$ automatically in the final cross section expressions. 
From \eref{OPEintro} above, it then appears that any nonperturbative modeling of large $\T{b}{}$ behavior can only impact the right hand side of \eref{intrel} through power suppressed terms that are mostly irrelevant.

However, \eref{OPEintro} is itself a restatement of \eref{intrel}, not a derivation of it. That is, a derivation  of~\eref{OPEintro} requires that one identify and factorize integrals over the small, nonperturbative region of parton transverse momentum (i.e., the core contribution to the TMD pdf) and identify them with contributions to a collinear pdf, just as in \eref{intrel}. Therefore, a derivation of \eref{intrel} that refers back to \eref{OPEintro} as its starting point is circular. Using \eref{OPEintro} assumes that a version of \eref{intrel} already holds nonperturbatively. 

There are practical consequences of this for setting up nonperturbative parametrizations of TMD correlations functions. One discards an important and useful constraint from the hadron structure interpretation of the nonperturbative $f_{i/p}(x,\T{k}{})$ if one assumes that \eref{intrel} holds automatically on the basis of the OPE alone. 
By doing so, one is effectively assuming that, from the outset, all important regions of the integrand on the left side of \eref{intrel} are at such large $\Tsc{k}{}$ that they are expressible entirely in terms of collinear pdfs, and that the intrinsic transverse momentum is completely irrelevant insofar as the integral is concerned. But there are many practical situations in TMD physics where that is obviously not the case. For example, at moderate scales where $f_{i/p}(x,\T{k}{})$ can be successfully modeled phenomenologically by a Gaussian distribution, with no collinear perturbative tail present at all, the entire range of the integration is described by a nonperturbative model. 
Thus, it is a requirement that \eref{intrel} be imposed explicitly as a constraint from the outset if the goal is to maintain a connection between the nonperturbative TMD parametrizations of hadron structure interpretations and the corresponding collinear functions.  

Many past TMD parton model approaches to Type I phenomenology do of course impose a version of \eref{intrel} on TMD pdfs, but they typically do not include the transition to perturbative transverse momentum at large transverse momentum (or small $\T{b}{}$ as in \eref{OPEintro}). Gaussian fits in momentum space are the most prominent examples.  
The reverse limitation is true of many CSS and similar approaches; these approaches account appropriately for large transverse momentum behavior, but they rarely (if ever) impose conditions analogous to \eref{intrel} directly on the models of nonpeturbative transverse momentum dependence.  
We will show in this paper how adopting a bottom-up perspective makes it clearer how to do both simultaneously, and we will give special attention to integral relations like \eref{intrel} and expand on the discussions above. 

It is worth noting that similar issues also appear in other formalisms that deal with parton transverse momentum, such as the small-$x$ oriented unintegrated pdfs of the Kimber-Martin-Ryskin-Watt~\cite{Kimber:2001sc,Watt:2003mx} treatments~\cite{Guiot:2022psv,Golec-Biernat:2018hqo,Guiot:2019vsm}.

The issues that are clarified by a bottom-up perspective are interrelated, and the two just discussed will turn out to be examples of that. 
On page 2, we mentioned how failing to interpolate explicitly between nonperturbative and perturbative transverse momentum dependence can lead to undesired $\bmax$-dependence in perturbative calculations. But, as we will see, setting up that interpolation involves the use of relations like~\eref{intrel}. Conversely, imposing a version of \eref{intrel} on small transverse momentum descriptions, in a way that accounts for a cutoff around $\Tsc{k}{} \approx Q_0$, naturally requires a careful treatment of the large $\Tsc{k}{} \approx Q_0$ region, and thus the interpolation to perturbatively large $\Tsc{k}{}$ becomes relevant.

Very broadly speaking, switching to a bottom-up approach helps with the inverse problem that arises from transforming between transverse momentum and coordinate space (see also~\cite{Ebert:2022cku}).  
Performing exact Fourier transformations to momentum space requires exact knowledge about the full range of the transverse coordinate space correlation functions. But actual measurements only probe specific limited ranges of $\Tsc{b}{}$. Very large $Q$ measurements are only sensitive to small transverse sizes and so they poorly constrain the nonpertrubative transverse coordinate dependence at very large $\Tsc{b}{}$. As a consequence, fits evolved from large $Q$ down to moderate $Q$ can be unstable. For instance,
\cite{Sun:2013dya,Sun:2013hua} pointed out that the nonperturbative evolution extracted from the earlier CSS fits, which were frequently dominated by very large $Q$ data, is too rapid when it is used to evolve down to the $Q$ of only a few GeVs that are relevant for hadron structure studies. Conversely, measurements done at a more moderate $Q \approx Q_0$ might be dominated by nonperturbatively large $\Tsc{b}{}$ (or small $\Tsc{k}{}$), but they are usually insensitive to the TMD correlation functions' very small $\Tsc{b}{} \ll 1/Q_0$ (or large $\Tsc{k}{} \gg Q_0$) behavior. The latter, however, do not generally need to be extracted but are instead calculable in terms of collinear correlation functions.
So, a more stable phenomenological strategy is to first emphasize the role of more moderate $Q$ measurements for constraining input nonperturbative transverse momentum and then rely on perturbation theory to evolve to larger $Q$.  

Another aspect of TMD phenomenology that will be easier to control in a bottom-up approach is the matching to a momentum space, large transverse momentum asymptotic term. That step is important for relating the small $\Tsc{q}{} \ll Q$ treatment supplied by TMD factorization to the standard momentum space $\Tsc{q}{} \approx Q$ treatments in collinear factorization. Agreement between the two types of calculations at large $\Tsc{q}{}$ provides an important consistency test.
In practice, however, ensuring that the asymptotic term and the TMD factorization calculations match can be difficult at moderate $Q$. That is especially the case near the lowest scales, around $Q_0 \approx 1-2$~GeV, that are conventionally accepted as reasonable input scales.
The difficulties can be largely traced back to those that we have discussed so far in this introduction. Errors that grow as $Q$ evolves downward propagate, at least to some degree, to all values of transverse momentum, including the large transverse momentum. The range of accessible $\Tsc{q}{}$ shrinks, so there is a less well-defined separation between different transverse momentum regions. 

We have so far emphasized a dichotomy between bottom-up/top-down approaches to phenomenology to highlight steps that continue to be a challenge when merging nonperturbative TMD parton model descriptions with TMD factorization and evolution in phenomenology, and to provide readers a basic foundation from which to read the rest of this paper. We reassure readers who find the discussion to be rather vague so far that it likely will become much clearer in the main body of this paper; some of the points noted above require explicit examples in order to be made entirely clear. We will provide some such examples within the main text. 

Another way to understand the goals of this paper is to recall that
nonperturbative transverse momentum dependence enters the standard  CSS implementations in the form of extra coordinate space factors, often expressed as exponential factors $e^{-g_{j/A}(x,\Tsc{b}{})}$ for a flavor $j$ in hadron $A$ or $e^{-g_K(\Tsc{b}{}) \ln \frac{Q}{Q_0}}$ for the nonperturbative TMD evolution. (For example, consider again Eq.~(22) of Ref.~\cite{Collins:2014jpa}.) Our purpose is to explain, in much more detail than is usual, how to optimally parametrize these functions in phenomenological applications, such that they match to existing models or calculations of intrinsic nonperturbative transverse momentum dependence. Nothing about the standard TMD factorization formalism itself will change. The final outcome will be a recipe for merging arbitrary nonperturbative models of $g_{j/A}(x,\Tsc{b}{})$ and $g_K(\Tsc{b}{})$ with TMD evolution in the CSS style. To our knowledge, this is the first instance of such a discussion. We will connect the bottom-up strategy with more standard presentations that involve $g_{j/A}(x,\Tsc{b}{})$ and $g_K(\Tsc{b}{})$ in \sref{css}. Some of our goals overlap with those recently addressed in \cite{Ebert:2022cku}, which introduced techniques for disentangling large and small $\Tsc{b}{}$ contributions to cross sections, and these techniques will likely be helpful in future efforts to identify and separate perturbative and nonperturbative contributions. 

To further prepare the reader, we end this introduction by listing the 
main aspects of what we will advocate that differ from what is sometimes done:
\begin{itemize}
\item When we discuss phenomenology at an input scale $Q_0$, we will work mainly in transverse momentum space rather than in coordinate space.
\item We will nevertheless account for the perturbative behavior at $\Tsc{q}{} \gg Q_0$ in TMD correlation functions. We will do this even for $Q \approx Q_0$ where such contributions are often  phenomenologically irrelevant. 
\item For the region where $Q \approx Q_0$, $\Tsc{q}{} \approx Q_0$, all renormalization scales will be fixed at order $Q_0$ and not $1/\Tsc{b}{}$ since there is no need for resummation of logarithms of the type $\ln (\Tsc{q}{}/Q_0)$ or $\ln (\Tsc{b}{} Q_0)$. 
\item We will nevertheless perform evolution to $Q \gg Q_0$ in coordinate space, and switch to a $\sim 1/\Tsc{b}{}$ scale, but only at a later step. 
\item Also for $Q \approx Q_0$, we will explicitly impose approximate matching to the fixed order asymptotic behavior at $\Tsc{q}{} \approx Q_0$.
\item In nonperturbative parametrizations of TMD correlation functions at $Q = Q_0$, we will explicitly impose a version of \eref{intrel}. We will do this in momentum space with a large transverse momentum cutoff to regulate UV divergences and to match with what is typically done in phenomenological models.
\item Our approach will include an explicit interpolation between purely nonperturbative and purely perturbative descriptions of transverse momentum dependence in the TMD correlation functions.
\end{itemize}
Along the way, we will highlight the advantages of these choices by using explicit examples. At the end, we will translate the expressions into forms that are more familiar from standard TMD factorization implementations in the CSS formalism. 

We will start by reminding the reader of the basic setup of TMD factorization and evolution in \sref{tmd_fact}. Section~\ref{s.tmd_integrals} discusses the role of integral 
constraints for nonperturbative parametrizations of TMD correlation functions in more detail. 
Sections~\ref{s.cskernel} and \ref{s.parameterization} will focus on the details of modeling nonperturbative parts of TMD functions at an input scale $Q_0$ and in momentum space.
The steps for using input parametrizations to evolve to higher scales are summarized in a practical phenomenological recipe  in~\sref{summary}.  
In~\sref{low_order} we use concrete toy examples to illustrate the steps, including plots. In~\sref{int_rels_2} we return the integral relation and discuss it in light of the bottom-up approach. In~\sref{css} we explain how to connect the bottom-approach to standard CSS expressions. We offer our concluding remarks in~\sref{conclusion}.

Explaining some intermediate steps will require a more pedantic system of notation than what is normally necessary. 
To help the reader keep track of symbols and conventions, we have therefore included a notation glossary in \aref{notglossary}. 

\section{TMD Factorization and Evolution}
\label{s.tmd_fact}

We begin by reviewing some of the basic setup of TMD factorization to establish context and introduce 
notation for later sections. While all of what we discuss is meant to apply to any of the basic processes for which there are TMD factorization theorems, it will be instructive to work within a specific example. For this we will 
use semi-inclusive annihilation (SIA) of a lepton-antilepton pair (usually electron-positron) into a pair 
of nearly back-to-back hadrons with a sum over all other final state particles X,
\begin{eqnarray}
e^-(l)+e^+(\bar{l})\to H_A(p_A)+H_B(p_B)+X \, .
\end{eqnarray}
A quark-antiquark pair is produced in the hard vertex, and hadrons $H_A$ and $H_B$ are measured in the final state.
This is among the simplest processes to work with theoretically, and it is ideal for illustrating the basics of TMD factorization.
(See, for example, the discussion in chapter 13 of Ref.~\cite{Collins:2011qcdbook}.)

\begin{figure}[h!]
\centering
\includegraphics[width=8cm]{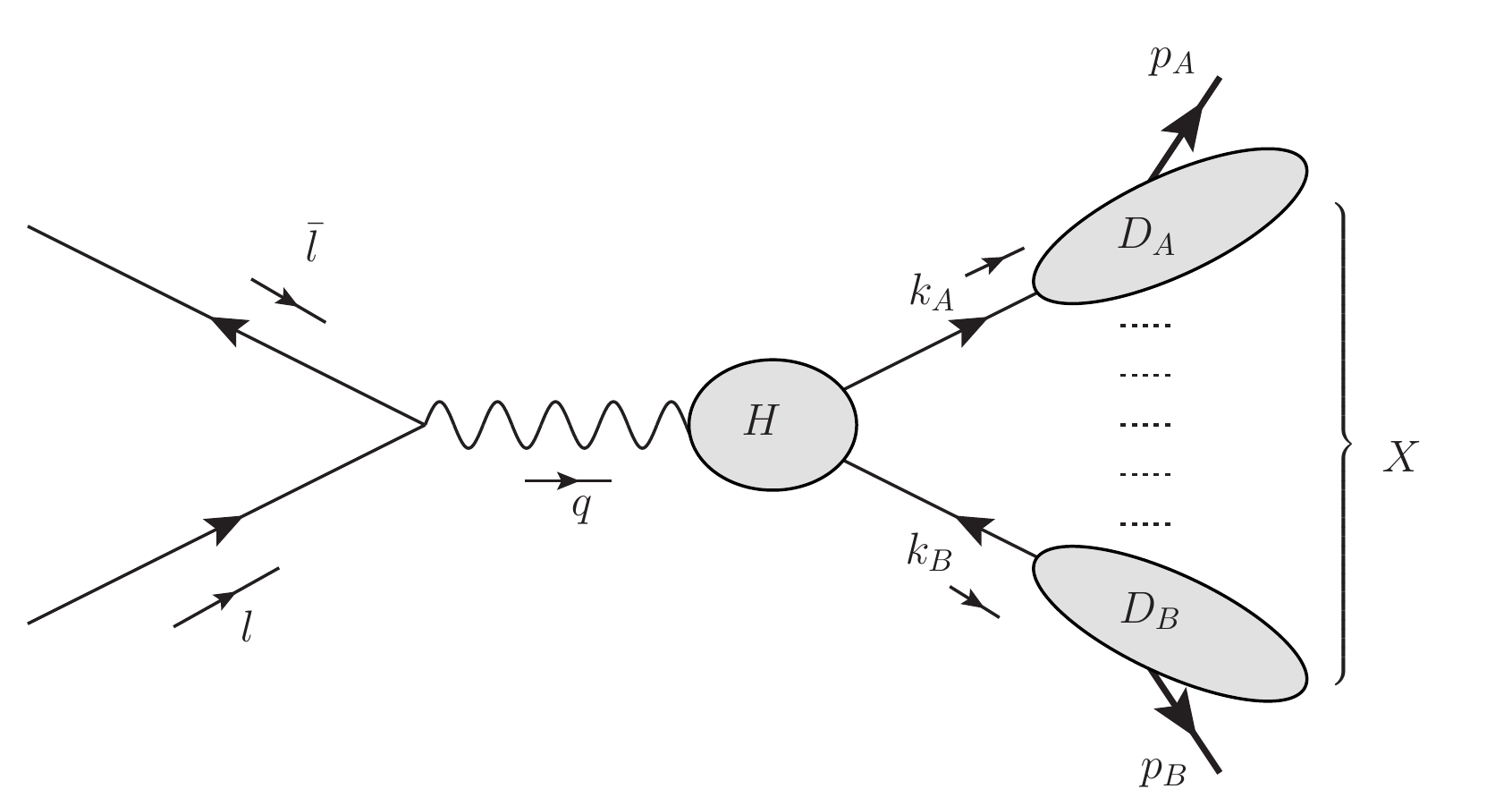}
\caption{Schematic of the semi-inclusive $e^+e^-$-annihilation process. }
\label{fig.Process}
\end{figure}

The process is illustrated graphically in Figure \ref{fig.Process}:  An electron ($l$) and a positron ($\bar{l}$) annihilate to create a virtual photon of momenta $q$, which creates a quark-antiquark pair. The  two hadrons measured in the final state with momenta $p_A$ and $p_B$ are then produced when partons A and B fragment. The momentum of the virtual photon sets the hard scale of the process $Q$, with $q^2\equiv Q^2$. See also Refs.\cite{Boer:2008fr,Boer:1997mf,Moffat:2019pci} for more details about the general kinematical setup.

In a reference frame where the hadrons are back-to-back, the transverse momentum of the photon $\T{q}{}$ is the relevant observed final state transverse momentum. When it is small relative to the hard scale, $\Tsc{q}{} \ll Q$, it is sensitive to intrinsic transverse momenta of the hadronizing quark and antiquark respectively.
The usual Lorentz invariant kinematic variables related to collinear momentum fractions are 
\begin{eqnarray}
z_A=\dfrac{p_A\cdot p_B}{q\cdot p_B} \approx \dfrac{p_{A,h}^+}{q_h^+},
\hspace{.5cm}
z_B=\dfrac{p_A\cdot p_B}{q\cdot p_A} \approx \dfrac{p_{B,h}^-}{q_h^-} \, 
\end{eqnarray}
where the ``$\approx$'' means we drop terms that are power suppressed in the current fragmentation region (by which we mean $z_A$ and $z_B$ are fixed and not too small relative to 1). The ``$h$'' subscripts on lightcone momentum components indicate that they are with respect to the hadron frame. 

In TMD factorization, the unpolarized cross section differential in 
$z_A$, $z_B$ and $\Tscsq{q}{}$ is written~\cite{Collins:2011qcdbook}
\begin{widetext}
\begin{align}
Q^2 & \frac{\diff{\sigma^{A,B}}}{\diff{z_A}\diff{z_B}\diff{\Tscsq{q}{}}} \no &{}= H_{j\bar{j}} (\mu_Q;C_2) \int \diff{^2\T{k}{A}}{}
\diff{^2\T{k}{B}}{}
D_{j/A}\parz{z_A,z_A \T{k}{A};\mu_Q,Q^2} 
D_{\bar{j}/B}\parz{z_B,z_B \T{k}{B};\mu_Q,Q^2}
\delta^{(2)}\parz{\T{q}{} - \T{k}{A} - \T{k}{B}} + \no &{} \qquad  Y^{A,B}(\Tsc{q}{},Q;\mu_Q) + \order{m/Q} \, .
    \label{e.tmd_factorization_momspace_a}
\end{align}
\end{widetext} 
The second line has the familiar form from the TMD parton model, but with extra auxiliary arguments for evolution. The capital $D_{j/A}$ and $D_{\bar{j}/B}$ are the TMD ffs for a quark of flavor $j$ ($\bar{j}$) to fragment into hadron $A$ ($B$). A sum over flavors is implied. 

In addition to the longitudinal and transverse parton momentum arguments $z_{A,B}$ and $\T{k}{A,B}$, the TMD ffs also depend on a renormalization group scale $\mu$ and a rapidity evolution scale $\zeta$, which in \eref{tmd_factorization_momspace_a} we have already fixed equal to $\mu = \mu_Q \equiv C_2 Q$ and $\zeta = Q^2$ to optimize perturbation theory. Here, $C_2$ is an arbitrary numerical constant of order unity. (Throughout this paper, we will assume $C_2 =1$.) $H(\mu_Q;C_2)$ is a hard factor of the form $H = 1 + \order{\alpha_s\parz{\mu_Q}}$, up to an uninteresting overall constant. The $Y(\Tsc{q}{},Q)$-term on the last line is an abbreviation for the correction needed for the $\Tsc{q}{} \approx Q$ behavior, and it is calculable entirely in fixed order collinear factorization. The second line in \eref{tmd_factorization_momspace} is exactly the TMD parton model familiar from typical Type I applications if we drop the auxiliary $\mu$ and $\zeta$ arguments and set $H(\mu_Q;C_2) = 1$. 

We will focus on a very specific combination of physical observables in order to simplify later illustrative examples. Say that hadron $A$ is $h^+$ and hadron $B$ is its antiparticle $h^-$. Then we can consider the combination
\begin{align}
&{} \frac{\diff{\sigma^\text{NS}}}{\diff{z_A}\diff{z_B}\diff{\Tscsq{q}{}}}
 = \no
&{} \qquad \frac{\diff{\sigma^{h^+,h^-}}}{\diff{z_A}\diff{z_B}\diff{\Tscsq{q}{}}} +
\frac{\diff{\sigma^{h^-,h^+}}}{\diff{z_A}\diff{z_B}\diff{\Tscsq{q}{}}} \no
&{} \qquad
-
\frac{\diff{\sigma^{h^+,h^+}}}{\diff{z_A}\diff{z_B}\diff{\Tscsq{q}{}}} 
-
\frac{\diff{\sigma^{h^-,h^-}}}{\diff{z_A}\diff{z_B}\diff{\Tscsq{q}{}}} \, .
\end{align}
We will also consider only the $j = \text{``up quark"}$ contribution to \eref{tmd_factorization_momspace_a}.
Then, summing the corresponding terms on the right hand side of \eref{tmd_factorization_momspace_a} gives
\begin{widetext}
\begin{align}
 &{} H_{u\bar{u}} (\mu_Q;C_2) \int \diff{^2\T{k}{A}}{}
\diff{^2\T{k}{B}}{}
\left[ D_{u/h^+}\parz{z_A,z_A \T{k}{A};\mu_Q,Q^2} - D_{u/h^-}\parz{z_A,z_A \T{k}{A};\mu_Q,Q^2} \right] \no
&{} \qquad \times \left[ D_{\bar{u}/h^-}\parz{z_B,z_B \T{k}{B};\mu_Q,Q^2} - D_{\bar{u}/h^+}\parz{z_B,z_B \T{k}{B};\mu_Q,Q^2} \right]
\delta^{(2)}\parz{\T{q}{} - \T{k}{A} - \T{k}{B}} \no &{} \qquad  + Y^{\text{NS}}(\Tsc{q}{},Q;\mu_Q) + \order{m/Q} \, .
    \label{e.tmd_factorization_momspace_b}
\end{align}
Then we can define non-singlet TMD fragmentation functions,
\begin{align}
\label{e.nsff1}
D_A\parz{z_A,z_A \T{k}{A};\mu_Q,Q^2} &{} \equiv D_{j/h^+}\parz{z_A,z_A \T{k}{A};\mu_Q,Q^2} - D_{j/h^-}\parz{z_A,z_A \T{k}{A};\mu_Q,Q^2} \\
\label{e.nsff2}
D_B\parz{z_B,z_B \T{k}{B};\mu_Q,Q^2} &{}
\equiv
 D_{\bar{j}/h^-}\parz{z_B,z_B \T{k}{B};\mu_Q,Q^2} - D_{\bar{j}/h^+}\parz{z_B,z_B \T{k}{B};\mu_Q,Q^2} \, .
\end{align}
And, we can drop the $j$ index for the rest of this paper and rewrite \eref{tmd_factorization_momspace_a} in a more abbreviated way as
\begin{align}
Q^2 & \frac{\diff{\sigma^\text{NS}}}{\diff{z_A}\diff{z_B}\diff{\Tscsq{q}{}}} \no &{}= H(\mu_Q;C_2) \int \diff{^2\T{k}{A}}{}
\diff{^2\T{k}{B}}{}
D_{A}\parz{z_A,z_A \T{k}{A};\mu_Q,Q^2} 
D_{B}\parz{z_B,z_B \T{k}{B};\mu_Q,Q^2}
\delta^{(2)}\parz{\T{q}{} - \T{k}{A} - \T{k}{B}}\no &{} \qquad  + Y^\text{NS}(\Tsc{q}{},Q;\mu_Q) + \order{m/Q} \, .
    \label{e.tmd_factorization_momspace}
\end{align}
\end{widetext}
Our results are general and independent of the specific hadrons in the final state, but organizing the discussion around this channel will simplify illustrative example calculations later on by allowing us to drop explicit flavor indices and consider only non-singlet ffs in parts of calculations that involve collinear DGLAP evolution. 
Specifically, in our 
examples we will use the JAM20 collinear fragmentation functions for $\pi^+$ and its corresponding grid for $\alpha_s$ values \cite{Moffat:2021dji}. Note that because of charge conjugation, to construct the TMDs in \eref{nsff1} and \eref{nsff2}, only the non-singlet combination $u-\bar{u}$ are involved. We use the set$=0$, from LHAPDF\cite{Buckley:2014ana}, since we will not be making any comparison to data. All that is needed for our present purposes is that the collinear ffs employed in our numerical examples obey non-singlet DGLAP evolution equations. 

Since our focus for this paper is on the $\Tsc{q}{} \ll Q$ region at leading power, where TMD correlation functions are relevant, we will also drop any mention of the $Y(\Tsc{q}{},Q)$-term from here forward.  

The second line in \eref{tmd_factorization_momspace} is exactly the TMD parton model familiar from typical Type I applications if we drop the auxiliary $\mu$ and $\zeta$ arguments and set $H(\mu_Q;C_2) = 1$. This term is sometimes called the $W$-term. 

It can be convenient to write the $W$-term in terms of coordinate space TMD ffs,
\begin{widetext}
\begin{align}
W(\Tsc{q}{},Q) &{}=  
   H(\mu_Q;C_2) 
    \int \frac{\diff[2]{\T{b}{}}}{(2 \pi)^2}
    ~ e^{-i\T{q}{}\cdot \T{b}{} }
    ~ \tilde{D}_{A}(z_A,\T{b}{};\mu_Q,Q^2) 
    ~ \tilde{D}_{B}(z_B,\T{b}{};\mu_Q,Q^2) \, ,
    \label{e.tmd_factorization}
\end{align}
\end{widetext}
where the transverse coordinate and momentum space TMD ffs are related to one another via
\begin{equation}
\label{e.FTM}
D\parz{z,z \T{k}{};\mu,\zeta}
= \int \frac{\diff{^2\T{b}{}}}{(2 \pi)^2}
e^{-i \T{k}{} \cdot \T{b}{}}
\tilde{D}(z,\T{b}{};\mu,\zeta) \, .
\end{equation}
The TMD ffs $D_{A}$ and $D_{B}$ are hadron-vacuum correlation functions defined in a very particular way in terms of quark field operators. An explanation of these definitions and their origins in factorization theorems would involve topics like Wilson lines, rapidity divergences, soft factors, and other related issues that are beyond the scope of this article. In recent years, they have been reviewed in many places, so to avoid repetition we refer the reader to~\cite{Rogers:2015sqa} which includes an overview and a list references associated with TMD definitions derived in the style relevant for this article, and also to \cite{Diehl:2015uka} which includes  additional relevant references. 

With regard to the TMD definitions, what is important for our purposes is only that they satisfy an exact set of evolution equations for the auxiliary variables $\mu$ and $\zeta$. For a TMD ff in coordinate space, the evolution equations are
\begin{align}
\label{e.CSPDF}
\frac{\partial \ln \tilde{D}(z,\T{b}{};\mu,\zeta)}{\partial \ln \sqrt{\zeta} } &{}= \tilde{K}(\Tsc{b}{};\mu) \, , \\
\label{e.RG}
\frac{\diff{\tilde{K}(\Tsc{b}{};\mu)}}{\diff{\ln \mu}{}} &{}=  - \gamma_K(\alpha_s(\mu))  \, , \\
\label{e.RGgamma}
\frac{\diff{\ln \tilde{D}(z,\T{b}{};\mu,\zeta)}}{\diff{\ln \mu}} &{}=  \gamma(\alpha_s(\mu);\zeta /\mu^2) \no
= \gamma(\alpha_s(\mu);1) &- \gamma_K(\alpha_s(\mu)) \ln \parz{\frac{\sqrt{\zeta}}{\mu}} \, . 
\end{align}
The evolution kernels, $\gamma$, $\gamma_K$ and $\tilde{K}$ are known by many 
different names in the literature.  In keeping with our earlier work, we will refer to $\tilde{K}(\Tsc{b}{};\mu)$ as the Collins-Soper (CS) kernel. 
 
For large enough $\mu$, the anomalous dimensions $\gamma_K$ and $\gamma$ are both calculable in perturbation theory and they are independent of $\Tsc{b}{}$. $\tilde{K}(\Tsc{b}{};\mu)$ is also perturbatively calculable at small $\Tsc{b}{} \sim 1/\mu$, but it becomes nonperturbative over large distances, $\Tsc{b}{} \to \infty$. However, since it is independent of the identity of external hadrons, it has very strong universality properties related to the QCD vacuum. 

The evolution equations are first order linear differential equations that relate the TMD ffs in \eref{tmd_factorization} at an input scale $Q_0$ to a different scale $Q$, so they can be easily solved exactly and analytically. The general solutions to the evolution equations for $D_A$ and $D_B$, substituted into \eref{tmd_factorization}, allow us to write $W(\Tsc{q}{},Q)$ as
\begin{widetext}
\begin{align}
W(\Tsc{q}{},Q)
&{}=
   H(\alpha_s(\mu_Q);C_2) 
    \int \frac{\diff[2]{\T{b}{}}}{(2 \pi)^2}
    ~ e^{-i\T{q}{}\cdot \T{b}{} }
    ~ \tilde{D}_{A}(z_A,\T{b}{};\mu_{Q_0},Q_0^2) 
    ~ \tilde{D}_{B}(z_B,\T{b}{};\mu_{Q_0},Q_0^2) \nonumber\\&
  \,\times
  \exp\left\{  
        \tilde{K}(\Tsc{b}{};\mu_{Q_0}) \ln \parz{\frac{ Q^2 }{ Q_0^2}}
           +\int_{\mu_{Q_0}}^{\mu_Q}  \frac{ \diff{\mu'} }{ \mu' }
           \biggl[ 2 \gamma(\alpha_s(\mu'); 1) 
                 - \ln\frac{Q^2}{ {\mu'}^2 } \gamma_K(\alpha_s(\mu'))
           \biggr]
  \right\} \, .
\label{e.Wtermev0}
\end{align}
\end{widetext}
We could express the same evolved $W$ with TMD ffs in transverse momentum space, and in some cases that might help with intuition since it more closely matches the TMD parton model. 
But then the Fourier transforms of simple factors become transverse momentum convolution integrals of several functions. For the purposes of implementing evolution, we will continue to work with \eref{Wtermev0} in coordinate space, as is normally done. 

When $Q = Q_0$, \eref{Wtermev0} exactly reproduces the TMD parton model form of the factorization, 
\begin{widetext}
\begin{align}
W(\Tsc{q}{},Q_0) &{}=  
   H(\mu_{Q_0};C_2) 
    \int \frac{\diff[2]{\T{b}{}}}{(2 \pi)^2}
    ~ e^{-i\T{q}{}\cdot \T{b}{} }
    ~ \tilde{D}_{A}(z_A,\T{b}{};\mu_{Q_0},Q_0^2) 
    ~ \tilde{D}_{B}(z_B,\T{b}{};\mu_{Q_0},Q_0^2) \, \no
 &{}= H(\mu_{Q_0};C_2) \int \diff{^2\T{k}{A}}{}
\diff{^2\T{k}{B}}{}
D_{A}\parz{z_A,z_A \T{k}{A};\mu_{Q_0},Q_0^2} 
D_{B}\parz{z_B,z_B \T{k}{B};\mu_{Q_0},Q_0^2}
\delta^{(2)}\parz{\T{q}{} - \T{k}{A} - \T{k}{B}} \, .
    \label{e.tmd_factorization_input}
\end{align}
\end{widetext}
So far, \eref{Wtermev0} and \eref{tmd_factorization_input} still involve no approximations at all on the $W$-term. However, approximations are ultimately necessary, of course, for getting practical results. 

Rather different kinds of approximations to \eref{Wtermev0} can enter in a number of different ways, so the language can become confusing. We will be as precise as possible with our wording here. There are at least four different types of approximation that generally take place simultaneously, including
\begin{enumerate}
\item The choices of models, assumptions, or approximations used to describe nonperturbative contributions,
\item The neglect of power suppressed corrections to factorization, like the last term in \eref{tmd_factorization_momspace},
\item Truncation of high powers of $\alpha_s$ in perturbative parts of the calculation and, 
\item In phenomenological applications involving fits, whatever assumptions and approximations are used at the level of fit extractions. 
\end{enumerate}
When we discuss an ``$n^\text{th}$-order'' perturbative treatment, we will mean by this that all parts that are calculable in fixed order perturbative QCD have been included through order $n$, and they are optimized using the renormalization group and Collins-Soper evolution (\eref{CSPDF}). These ``perturbative parts'' include the right sides of \erefs{CSPDF}{RGgamma}, which in \eref{Wtermev0} appear as $\gamma_K(\alpha_s(\mu'))$, $\gamma(\alpha_s(\mu'); 1)$ 
and $\tilde{K}(\Tsc{b}{};\mu_{Q_0})$ when $\Tsc{b}{}$ (or $\Tsc{k}{}$) is small (large) enough that its $\Tsc{b}{}$-dependence is perturbative.  In \eref{Wtermev0} it also includes $H(\mu_Q;C_2)$ and the $\tilde{D}_A$, $\tilde{D}_B$ functions in regions of small $\Tsc{b}{}$ (or large $\Tsc{k}{}$).
An ``$(n)$'' superscript on a function means that it has been replaced by its truncated, fixed order perturbation theory calculation through order $n$. 
When we discuss nonperturbative parts in later sections, it should be understood to be in the context of phenomenological extractions. We will \emph{assume} for the sake of our discussion here that all nonperturbative parametrizations have been made flexible enough that the significant errors come only from the limitations of factorization and truncated perturbation theory, and not from a poor choice of nonperturbative parametrizations or artifacts of the fitting procedure. 
Thus, a function like $\tilde{D}_A^{(n)}$ should be read as ``a $\tilde{D}_A$ parametrization including nonperturbative parts extracted from measurements and using an $n^\text{th}$-order perturbation theory treatment for its perturbative ingredients.''

Also, when we use the phrase ``nonperturbative part,'' (e.g., the $g$-functions of the CSS formalism) it should generally be understood that we are not necessarily referring to  parts of a calculation that cannot ever be improved with small coupling techniques. It only refers to contributions that we choose to exclude from those factors that we explicitly identify as perturbative.    

Now consider how one might use \eref{Wtermev0} to do phenomenology in a Type I scenario and from a bottom-up perspective. Near the input scale, $Q \approx Q_0$, the evolution factor on the second line is nearly unity and we can, to a good approximation, just work with  \eref{tmd_factorization_input}. If $Q_0$ is only of order $\sim 1-2$~GeV, then a phenomenological measurement of $W(\Tsc{q}{},Q_0)$ is only sensitive to the region of nonperturbatively large $\Tsc{b}{}$ parts of $\tilde{D}_{A}$ and $\tilde{D}_{B}$ whereas the $\Tsc{b}{} \ll 1/Q_0$ contributions is strongly suppressed in cross sections. Therefore, measurements of the $Q \approx Q_0$ region place rigid constraints on the nonperturbative transverse momentum dependence while remaining mostly insensitive to perturbative tails from large $\Tsc{k}{A,B}$ (or small $\Tsc{b}{}$). Indeed, past phenomenology with \eref{tmd_factorization_input} has confirmed that the basic TMD parton model is quite successful  at describing moderate $Q$ data with entirely nonperturbative $\Tsc{k}{}$ 
parametrizations like Gaussians~\cite{Anselmino:2002pd,Collins:2005ie,Schweitzer:2010tt,Signori:2013mda}.
Work that is done in this style effectively replaces the exact \eref{tmd_factorization_input} with the approximation 
\begin{widetext}
\begin{align}
W^{(n)}(\Tsc{q}{},Q_0) &{}=  
   H^{(n)}(\mu_{Q_0};C_2) 
    \int \frac{\diff[2]{\T{b}{}}}{(2 \pi)^2}
    ~ e^{-i\T{q}{}\cdot \T{b}{} }
    ~ \tilde{D}_{A,\text{np}}(z_A,\T{b}{};\mu_{Q_0},Q_0^2) 
    ~ \tilde{D}_{B,\text{np}}(z_B,\T{b}{};\mu_{Q_0},Q_0^2) \, ,
    \label{e.tmd_factorization_input2}
\end{align}
\end{widetext}
where the ``$\text{np}$'' subscripts on the TMD ffs refer to Fourier-Bessel transforms of entirely nonperturbative model parametrizations for $D_{A}\parz{z_A,z_A \T{k}{A};\mu_{Q_0},Q_0^2}$ and $D_{B}\parz{z_B,z_B \T{k}{B};\mu_{Q_0},Q_0^2}$ (e.g., Gaussians, etc), with no matching to perturbative large-$\Tsc{k}{}$ tails.  
If most of the $0 \lesssim \Tsc{q}{} \lesssim Q_0$ region of transverse momentum dependence is nonperturbative at $Q \approx Q_0$, then there is nothing more to do at this scale, and replacing $H(\mu_{Q_0};C_2)$ by $H^{(n)}(\mu_{Q_0};C_2)$ in \eref{tmd_factorization_input2} is the maximum amount of perturbative input that is possible. 

Notice that, if we have perfectly constrained the input TMD ffs for all $\Tsc{k}{}$ at the input scale $Q_0$, then evolving them to larger $Q \gg Q_0$ with \eref{Wtermev0} is almost trivial. All that is necessary are 
$\gamma_K(\alpha_s(\mu'))$, $\gamma(\alpha_s(\mu'); 1)$ and $\tilde{K}(\Tsc{b}{};\mu_{Q_0})$ calculated to $n^\text{th}$ order in perturbation theory and a way to perform the inverse Fourier-Bessel transform. Thus, traditional Type I extractions of TMD ffs near an input scale 
might appear at first sight to be all the input that is necessary, aside from a treatment of the evolution kernels, for evolution to any other scale. 

 However, this will fail to work if applied directly to traditional Type I oriented styles of phenomenological extractions that neglect large $\Tsc{k}{}$ tails. The reason is that $D_{A}\parz{z_A,z_A \T{k}{A};\mu_{Q_0},Q_0^2}$ and $D_{B}\parz{z_B,z_B \T{k}{B};\mu_{Q_0},Q_0^2}$ need to be well-described over their \emph{entire} $\Tsc{k}{}$ ranges, not just over the $0 \lesssim \Tsc{k}{} \lesssim Q_0$ regions that moderate $Q_0$ measurements effectively probe, in order for the evolution to larger $Q$ to be accurate. Fortunately, the large-$\Tsc{k}{}$ behavior of TMD ffs is calculable in collinear factorization, so nothing prevents us from simply interpolating between phenomenological fits of the small $\Tsc{k}{}$ region and a large-$\Tsc{k}{}$ perturbative description of the input TMD correlation functions at $\Tsc{k}{} > Q_0$. We will discuss how to make this adjustment to standard Type II oriented approaches in later sections. 

If $Q_0$ is, as we intend, a lower bound on what is considered a reasonable hard scale, near the boundary between large and small coupling, then there is a very rapid scale dependence when a choice of $\mu \propto Q_0$ is varied downward.
Therefore, it is preferable to keep renormalization and rapidity scales fixed around $\mu_{Q_0}$ and $Q_0$ in the moderate $\Tsc{k}{} \approx Q_0$ or $\Tsc{b}{} \approx 1/Q_0$ regions.  
(We elaborate on this slightly subtle point in \aref{scale_setting} to avoid breaking the flow of our main discussion here.)

But in the $\Tsc{k}{} \gg Q_0$ region, large logarithms of the form $\ln(\Tsc{k}{}/Q_0)$ start to degrade the accuracy of such a treatment. Extending calculations to the very large $\Tsc{k}{} \gg Q_0$ region requires another iteration of the evolution equations to switch scales.  
(Actually, when we do this in later sections, it will be in coordinate space as is more common, and we will switch to the usual $\sim 1/\Tsc{b}{}$ scale.) We will explain the details of both the interpolation and the scale transformation in \sref{parameterization}.
The discussion of \eref{logsaa} in the introduction was in reference to these types of logarithmic errors. 

For readers accustomed to more standard presentations of the CSS formalism, the above discussion might seem like a  slightly odd starting point because normally one does not stop with \eref{Wtermev0}. Rather, one immediately performs the step of evolving $\tilde{D}_{A}(z_A,\T{b}{};\mu_{Q_0},Q_0^2)$ and $\tilde{D}_{B}(z_B,\T{b}{};\mu_{Q_0},Q_0^2)$ to  scales of order $ 1/\Tsc{b}{}$ to optimize collinear factorization calculations with the OPE in the region of $\Tsc{b}{} \sim 1/Q$.  
The resulting expressions (see for example Eq.(22) of \cite{Collins:2014jpa}) highlight the role of the $\Tsc{b}{} \sim 1/Q$ calculation. Ultimately, we will perform an equivalent step, but we choose to delay it for now for reasons that we hope will become clear later. The connection between \eref{Wtermev0} and the usual CSS presentation will be made in \sref{css}.

\section{Integral Relations}
\label{s.tmd_integrals}

As we will see in later sections, consistently interpolating between nonperturbatively small $\Tsc{k}{}$ and perturbatively large $\Tsc{k}{}$ will require integral relations like \eref{intrel}. In this section we will define and discuss a collinear ff that obeys an analogous equation with a cutoff on large transverse momentum. This section also introduces additional notation that will be necessary in later sections. We remind the reader of our notation glossary, \aref{notglossary}. 

We will use a superscript ``$(n,d_r)$'' on a TMD ff to indicate that it is calculated in a large ($\Tsc{k}{} \approx Q$) transverse momentum approximation. 
Thus,
\begin{align}
&{}D\parz{z,z \T{k}{} \approx Q;\mu_Q,Q^2} \equiv \no
&{}D^{(n,d_r)}\parz{z,z \T{k}{};\mu_Q,Q^2} + \order{\frac{m}{\Tsc{k}{}},\alpha_s(\Tsc{k}{})^{n+1}} \, . \label{e.colliner_ff_approx}
\end{align}
As usual, ``$n$'' is the order of collinear perturbation theory. Thus, $D^{(n,d_r)}\parz{z,z \T{k}{};\mu_Q,Q^2}$ is calculated \emph{through} order $\alpha_s^n$, with powers of $\alpha_s^{n+1}$ and $\sim m/\Tsc{k}{}$ errors neglected.  However, now we have also included a ``$d_r$'' in the superscript. This is to indicate that the collinear factorization calculation uses a renormalized collinear ff $d_r(z;\mu_Q)$. The subscript ``$r$'' on ``$d_r$'' in turn labels the UV renormalization or regularization scheme (such as, for example, $r = \msbar{}$ renormalization). 
We also define
\begin{equation}
D^{(n,d_r)}\parz{z,z \T{k}{};\mu_Q,Q^2}  \equiv 
0 \;\;\forall \;\;  n < 1 \, .
\end{equation}
The $m/\Tsc{k}{}$ in the error term of \eref{colliner_ff_approx}
symbolizes contributions that are power suppressed when $\Tsc{k}{} \approx Q$. Throughout this paper, an ``$m$'' will always represent any generic mass scale that is of order a small hadronic size like $\Lambda_\text{QCD}$ or an intrinsic transverse momentum. Also, to simplify notation, any power-suppressed contributions of the form $(m/Q)^{\beta}$ or 
$(m/\Tsc{k}{})^{\beta}$, with $\beta > 0$ will always simply be written as $\order{m/Q}$ or $\order{m/\Tsc{k}{}}$, regardless of the power $\beta$.

To summarize, the symbol $D^{(n,d_r)}\parz{z,z \T{k}{};\mu_Q,Q^2}$ is the approximation to an individual TMD ff wherein it is calculated in fixed order collinear perturbation theory, optimized to the region $\Tsc{k}{} \approx Q$ and $Q \to \infty$, and using $d_r$ collinear fragmentation functions. 
The fixed order perturbative expression for $D^{(n,d_r)}\parz{z,z \T{k}{};\mu_Q,Q^2}$ in collinear factorization has the form
\begin{equation}
D^{(n,d_r)}\parz{z,z \T{k}{};\mu_Q,Q^2} = \left[ \mathcal{C}^{(n)}_D(z\Tsc{k}{}) \otimes d_r \right](z;\mu_Q) \, . \label{e.D_fact}
\end{equation}
The ``$\otimes$'' here symbolizes the usual collinear convolution integral,
\begin{equation}
(f\otimes{g})(z;\mu) \equiv \int_{z}^1\frac{\diff\xi}{\xi}f(z/\xi)g(\xi;\mu) \, .
\end{equation}
In \eref{D_fact}, $\mathcal{C}^{(n)}_D(z\Tsc{k}{})$ is a hard coefficient. We have written its $z\Tsc{k}{}$ argument explicitly as a reminder that this particular hard factor has $\Tsc{k}{}$-dependence. 
Approximations to $D\parz{z,z \T{k}{};\mu_Q,Q^2}$ appropriate to regions other than $\Tsc{k}{} \approx Q$ will be left unaddressed for now. They will be discussed in \sref{parameterization}.

The TMD ff is related to a collinear ff by an integral over transverse momentum,
\begin{align}
&{}2 \pi z^2 \int_0^{\mu_Q} \diff{\Tsc{k}{}}{} \Tsc{k}{} 
D(z,z \T{k}{};\mu_Q,Q^2) = d_r(z;\mu_Q)  \no 
&{}
+ \Delta^{(n,d_r)}(\alpha_s(\mu_Q)) + \order{\frac{m}{Q},\alpha_s(\mu_Q)^{n+1}} \, ,
\label{e.cutff}
\end{align}
with the details of the notation to be explained below. 
In a literal probability density interpretation, $\mu_Q$ would be set equal to infinity and the second two terms on the right-hand side would be zero.  
In a renormalizable theory like QCD, the integral needs to be regulated, and corrections are necessary to relate the cutoff integral to collinear ffs defined in standard schemes. The $\Delta^{(n,d_r)}(\alpha_s(\mu_Q))$ term on the right-hand side of \eref{cutff} is our notation for the perturbative correction through $n^\text{th}$-order that relates the cutoff integral to the collinear ff $d_r(z;\mu_Q)$ in scheme $r$. There are also, in general, power-suppressed corrections, as indicated by the error term in \eref{cutff}. The correction term $\Delta^{(n,d_r)}(\alpha_s(\mu_Q))$ is related to collinear ffs via another factorization theorem,
\begin{equation}
\Delta^{(n,d_r)}(\alpha_s(\mu_Q)) = \left[ \mathcal{C}^{(n)}_\Delta \otimes d_r \right](z;\mu_Q) \, . \label{e.delta_fact}
\end{equation}
and $C_\Delta^{(n)}$ is an order-$\alpha_s(\mu_Q)^n$ hard coefficient, with a $\Delta$ subscript included here to distinguish it from the $\Tsc{k}{}$-dependent hard coefficient in \eref{D_fact}. The ``$(n,d_r)$'' superscript in $\Delta^{(n,d_r)}(\alpha_s(\mu_Q))$ is to  symbolize that \eref{delta_fact} is to be calculated through order $\alpha_s(\mu_Q)^n$, and that the collinear ff is defined in the $r$ renormalization and/or regularization scheme. Note that $\Delta^{(n,d_r)}(\alpha_s(\mu_Q))$ is also a function of $z$ and $Q$, but we have dropped explicit dependence on those variables to maintain as compact a notation as possible. We also define
\begin{equation}
\Delta^{(n,d_r)}(\alpha_s(\mu_Q)) \equiv 
0 \;\;\forall \;\;  n < 1 \, .
\end{equation}

To make the above more explicit, let us \emph{define} a new collinear ff that is the transverse momentum integral of a TMD ff regulated with a cutoff on all $\Tsc{k}{} > \mu_Q$:
\begin{equation}
d_c(z;\mu_Q) \equiv 2 \pi z^2 \int_0^{\mu_Q} \diff{\Tsc{k}{}}{} \Tsc{k}{} D(z,z \T{k}{};\mu_Q,Q^2) \, . \label{e.dc_deff}
\end{equation}
The $r = c$ subscript on the left indicates that this is an ff defined in the ``cutoff'' scheme.\footnote{For the renormalization of the TMD ff in the integrand of \eref{dc_deff}, it is to be understood that the scheme is a standard one like $\msbar$.} 
Equation~\eqref{e.dc_deff} is just the left side of \eref{cutff}. Dropping the power-suppressed and order-$\alpha_s^{n+1}(\mu_Q)$ terms on the right side of \eref{cutff} gives an equation that is satisfied only approximately. 
To give this a notation, we also define
\begin{equation}
d^{(n,d_r)}_c(z;\mu_Q) \equiv d_r(z;\mu_Q) 
+ \Delta^{(n,d_r)}(\alpha_s(\mu_Q)) \, . \label{e.dnr_rel}
\end{equation}
Then, \eref{cutff} is
\begin{equation}
d^{(n,d_r)}_c(z;\mu_Q) - d_c(z;\mu_Q) =  \order{\frac{m}{Q},\alpha_s(\mu_Q)^{n+1}} \, , \label{e.dc_err}
\end{equation} 

If the scheme for dealing with UV transverse momentum divergences is the cutoff scheme itself, $r = c$, then 
\begin{equation}
\Delta^{(n,d_c)}(\alpha_s(\mu_Q)) = 0
\end{equation}
exactly by definition.
If the scheme is modified minimal subtraction, $r = \msbar$, then 
\begin{align}
&{} \Delta^{(n,d_\text{\msbar})}(\alpha_s(\mu_Q)) = \frac{\alpha_s(\mu_Q)}{2 \pi}  \times \no
&{} \times \int_{z}^1 \frac{\diff{z'}{}}{z^{'}}
d_{\msbar}(z/z';\mu_Q) \left[ 2 P_{qq}(z') \ln z'  + C_F (1 - z')  \right] \, \no
&{} \qquad + \order{\alpha_s(\mu_Q)^2} \, ,
\label{e.msbar_trans}
\end{align}
where we have used the standard splitting function notation
\begin{equation}
P_{qq}(z)=P_{\bar{q}\bar{q}}(z)=C_F\left[\frac{1+z^2}{\parz{1-z}_+}+\frac{3}{2}\delta\parz{1-z}\right] \, .  \label{e.pdists}
\end{equation}

The quantities in \erefs{colliner_ff_approx}{dc_err} are so far intended to be exact. That is, $D(z,z \T{k}{};\mu_Q,Q^2)$, $d_r(z;\mu_Q)$, etc are the operator definitions. 

Equation \eqref{e.dc_deff} is a version of \eref{intrel} wherein the collinear ff $d_c(z;\mu_Q)$ is defined such that the integral relation is satisfied exactly when the UV divergent integral is regulated with a cutoff. A collinear pdf or ff defined in this way does have some practical advantages. 
It is a natural definition when regions are categorized as perturbative or nonperturbative in transverse momentum space, and it is the preferred definition in some work associated with hadron structure (see, for example, the work of Brodsky and collaborators~\cite{Brodsky:2000ii}). 
It is also a natural definition for matching to fixed order large-$\Tsc{q}{}$ calculations, which of course are almost always performed in momentum space.
However, it also comes with significant disadvantages as well; see, for example, reference~\cite{Collins:2003fm} for a discussion of the different ways of handling UV divergences in definitions of collinear pdfs and the pitfalls of each. An important point is that the correction terms relating cut-off scheme and renormalized pdfs are not just perturbative, but include also the power-suppressed errors in \eref{cutff}. 

\section{The Collins-Soper Kernel near the Input Scale $Q_0$}
\label{s.cskernel}

Up to now, our discussion has mainly focused on establishing context and defining notation. In the bulleted list of \sref{intro}, we explained that we will explicitly interpolate between nonperturbatively small $\Tsc{k}{}$ and perturbatively large $\Tsc{k}{}$ in our parametrizations of TMD functions at an input scale. With the language and notation of \sref{tmd_fact} and \sref{tmd_integrals}, we are ready to step through the interpolation details.

Inspection of \eref{Wtermev0} shows that there are actually two separate types of functions in which we will need to interpolate between nonperturbative parametrizations and perturbative regions of transverse momentum when we do calculations for phenomenology. First, there are the TMD ffs themselves, $\tilde{D}_{A}(z_{A},\T{b}{};\mu_{Q_0},Q_0^2)$ and $\tilde{D}_{B}(z_{B},\T{b}{};\mu_{Q_0},Q_0^2)$, at the input scale $Q_0$. Second, there is the CS kernel $\tilde{K}(\Tsc{b}{};\mu_{Q_0})$, also at the input scale $Q_0$, which will be necessary for evolving away from $Q = Q_0$. 

Between these two, it is actually the CS kernel that is the simpler to handle, so we will begin by focusing on it. In fact, the steps for interpolating a full parametrization of an input TMD ff  $D(z,z\T{k}{};\mu_{Q_0},Q_0^2)$ between large and small $\Tsc{k}{}$ will turn out to be very analogous to what we do for the CS kernel. 

It is straightforward to calculate $K(\Tsc{k}{};\mu_{Q_0})$ in fixed order perturbative theory if $\Tsc{k}{} \approx Q_0$. Using notation analogous to that of \sref{tmd_integrals}, we express this as 
\begin{align}
\label{e.Kexp}
&{} K(\Tsc{k}{};\mu_{Q_0}) = \no &{} \qquad K^{(n)}\parz{\Tsc{k}{};\mu_{Q_0}} + \order{\alpha_s(\mu_{Q_0})^{n+1},\frac{m}{\Tsc{k}{}}} \, .
\end{align}
$K^{(n)}\parz{\Tsc{k}{};\mu_{Q_0}}$ is our notation for the fixed order perturbative calculation through order $n$ and with all nonperturbative effects ignored. It takes the form 
\begin{equation}
K^{(n)}\parz{\Tsc{k}{};\mu_{Q_0}} = \frac{1}{\Tscsq{k}{}}  \kappa\parz{\alpha_s(\mu_{Q_0}),\frac{\Tsc{k}{}}{\mu_{Q_0}}} \, ,
\label{e.kappa_def}
\end{equation}
where $\kappa$ is a function of $\alpha_s(\mu_{Q_0})$ and logarithms involving $\Tsc{k}{}/\mu_{Q_0}$. For example, in $\msbar$ renormalization it is
\begin{equation}
K^{(1)}(\Tsc{k}{};\mu_{Q_0}) = \frac{\alpha_s(\mu_{Q_0}) C_F}{\pi^2}  \frac{1}{\Tscsq{k}{}} \, . \label{e.K_pform}
\end{equation}

Away from $\Tsc{k}{} \approx \mu_{Q_0}$, the accuracy of $K^{(n)}\parz{\Tsc{k}{};\mu_{Q_0}}$  as an approximation to $K\parz{\Tsc{k}{};\mu_{Q_0}}$ degrades. The reason 
for $\Tsc{k}{} < \mu_{Q_0}$ is simply that the $\Tsc{k}{}$-dependence of the zero momentum soft radiation is nonperturbative. Conversely, the reason for the approximation failing at $\Tsc{k}{} \gg Q_0$ is that the error terms in \eref{Kexp} include logarithms of $\Tsc{k}{}/\mu_{Q_0}$ that diverge in the large $\Tsc{k}{}/\mu_{Q_0}$ limit.

The $\Tsc{k}{} \gg Q_0$ contribution is numerically unimportant in \eref{tmd_factorization_input} or \eref{tmd_factorization_input2} because the region of transverse momenta far above $Q_0$ is suppressed at the input scale (for reasonable fixed values of $z_A$ and $z_B$). With regard to $Q \approx Q_0$ cross sections, therefore, a suitable parametrization of  $K\parz{\Tsc{k}{};\mu_{Q_0}}$ only requires that we deal with the region $0 < \Tsc{k}{} \lesssim Q_0$. That is to say, near the input scale we only need to extend the perturbative $\Tsc{k}{} \approx Q_0$ treatment in $K^{(n)}(\Tsc{k}{};\mu_{Q_0})$ downward into the nonperturbative $\Tsc{k}{} < \mu_{Q_0}$ region in order to have a parametrization of $K(\Tsc{k}{};\mu_{Q_0})$ that is useful for all $\Tsc{k}{}$ accessible at the input scale.
Let us therefore start by defining an ``input'' parametrization for $K^{(n)}\parz{\Tsc{k}{};\mu_{Q_0}}$ that equals the fixed order calculation, \eref{K_pform}, when $\Tsc{k}{} \approx \mu_{Q_0}$ or larger, but interpolates to a nonperturbative parametrization for $\Tsc{k}{} < \mu_{Q_0}$. 
\begin{widetext}
\begin{equation}
\inpt{K}^{(n)}\parz{\T{k}{};\mu_{Q_0}} \equiv 
\begin{cases}
K^{(n)}\parz{\T{k}{};\mu_{Q_0}} \qquad \text{if} \; \; \Tsc{k}{} \gtrsim \mu_{Q_0} \, , \\ 
\\
\text{nonperturbative parametrization otherwise}
\end{cases} \, . \label{e.K_conds}
\end{equation}
\end{widetext}
Now the ``$(n)$'' superscript on the left side of this equation refers to the perturbative order of the large $\Tsc{k}{}$ tail in this input nonperturbative parametrization. 
When we work with \eref{Wtermev0}, we will need its coordinate space version of the CS kernel,
\begin{equation}
\label{e.Kinput_def}
\inpt{\tilde{K}}^{(n)}(\Tsc{b}{};\mu_{Q_0}) \equiv
\int \diff{^2\T{k}{}}{} e^{i \T{k}{} \T{b}{}}
\inpt{K}^{(n)}(\Tsc{k}{};\mu_{Q_0}) \, .
\end{equation}
The scale-dependence of the exact $\tilde{K}$ is exactly $\Tsc{b}{}$-independent by the RG equation \eref{RG}, so we will enforce the condition that an $n^\text{th}$-order parametrization satisfies \eref{RG} to order $\alpha_s(\mu)^n$, with only $\order{\alpha_s(\mu)^{n+1}}$ errors,
\begin{equation}
\label{e.param_K_RG}
\frac{\diff{\inpt{\tilde{K}}^{(n)}(\Tsc{b}{};\mu)}}{\diff{\ln \mu}{}} =  - \gamma^{(n)}_K(\alpha_s(\mu)) + \order{\alpha_s(\mu)^{n+1}} \, .
\end{equation}

In \sref{low_order_cs} we will provide an example of a specific trial functional form for 
\eref{K_conds}. In general, however, any phenomenologically successful parametrization that satisfies \eref{K_conds} and \eref{param_K_RG} is allowed. 
The parametrizations in \eref{K_conds} and \eref{Kinput_def} are appropriate specifically when $Q \approx Q_0$ such that only the region of $0 < \Tsc{k}{} \lesssim Q_0$ is important.  

However, it is a poor approximation to the true 
$\tilde{K}(\Tsc{b}{};\mu_{Q_0})$ in the $\Tsc{k}{} \gg Q_0$ region, and this matters if we evolve to large enough $Q$ for contributions from $\Tsc{k}{} \gg Q_0$ to become significant. In \eref{Kinput_def}, the large errors manifest themselves as higher order terms logarithmic in 
$\Tsc{b}{} \mu_{Q_0}$, which diverge in the $\Tsc{b}{} \to 0$ limit. There needs to be a change in renormalization scale. Thus, in coordinate space the more common choice for the RG scale is $\mu = C_1/\Tsc{b}{}$, with $C_1$ being an order unity proportionality constant. The truncated RG improved perturbation theory then increases in accuracy as $\Tsc{b}{} \to 0$.

To obtain a $K\parz{\T{k}{};\mu_{Q_0}}$ parametrization that works well for all $Q$, we need steps that combine the stability of fixed scale calculations in the $Q \approx Q_0$, $\Tsc{k}{} \approx Q_0$ region with the RG-improved calculations that optimize for the $\Tsc{b}{} \to 0$ limit. Specifically, we need to perform a scale transformation on the above parametrization using the RG equation at a $\Tsc{b}{}$ somewhat below $1/Q_0$. If we implement this scale transformation at small enough $\Tsc{b}{}$, it will have a negligible effect on phenomenology that uses the above parametrization near $Q \approx Q_0$ where the $\Tsc{b}{} \ll 1/Q_0$ is strongly suppressed.
Therefore, fits that use \eref{K_conds} will be largely unaffected. And, if the transformation takes place in a range of $\Tsc{b}{}$ at least comparable to $\lesssim 1/Q_0$, then its overall effect will only appear at order $n+1$ or higher, so the effect of the transformation will always be one order higher in perturbation theory than the working order. So, the transformation will ensure an accurate treatment of evolution to large $Q$ in any subsequent steps. We will show how this works in detail below.   

The first step in implementing the scale transformation is to define a $\Tsc{b}{}$-dependent mass scale, which we will call $\overline{Q}_0(\Tsc{b}{})$, that smoothly transitions between $Q_0$ and a $1/\Tsc{b}{}$-dependence in the region just below $\Tsc{b}{} \approx 1/Q_0$. Specifically,   
\begin{equation}
\overline{Q}_0(\Tsc{b}{}) =
\begin{dcases}
C_1/\Tsc{b}{} & \Tsc{b}{} \ll C_1/Q_0 \, , \\
Q_0 &  \text{otherwise} \, ,  
\label{e.bdef}
\end{dcases}
\end{equation}
where $C_1$ is an order unity numerical constant, typically taken to be $C_1 = 2 e^{-\gamma_E}$. When $\Tsc{b}{}$ is comparable to $C_1/Q_0$, the scales $Q_0$ and $C_1/\Tsc{b}{}$ are numerically similar, so any sensitivity to the difference between the two scale 
is a higher order effect that can be reduced by 
including higher orders in perturbation theory. Therefore, the exact form of $\overline{Q}_0(\Tsc{b}{})$ is arbitrary so long as it provides a reasonably smooth interpolation between the $Q_0$ and $C_1/\Tsc{b}{}$ behavior at large and small $\Tsc{b}{}$. 
Some example suggestions for $\overline{Q}_0(\Tsc{b}{})$, which we will call the transformation function, are shown in \aref{interp}.  

Next, we need to combine this with the RG equation \eref{RG}, whose exact solution is
\begin{equation}
\label{e.RG_solna}
\tilde{K}(\Tsc{b}{};\mu) = \tilde{K}(\Tsc{b}{};\mu_i) -
\int_{\mu_i}^{\mu} \frac{\diff{\mu'}}{\mu'}  \gamma_K(\alpha_s(\mu')) \, .
\end{equation} 
Here, $\mu_i$ is an arbitrary initial scale.
To make it useful in applications of \eref{Wtermev0}, let us evolve from an initial scale $\mu_i = \mu_{\overline{Q}_0}$ (where $\mu_{\overline{Q}_0} = C_2 \overline{Q}_0$) so that the right side contains $\tilde{K}(\Tsc{b}{};\mu_{\overline{Q}_0})$:
\begin{equation}
\label{e.RG_solnb}
\tilde{K}(\Tsc{b}{};\mu) = \tilde{K}(\Tsc{b}{};\mu_{\overline{Q}_0}) -
\int_{\mu_{\overline{Q}_0}}^{\mu} \frac{\diff{\mu'}}{\mu'}  \gamma_K(\alpha_s(\mu')) \, .
\end{equation}
For any $\mu \approx Q_0$, the second term in \eref{RG_solnb} is calculable in perturbation theory with 
the $n^\text{th}$-order anomalous dimension, $\gamma_K(\alpha_s(\mu')) \to \gamma_K^{(n)}(\alpha_s(\mu'))$, and it vanishes for $\Tsc{b}{} \approx C_1/Q_0$ or larger. 

The original parametrization in \eref{Kinput_def} was designed to provide an accurate perturbative description of $\tilde{K}(\Tsc{b}{};\mu_{Q_0})$ in the region of $\Tsc{b}{} \approx 1/Q_0$ and larger. 
Now if we replace the first term on the right side of \eref{RG_solnb} with $\inpt{\tilde{K}}^{(n)}(\Tsc{b}{};\mu_{\overline{Q}_0})$, it continues to describe the $\Tsc{b}{} \gtrsim 1/Q_0$ region, by our construction of $\overline{Q}_0(\Tsc{b}{})$. However, now the RG-improved perturbative contribution to $\tilde{K}(\Tsc{b}{};\mu_{\overline{Q}_0})$ also remains accurate into the $\Tsc{b}{} \ll 1/Q_0$ region. 

Therefore, we obtain an optimal parametrization 
by replacing the exact $\tilde{K}(\Tsc{b}{};\mu_{\overline{Q}_0})$ on the right side of \eref{RG_solnb} by the approximate $\inpt{\tilde{K}}^{(n)}(\Tsc{b}{};\mu_{\overline{Q}_0})$ and the exact $\gamma_K(\alpha_s(\mu'))$ by $\gamma_K^{(n)}(\alpha_s(\mu'))$. We define this this as 
\begin{align}
\label{e.evol_parama}
&{} \tilde{\param{K}}^{(n)}(\Tsc{b}{};\mu) \no 
&{} \equiv \inpt{\tilde{K}}^{(n)}(\Tsc{b}{};\mu_{\overline{Q}_0}) -
\int_{\mu_{\overline{Q}_0}}^{\mu} \frac{\diff{\mu'}}{\mu'}  \gamma_K^{(n)}(\alpha_s(\mu')) \, .
\end{align}
The underline on $\tilde{\param{K}}^{(n)}(\Tsc{b}{};\mu)$ is our notation for the final parametrization to be used with evolution. 
The above applies to the cases where $\mu \approx Q_0$, so as a final step we set $\mu = \mu_{Q_0}$ and write the underlined parametrization as
\begin{align}
\label{e.evol_paramb}
&{} \tilde{\param{K}}^{(n)}(\Tsc{b}{};\mu_{Q_0}) \no &{}\equiv \inpt{\tilde{K}}^{(n)}(\Tsc{b}{};\mu_{\overline{Q}_0}) -
\int_{\mu_{\overline{Q}_0}}^{\mu_{Q_0}} \frac{\diff{\mu'}}{\mu'}  \gamma_K^{(n)}(\alpha_s(\mu')) \, .
\end{align}
This is the form of the parametrization for the CS kernel that we will ultimately use in \eref{Wtermev0}.
The errors in $\tilde{\param{K}}^{(n)}(\Tsc{b}{};\mu_{Q_0})$, as an approximation to $\tilde{K}(\Tsc{b}{};\mu_{Q_0})$, are suppressed by at least $\alpha_s(\mu_{Q_0})^{n+1}$ point-by-point for all $\Tsc{b}{}$.

A final constraint on parametrizations of $\inpt{K}^{(n)}\parz{\T{k}{};\mu_{Q_0}}$ is 
obtained by recalling that soft gluon effects cancel in collinear factorization when we integrate over all transverse momentum. Thus, after an integration of  $\inpt{K}^{(n)}\parz{\T{k}{};\mu_{Q_0}}$ over $\T{k}{}$ up to a cutoff $k_{\rm max}$ of order $\mu_{Q_0}$, sensitivity to any nonperturbative mass parameters must vanish as $m/\mu_{Q_0} \to 0$. We may express this by demanding that 
\begin{align}
&{} \pi \int_0^{k_{\rm{max}}^2} \diff{\Tscsq{k}{}}{} \inpt{K}^{(n)}\parz{\T{k}{};\mu_{Q_0}} \no 
&{} \qquad = \chi^{(n)}(k_{\rm{max}}/\mu_{Q_0},\alpha_s(\mu_{Q_0})) + \order{\frac{m}{\mu_{Q_0}},\frac{m}{k_{\rm{max}}}} \, , \label{e.K_closure}
\end{align}
where $\chi^{(n)}(k_{\rm{max}}/\mu,\alpha_s(\mu))$ is either zero or a perturbatively calculable function, independent of any nonperturbative mass parameters in $\inpt{K}^{(n)}\parz{\T{k}{};\mu_{Q_0}}$. 

Before continuing, let us summarize the basic properties of the parametrization, $\tilde{\param{K}}^{(n)}(\Tsc{b}{};\mu_{Q_0})$:
\begin{itemize}
\item For $\Tsc{b}{} \approx 1/Q_0$ or larger, it differs negligibly  
from $\inpt{\tilde{K}}^{(n)}(\Tsc{b}{};\mu_{Q_0})$, by construction. Therefore, both $\tilde{\param{K}}^{(n)}(\Tsc{b}{};\mu_{Q_0})$ and $\inpt{\tilde{K}}^{(n)}(\Tsc{b}{};\mu_{Q_0})$ are equally appropriate for describing the $Q \approx Q_0$ region phenomenologically. 
\item For small $\Tsc{b}{}$, the RG scale transitions to the usual $\mu \sim 1/\Tsc{b}{}$ RG-improved form, but only when $\Tsc{b}{}$ is very small relative to the input scale, $\Tsc{b}{} \ll 1/Q_0$. 
\item The parametrization in \eref{evol_parama} obeys an exact RG equation,
\begin{equation}
\label{e.param_K_RG2}
\frac{\diff{\tilde{\param{K}}^{(n)}(\Tsc{b}{};\mu)}}{\diff{\ln \mu}{}} =  - \gamma^{(n)}_K(\alpha_s(\mu)) \, ,
\end{equation}
with no error terms present.
\item By contrast, the ``input'' parametrization defined in \erefs{K_conds}{param_K_RG} obeys the approximate RG equation in \eref{param_K_RG} with possible perturbative error terms, as shown in the equation. 
\item Both \eref{param_K_RG} and \eref{param_K_RG2} are satisfied for all $\Tsc{b}{}$.
\end{itemize}
The resulting $\tilde{\param{K}}^{(n)}(\Tsc{b}{};\mu_{Q_0})$ is an accurate representation of the exact $\tilde{K}(\Tsc{b}{};\mu_{Q_0})$ up to at most (non-logarithmic) order $\alpha_s(\mu_{Q_0})^{n+1}$ errors. 

There is an ambiguity in the exact choice of functional form for $\overline{Q}_0(\Tsc{b}{})$ in \eref{bdef} in the region of $\Tsc{b}{} \approx 1/Q_0$, but this is just the usual scale uncertainty that appears in any truncated perturbation theory, akin to the dependence on the exact numerical choices for $C_1$ and $C_2$. Since $Q_0$ and $C_1/\Tsc{b}{}$ are of similar size when $\Tsc{b}{} \approx 1/Q_0$ the effect of the transformation is under perturbative control and the ambiguity diminishes as one incorporates higher orders. 

To state this more explicitly, consider a family of different choices for $\overline{Q}_0(\Tsc{b}{})$ smoothly connected by an extra parameter $a$:
\begin{equation}
\overline{Q}_0(\Tsc{b}{}) \to \overline{Q}_0(\Tsc{b}{},a) \, . \label{e.Qbar_params}
\end{equation}
The only requirement is that \eref{Qbar_params} satisfies \eref{bdef} for all the $a$ one might consider.
Then, 
\begin{align}
&{} \frac{\diff{}}{\diff{a}} \tilde{\param{K}}^{(n)}(\Tsc{b}{};\mu) 
 =\frac{1}{\mu_{\overline{Q}_0}} 
\frac{\diff{\mu_{\overline{Q}_0}}}{\diff{a}}
\frac{\diff{}}{\diff{\ln \mu_{\overline{Q}_0}}} \tilde{\param{K}}^{(n)}(\Tsc{b}{};\mu) \nonumber \\
&{} = \frac{1}{\mu_{\overline{Q}_0}} 
\frac{\diff{\mu_{\overline{Q}_0}}}{\diff{a}} 
\left[ \frac{\diff{}}{\diff{\ln \mu_{\overline{Q}_0}}} \inpt{\tilde{K}}^{(n)}(\Tsc{b}{};\mu_{\overline{Q}_0}) + \gamma_K(\alpha_s(\mu_{\overline{Q}_0})) \right] 
\nonumber \\
&{} \sim \frac{1}{\mu_{\overline{Q}_0}} 
\frac{\diff{\mu_{\bar{Q}_0}}}{\diff{a}} 
\alpha_s(\mu_{\overline{Q}_0})^{n+1} \ln^{(n+1)} \parz{\Tsc{b}{} \mu_{\overline{Q}_0}} \, . 
\label{e.adiff}
\end{align}
In the second line, we have substituted \eref{evol_parama} and in the last 
line we have used \eref{param_K_RG} while noting 
that at small $\Tsc{b}{}$ the suppressed errors are enhanced by terms logarithmic in $\Tsc{b}{} \mu_{\overline{Q}_0}$. However, by construction $\frac{1}{\mu_{\overline{Q}_0}}
\frac{\diff{\mu_{\overline{Q}_0}}}{\diff{a}}$ is only allowed to be nonzero in a region of $\Tsc{b}{}$ where $1/\Tsc{b}{}$, $Q_0$, and $\overline{Q}_0$ are all of
comparable size. 
So \eref{adiff} is just 
\begin{equation}
\frac{\diff{}}{\diff{a}} \tilde{\param{K}}^{(n)}(\Tsc{b}{};\mu) = 
\order{\alpha_s(\mu_{Q_0})^{n+1}} \, ,
\label{e.adiff2}
\end{equation}
So effects from varying the precise choice of transition function $\overline{Q}_0(\Tsc{b}{})$  
are always one order higher in $\alpha_s(Q_0)$ than the working order $n$.

We will illustrate the steps above more concretely with specific examples in \sref{low_order_cs}.

\section{Parametrization of the TMD ffs at $Q = Q_0$}
\label{s.parameterization}

Now we turn to the parametrizations of the input TMD ffs themselves. Following the strategy outlined in the introduction, we will categorize regions as perturbative or nonperturbative for the input scale TMD ff in transverse momentum space. The steps will be very analogous to those just described in the previous section for $\tilde{K}(\Tsc{b}{};\mu)$. 
As in that case, we will use an ``$\text{input}$'' subscript to label the TMD ff parametrization that applies phenomenologically at the input scale $Q = Q_0$, and which is to be used in \eref{tmd_factorization_input2}. 
For $\Tsc{k}{} < Q_0$, the input parametrization will be defined to have a mainly nonperturbative transverse momentum dependence while for $\Tsc{k}{} \approx Q_0$ or larger it will transition into its $n^\text{th}$-order perturbative description, the first term in \eref{colliner_ff_approx}.  Specifically, we define
\begin{widetext}
\begin{equation}
\inpt{D}^{(n,d_r)}\parz{z,z \T{k}{};\mu_{Q_0},Q_0^2} \equiv 
\begin{cases}
D^{(n,d_r)}\parz{z,z \T{k}{};\mu_{Q_0},Q_0^2} \qquad \text{if} \; \; \Tsc{k}{} \gtrsim Q_0 \,  \\ 
\\
\text{nonperturbative parametrization otherwise}
\end{cases} \, . \label{e.param_conds}
\end{equation}
\end{widetext}
The only condition on the intermediate region between $\Tsc{k}{} \ll Q_0$ and $\Tsc{k}{} \approx Q_0$ is that it should be reasonably smooth.
The input parametrization in \eref{param_conds} has the coordinate space representation,
\begin{align}
&{} \inpt{\tilde{D}}^{(n,d_r)}(z,\T{b}{};\mu_{Q_0},Q_0^2) \no
&{} \equiv \int \diff{^2\T{k}{}}{} e^{i \T{k}{} \cdot \T{b}{}} \inpt{D}^{(n,d_r)}\parz{z,z \T{k}{};\mu_{Q_0},Q_0^2} \, .
\label{e.mu0_coord}
\end{align}
We will require that the input coordinate space parametrizations satisfy the evolution equations through $n^\text{th}$-order in the evolution kernels, at least for the $\Tsc{b}{} \approx 1/Q_0$ region:
\begin{align}
\label{e.CSPDFinpt}
&{} \frac{\partial \ln \inpt{\tilde{D}}^{(n,d_r)}(z,\T{b}{};\mu,Q^2)}{\partial \ln Q } \no  &{} \; =  \inpt{\tilde{K}}^{(n)}(\Tsc{b}{};\mu) + \order{\alpha_s(\mu)^{n+1}} + \order{\Tsc{b}{} m} \, , \\
\label{e.RGinpt}
&{} \frac{\diff{\ln \inpt{\tilde{D}}^{(n,d_r)}(z,\T{b}{};\mu,Q^2)}}{\diff{\ln \mu}} \no &{} \; = \gamma^{(n)}(\alpha_s(\mu);Q^2 /\mu^2) + \order{\alpha_s(\mu)^{n+1}} + \order{\Tsc{b}{} m} \, . 
\end{align}
Usually, these will be satisfied automatically if the parametrization follows \eref{param_conds}.
Note the analogy between \erefs{param_conds}{RGinpt} above and \erefs{K_conds}{param_K_RG} for the CS kernel. 

Finally, for the integrated TMD ff to be consistent with the definition in \eref{dc_deff} we must impose it directly on the parametrization, 
\begin{align}
&{} 2 \pi z^2 \int_0^{\mu_{Q_0}} \diff{\Tsc{k}{}}{} \Tsc{k}{} \inpt{D}^{(n,d_r)}(z,z \T{k}{};\mu_{Q_0},Q_0^2) \no  
&{} \qquad \equiv \param{d}^{(n,d_r)}_c(z;\mu_{Q_0}) \, .
\label{e.cutff1}
\end{align} 
Here we have introduced new notation and another definition. 
The underline on $\param{d}^{(n,d_r)}_c(z;\mu_{Q_0})$ is meant to indicate that 
this is a specific \emph{parametrization} (one determined by the $\inpt{D}^{(n,d_r)}(z,z \T{k}{};\mu_{Q_0},Q_0^2)$ parametrization) of the $d_c(z;\mu_{Q_0})$ defined in \eref{dc_deff}. In accordance with \eref{dc_err}, it is to have, by its definition, the property that
\begin{equation}
\param{d}^{(n,d_r)}_c(z;\mu_{Q_0}) = d^{(n,d_r)}_c(z;\mu_{Q_0}) + \order{\frac{m}{\mu_{Q_0}}} \, . \label{e.d_little_underline}
\end{equation}
The parametrization $\param{d}^{(n,d_r)}_c(z;\mu_{Q_0})$ is a description of the definition $d_c(z;\mu_{Q_0})$, with an $n^\text{th}$-order collinear treatment of the high transverse momentum region.
It is simply the $d^{(n,d_r)}_c(z;\mu_{Q_0})$ from \eref{dnr_rel}, but with account taken of the  power-suppressed behavior in \eref{cutff} that vanishes as $m/\mu_{Q_0} \to 0$. 

It is worth pausing to review the different types of cutoff collinear ffs that we have introduced so far, given that there are now at least three. First, the version in \eref{dc_deff} with no underlines or superscripts is the exact $d_c(z;\mu_{Q_0})$ that follows from the abstract operator definitions. Second,
the $\param{d}^{(n,d_r)}_c(z;\mu_{Q_0})$ above is a specific parametrization of that definition, with the only requirement being that in the limit of large $\mu_{Q_0}$ it reduces to $n^\text{th}$-order collinear perturbation theory in terms of renormalized ffs with scheme $r$.
Finally, there is the $d^{(n,d_r)}_c(z;\mu_{Q_0})$ defined in \eref{dnr_rel}, which is just the limit of $\param{d}^{(n,d_r)}_c(z;\mu_{Q_0})$ where power-suppressed terms are dropped. An equally valid definition is
\begin{equation}
d^{(n,d_r)}_c(z;\mu_{Q_0}) \equiv \lim_{\frac{m}{\mu_{Q_0}} \to 0} \param{d}^{(n,d_r)}_c(z;\mu_{Q_0}) \, .
\end{equation}
Because \eref{cutff1} is just a definition, it contains no constraint by itself. The constraint is in \eref{d_little_underline}.

So far, the steps for constructing the TMD ff parametrizations are very analogous to those of \sref{cskernel} for $K(\Tsc{k}{};\mu_{Q_0})$, but there are some differences. The most significant is that the ``perturbative'' large $\Tsc{k}{}$ part of \eref{param_conds} is not \emph{entirely} perturbative because it involves non-perturbative collinear ffs as input in \eref{D_fact}. The perturbative contribution to large transverse momentum dependence only enters in the coefficient function $C_D^{(n)}(z \Tsc{k}{})$. By contrast, the only input to the perturbative calculation in \eref{kappa_def} is the strong coupling $\alpha_s$. 

The conditions in \erefs{param_conds}{cutff1} are all that we need for constructing phenomenologically useful parametrizations in the $Q \approx Q_0$ region. Any model or parametrization that satisfies them is acceptable, but we will give some explicit examples in later sections. 

However, the perturbative part of the parametrization in \eref{param_conds}
does not provide an accurate description in the 
region of $\Tsc{k}{} \gg Q_0$, where ratios of $\Tsc{k}{}$ and $Q_0$ diverge. In coordinate space, the same issue arises at $\Tsc{b}{} \ll 1/Q_0$ in the form of large logarithms of $\mu \Tsc{b}{}$. That does not create a problem for phenomenological applications near $Q \approx Q_0$ where the $\Tsc{k}{} \gg Q_0$ contributions are suppressed in the integral of \eref{tmd_factorization_input}.  However, it becomes important 
as one evolves to $Q \gg Q_0$ and the $\Tsc{k}{} \gg Q_0$ region starts to contribute more significantly.

Therefore, there needs to be a scale transformation from $\mu = \mu_{Q_0}$ to $\mu = C_1/\Tsc{b}{}$ in the coordinate space TMD ff in the region of $\Tsc{b}{}$ just below $\Tsc{b}{} \approx 1/Q_0$. This of course is exactly the same issue that we faced in the case of $\inpt{\tilde{K}}^{(n)}(\Tsc{b}{};\mu_{Q_0})$ in the previous section. For the TMD ff, it also implies that we have to evolve the CS scale $\sqrt{\zeta}$ from $Q_0$ to $C_1/\Tsc{b}{}$.
For the scale change we can reuse the same scale  transformation function from \eref{bdef}. 

The exact solution to the TMD evolution equations for an individual TMD ff evolving from scales $\mu_i$, $Q_i$ to $\mu_{Q_0}$, $Q_0$ is
\begin{widetext}
\begin{align}
&\tilde{D}(z,\T{b}{};\mu_{Q_0},Q_0^2)= 
\tilde{D}(z,\T{b}{};\mu_i,Q_i^2)
\exp \left\{
\int_{\mu_i}^{\mu_{Q_0}} \frac{d \mu^\prime}{\mu^\prime} \left[\gamma(\alpha_s(\mu^\prime);1) 
- \ln \frac{Q_0}{\mu^\prime} \gamma_K(\alpha_s(\mu^\prime))
  \right] +\ln \frac{Q_0}{Q_i} \tilde{K}(\Tsc{b}{};\mu_i) \right\} \, . \label{e.evolvedd}
\end{align} 
Or, if we use $\mu_i = \mu_{\overline{Q}_0}$, $Q_i = \overline{Q}_0$ for the input scale,
\begin{align}
&\tilde{D}(z,\T{b}{};\mu_{Q_0},Q_0^2)= 
\tilde{D}(z,\T{b}{};\mu_{\overline{Q}_0},\overline{Q}_0^2)
\exp \left\{
\int_{\mu_{\overline{Q}_0}}^{\mu_{Q_0}} \frac{d \mu^\prime}{\mu^\prime} \left[\gamma(\alpha_s(\mu^\prime);1) 
- \ln \frac{Q_0}{\mu^\prime} \gamma_K(\alpha_s(\mu^\prime))
  \right] +\ln \frac{Q_0}{\overline{Q}_0} \tilde{K}(\Tsc{b}{};\mu_{\overline{Q}_0}) \right\} \, . \label{e.evolvedd2}
\end{align} 
\end{widetext}
As of yet, there are still no approximations on $\tilde{D}(z,\T{b}{};\mu_{Q_0},Q_0^2)$.  
The left side has no dependence on $\overline{Q}_0$; any $\overline{Q}_0$-dependence in 
$\tilde{D}(z,\T{b}{};\mu_{\overline{Q}_0},\overline{Q}_0^2)$ is exactly canceled by an opposite $\overline{Q}_0$-dependence in 
the exponential factor. 

Now we can substitute approximations into the right side of \eref{evolvedd2} in a way that is again very analogous to the way we handled 
$\tilde{K}(\Tsc{b}{};\mu)$ in the previous section by making substitutions on the right side of \eref{RG_solnb}. We approximate 
$\tilde{D}(z,\T{b}{};\mu_{\overline{Q}_0},\overline{Q}_0^2)$ 
on the right side of \eref{evolvedd2} by replacing it with the $\inpt{D}^{(n,d_r)}\parz{z,z \T{k}{};\mu_{\overline{Q}_0},\mu_{\overline{Q}_0}^2}$ from \eref{param_conds}. Because of the scale transformation, the result is a parametrization of $\tilde{D}(z,\T{b}{};\mu_{\overline{Q}_0},\overline{Q}_0^2)$ that is an accurate description not only for $\Tsc{b}{} \approx 1/Q_0$ and larger but also for all $\Tsc{b}{} \ll 1/Q_0$. For the  $\tilde{K}(\Tsc{b}{};\mu_{\overline{Q}_0})$ in the exponent on the right side of \eref{evolvedd2}, we already have the analogous result from \eref{evol_paramb} in \sref{cskernel}, and we can reuse it here. All that remains then is to substitute 
$\gamma(\alpha_s(\mu^\prime);1)$ and $\gamma_K(\alpha_s(\mu^\prime))$ by their truncated $n^\text{th}$-order perturbation theory approximations. Thus, our final parametrization of the input TMD ff is 
\begin{widetext}
\begin{align}
&\tilde{\param{D}}^{(n,d_r)}(z,\T{b}{};\mu_{Q_0},Q_0^2) \no
&= \inpt{\tilde{D}}^{(n,d_r)}(z,\T{b}{};\mu_{\overline{Q}_0},\overline{Q}_0^2)
\exp \left\{
\int_{\mu_{\overline{Q}_0}}^{\mu_{Q_0}} \frac{d \mu^\prime}{\mu^\prime} \left[\gamma^{(n)}(\alpha_s(\mu^\prime);1) 
- \ln \frac{Q_0}{\mu^\prime} \gamma_K^{(n)}(\alpha_s(\mu^\prime))
  \right] +\ln \frac{Q_0}{\overline{Q}_0} \inpt{\tilde{K}}^{(n)}(\Tsc{b}{};\mu_{\overline{Q}_0}) \right\} \, . \label{e.evolvedd3p}
\end{align} 
\end{widetext}
The underline on $\tilde{\param{D}}^{(n,d_r)}(z,\T{b}{};\mu_{Q_0},Q_0^2)$ is our notation for the final parametrization of the 
TMD ff at the input scale $\mu = \mu_{Q_0}$, $\zeta = Q_0^2$. 

To summarize, $\tilde{\param{D}}(z,\T{b}{};\mu_{Q_0},Q_0^2)$ has the following properties:
\begin{itemize}
\item Each factor in \eref{evolvedd3p} is an 
accurate approximation to the corresponding factor in the exact \eref{evolvedd2} point-by-point in $\Tsc{b}{}$.  
\item For $\Tsc{b}{} \approx 1/Q_0$ or larger, the exponential evolution factor deviates from unity by a negligible amount. This is by our construction of $\overline{Q}_0(\Tsc{b}{})$. $\tilde{\param{D}}^{(n,d_r)}(z,\T{b}{};\mu_{Q_0},Q_0^2)$ therefore deviates negligibly from $\inpt{\tilde{D}}^{(n,d_r)}(z,\T{b}{};\mu_{Q_0},Q_0^2)$ when $Q \approx Q_0$. Both $\tilde{\param{D}}^{(n,d_r)}(z,\T{b}{};\mu_{Q_0},Q_0^2)$ and $\inpt{\tilde{D}}^{(n,d_r)}(z,\T{b}{};\mu_{Q_0},Q_0^2)$ with \eref{tmd_factorization_input2} are equally appropriate for applications to phenomenology when $Q \approx Q_0$. Also, recall that the 
$\inpt{\tilde{D}}^{(n,d_r)}(z,\T{b}{};\mu_{Q_0},Q_0^2)$ was originally formulated in transverse momentum space.
\item There is a smooth scale transformation to $\mu \sim \sqrt{\zeta} \sim 1/\Tsc{b}{}$ at small $\Tsc{b}{}$, but only when $\Tsc{b}{} \ll 1/Q_0$. Therefore, $\tilde{\param{D}}^{(n,d_r)}(z,\T{b}{};\mu_{Q_0},Q_0^2)$ continues to provide an accurate approximation to the exact TMD ff after $Q \gg Q_0$ where the $\Tsc{b}{} \ll 1/Q_0$ region starts to be relevant. 
\item Thus, at small $\Tsc{b}{}$ we may express the TMD ff in terms of collinear ffs $d_r$ using the usual OPE methods. 
\item $\tilde{\param{D}}^{(n,d_r)}(z,\T{b}{};\mu_{Q_0},Q_0^2)$ satisfies the exact evolution equations
\begin{align}
\label{e.CSPDF_underline}
&{} \frac{\partial \ln \tilde{\param{D}}^{(n,d_r)}(z,\T{b}{};\mu_{Q_0},Q_0^2)}{\partial \ln Q_0 } \no
&{} 
\qquad \qquad = \tilde{\param{K}}^{(n)}(\Tsc{b}{};\mu_{Q_0}) \, , \\
\label{e.RGgamma_underline} 
&{} \frac{\diff{\ln  \tilde{\param{D}}^{(n,d_r)}(z,\T{b}{};\mu_{Q_0},Q_0^2)}}{\diff{\ln \mu_{Q_0}}} \no
&{} \qquad \qquad = \gamma^{(n)}(\alpha_s(\mu_{Q_0});1) 
\no
&{} \qquad \qquad \; - \gamma_K^{(n)}(\alpha_s(\mu_{Q_0})) \ln \parz{\frac{Q_0}{\mu_{Q_0}}} \, . 
\end{align}
There are no error terms in either of these equations, and
both~\eref{CSPDF_underline} and \eref{RGgamma_underline} are valid for all $\Tsc{b}{}$.
\item By contrast, the evolution equations for the ``input'' subscript TMD ffs in \erefs{CSPDFinpt}{RGinpt} do come with explicit error terms.
\end{itemize}

As was the case for $\tilde{\param{K}}^{(n)}(\Tsc{b}{};\mu_{Q_0})$,  sensitivity to the choice of functional form for $\overline{Q}_0(\Tsc{b}{})$ is the standard scale uncertainty in truncated perturbation theory, and it vanishes in the limit that $Q_0$ is large and/or high enough orders in $\alpha_s(Q_0)$ are included. To see this, we may repeat steps analogous to those after \eref{Qbar_params}:
\begin{align}
&{} \frac{\diff{}}{\diff{a}} \ln \tilde{\param{D}}^{(n,d_r)}(z,\T{b}{};\mu_{Q_0},Q_0^2) 
  \no
 &{}\; = \frac{1}{\mu_{\overline{Q}_0}} 
\frac{\diff{\mu_{\overline{Q}_0}}}{\diff{a}}
\frac{\diff{}}{\diff{\ln \mu_{\overline{Q}_0}}} \ln \tilde{\param{D}}^{(n,d_r)}(z,\T{b}{};\mu_{Q_0},Q_0^2) \no
&{}\; 
+ \frac{1}{\overline{Q}_0} 
\frac{\diff{\overline{Q}_0}}{\diff{a}}
\frac{\partial}{\partial \ln \overline{Q}_0} \ln \tilde{\param{D}}^{(n,d_r)}(z,\T{b}{};\mu_{Q_0},Q_0^2)
\no
&{}\; = \frac{1}{\mu_{\overline{Q}_0}} 
\frac{\diff{\mu_{\overline{Q}_0}}}{\diff{a}} 
\left[ \inpt{\tilde{K}}^{(n)}(\Tsc{b}{};\mu_{\overline{Q}_0}) + \gamma^{(n)}(\alpha_s(\mu_{\overline{Q}_0});1) \right. \no
&{}\; \left.  - \inpt{\tilde{K}}^{(n)}(\Tsc{b}{};\mu_{\overline{Q}_0}) - \gamma^{(n)}(\alpha_s(\mu_{\overline{Q}_0});1)
\right. \no 
&{}\; \left. + \order{\alpha_s(\mu_{\overline{Q}_0})^{n+1}} + \order{\Tsc{b}{} m} \right] \, .
\nonumber \\
&{}\; = \order{\alpha_s(\mu_{Q_0})^{n+1}} + \order{\frac{m}{Q_0}} \, . 
\label{e.adiff_Dunderline}
\end{align}
After the second equality, we have substituted the 
expression for $\tilde{\param{D}}^{(n,d_r)}(z,\T{b}{};\mu_{Q_0},Q_0^2)$ from \eref{evolvedd3p} and applied \erefs{CSPDFinpt}{RGinpt} to $\inpt{\tilde{D}}^{(n,d_r)}(z,\T{b}{};\mu_{\overline{Q}_0},\overline{Q}_0^2)$. We have also used that $\mu_{\overline{Q}_0} = \overline{Q}_0$ when $C_2 = 1$ to simplify expressions. On the last line, we have used the fact that $\frac{1}{\mu_{\bar{Q}_0}}
\frac{\diff{\mu_{\bar{Q}_0}}}{\diff{a}}$ vanishes by construction everywhere except where $\Tsc{b}{} \sim 1/Q_0$. 

Thus, sensitivity to the choice of $\overline{Q}_0(\Tsc{b}{})$, at any order $n$, vanishes as $m/Q_0 \to 0$. 

Since our notation has now grown rather extensive, we remind the reader that it is summarized in \aref{notglossary}.

\section{Summary of Steps}
\label{s.summary}

So far, we have focused on describing $\tilde{D}(z,\T{b}{};\mu_{Q_0},Q_0^2)$ only at a fixed input scale $Q_0$. Now all that is necessary to calculate $W(\Tsc{q}{},Q)$ at any other scale using  $\tilde{\param{D}}^{(n,d_r)}(z,\T{b}{};\mu_{Q_0},Q_0^2)$ and $\tilde{\param{K}}^{(n)}(\Tsc{b}{};\mu_{Q_0})$ is to substitute them into the right side of \eref{Wtermev0}, along with the $n^\text{th}$-order perturbative expressions for $H(\alpha_s(\mu_Q);C_2)$, $\gamma(\alpha_s(\mu'); 1)$, and $\gamma_K(\alpha_s(\mu'))$.  The result is an approximation for $W(\Tsc{q}{},Q)$ that includes evolution and is accurate for $Q \geq Q_0$,
\begin{widetext}
\begin{align}
W^{(n)}(\Tsc{q}{},Q)
&{}\equiv
   H^{(n)}(\alpha_s(\mu_Q);C_2) 
    \int \frac{\diff[2]{\T{b}{}}}{(2 \pi)^2}
    ~ e^{-i\T{q}{}\cdot \T{b}{} }
    ~ \tilde{\param{D}}_{A}^{(n,d_r)}(z_A,\T{b}{};\mu_{Q_0},Q_0^2) 
    ~ \tilde{\param{D}}_{B}^{(n,d_r)}(z_B,\T{b}{};\mu_{Q_0},Q_0^2) \nonumber\\&
  \,\times
  \exp\left\{  
        \tilde{\param{K}}^{(n)}(\Tsc{b}{};\mu_{Q_0}) \ln \parz{\frac{ Q^2 }{ Q_0^2}}
           +\int_{\mu_{Q_0}}^{\mu_Q}  \frac{ \diff{\mu'} }{ \mu' }
           \biggl[ 2 \gamma^{(n)}(\alpha_s(\mu'); 1) 
                 - \ln\frac{Q^2}{ {\mu'}^2 } \gamma_K^{(n)}(\alpha_s(\mu'))
           \biggr]
  \right\} \, .
\label{e.Wtermev0_finalversion}
\end{align}
\end{widetext}
The approximation, notated by the ``$(n)$'' superscript on $W^{(n)}(\Tsc{q}{},Q)$, is such that the scale dependence given by the evolution equations in \erefs{CSPDF}{RGgamma} is accurate point-by-point for $Q \geq Q_0$ with errors at most of order $\order{\alpha_s(Q_0)^{n+1}}$ -- see \erefs{CSPDF_underline}{RGgamma_underline}. 
When $Q = Q_0$ the $W^{(n)}(Q_0,\Tsc{q}{})$ defined in \eref{Wtermev0_finalversion} reduces to the TMD parton model up to the overall factor of $H^{(n)}(\alpha_s(\mu_Q);C_2)$ and 
\begin{align}
&{}W(\Tsc{q}{},Q_0) - W^{(n)}(\Tsc{q}{},Q_0) \no 
&{} \qquad = \order{\alpha_s(\mu_{Q_0})^{n+1}} + \order{\frac{m}{Q_0}} \, .
\end{align}

While it might appear that we have only succeeded at introducing an excessive amount of notation, the end result is a fairly simple recipe for combining any arbitrary model of nonperturbative transverse momentum dependence with full TMD factorization and evolution. After some basic initial decisions like choosing a value for $Q_0$ and fixing renormalization schemes, the steps are as follows:\\

\vspace{0.0in}
\textbf{A) Model Building} 

\begin{stepM}{}
\label{m.choice_model}
Choose a nonperturbative model, or a nonperturbative technique more generally, to phenomenologically parametrize the small  transverse momentum dependence in the TMD ff $D\parz{z_A,z_A\T{k}{A};\mu_{Q_0},Q_0^2}$ and in $K(\Tsc{k}{};\mu_{Q_0})$ at the input scale. (See, for example, the list of models in the introduction. These can likely be used here.)
\end{stepM}

\begin{stepM}{}
\label{m.model_mods}
For step \mref{choice_model}, make any modifications to the models that are necessary to ensure that they satisfy \eref{K_conds}, \eref{param_K_RG}, \eref{K_closure}, \eref{param_conds}, and  \erefs{CSPDFinpt}{d_little_underline}. This step mostly amounts to extrapolating existing models to low order perturbative descriptions of $\Tsc{k}{} \approx Q_0$ behavior. The result is a set of parametrizations for ${\tilde{D}^{(n,d_r)}_{\text{input},A}}\parz{z_A, \T{b}{};\mu_{Q_0},Q_0^2}$, ${\tilde{D}^{(n,d_r)}_{\text{input},B}}\parz{z_B, \T{b}{};\mu_{Q_0},Q_0^2}$, and $\inpt{\tilde{K}}^{(n)}(\Tsc{b}{};\mu_{Q_0})$.
\end{stepM}

\begin{stepM}{}
	\label{m.tilde}
Choose a functional form for the $\overline{Q}_0(\Tsc{b}{})$ in \eref{bdef} to implement the transition between scales. Use the ``input'' functions from step \mref{model_mods} to
	construct $\tilde{\param{K}}^{(n)}(\Tsc{b}{};\mu_{Q_0})$ and $\tilde{\param{D}}^{(n,d_r)}(z,\T{b}{};\mu_{Q_0},Q_0^2)$ via
	 \eref{evol_parama} and \eref{evolvedd3p}.
\end{stepM}

\vspace{0.2in}
\textbf{B) Phenomenology at $\bm{Q\approx Q_0}$} 
\begin{stepNP}{} 
\label{ph.do_pheno_a}
Apply factorization phenomenologically to $Q = Q_0$, Type I processes by taking
\begin{align}
\tilde{\param{D}}^{(n,d_r)}(z,\T{b}{};\mu_{Q_0},Q_0^2) &\to \inpt{\tilde{D}}^{(n,d_r)}(z,\T{b}{};\mu_{Q_0},Q_0^2)\nonumber
\end{align}
in \eref{Wtermev0_finalversion}. This corresponds to the TMD parton model formula in \eref{tmd_factorization_input2} with the input function of step \mref{model_mods}. Fix any parameters in the nonperturbative model. This step is essentially no different from traditional TMD parton model motivated approaches to describing Type I processes. Thus, prior existing phenomenological results can likely be reused here.
\end{stepNP}

\begin{stepNP}{}
\label{ph.do_pheno_b}
Consider the phenomenological behavior of cross sections in a region of $Q$ around $Q \approx Q_0$.  
Take
\begin{align}
\tilde{\param{D}}^{(n,d_r)}(z,\T{b}{};\mu_{Q_0},Q_0^2) &\to \inpt{\tilde{D}}^{(n,d_r)}(z,\T{b}{};\mu_{Q_0},Q_0^2)
\nonumber\\
\tilde{\param{K}}^{(n)}(\Tsc{b}{};\mu_{Q_0}) &\to \inpt{\tilde{K}}^{(n)}(\Tsc{b}{};\mu_{Q_0}) \nonumber
\end{align}
in \eref{Wtermev0_finalversion}, and use the resulting formula in phenomenological fits to fix any nonperturbative parameters in $\inpt{\tilde{K}}^{(n)}(\Tsc{b}{};\mu_{Q_0})$ in the $Q \approx Q_0$ region. 
\end{stepNP}

\begin{stepNP}{}
\label{ph.check_tilde}
Verify that the effect of replacing $\tilde{\param{K}}^{(n)}(\Tsc{b}{};\mu_{Q_0})$ and $\tilde{\param{D}}^{(n,d_r)}(z,\T{b}{};\mu_{Q_0},Q_0^2)$ by 
$\inpt{\tilde{K}}^{(n)}(\Tsc{b}{};\mu_{Q_0})$
and ${\tilde{D}^{(n,d_r)}_{\text{input}}}\parz{z,\T{b}{};\mu_{Q_0},Q_0^2}$ respectively is negligible for numerical calculations around $Q \approx Q_0$. 
\end{stepNP}

\vspace{0.1in}
\textbf{C) Phenomenology at large $\bm{Q}$}
\begin{stepEP}{}
\label{ev.evolve_W}
Use \eref{Wtermev0_finalversion} to evolve to significantly larger $Q$ and make predictions for Type II observables. 
Then refit and/or tune the nonperturbative parameters and improve the agreement with the higher $Q$ observables. The adjustment of parameters should be expected to be minimal since the larger $Q$ measurements are less sensitive to large $\Tsc{b}{}$. 
\end{stepEP}
\begin{stepEP}{}
\label{ev.evolve_W_largerQ}
Continue to repeat step \evref{evolve_W} with even higher $Q$. One should expect the accuracy of predictions to increase, both because of the growing constraints on nonperturbative parameters from previous steps and because larger $Q$ is less sensitive to large $\Tsc{b}{}$ and more sensitive to small $\alpha_s(Q)$ perturbative contributions. 
\end{stepEP}

It is possible to transform \eref{Wtermev0_finalversion} into a form more familiar from traditional implementations of the CSS formalism. While that is not necessary, and indeed we would advocate for using \eref{Wtermev0_finalversion} directly, at least for hadron structure applications, there may still be situations where it is desirable. The steps for transforming \eref{Wtermev0_finalversion} into the more familiar form will be explained later in \sref{css}. 

We have used the example of SIA and TMD ffs to have a concrete factorization formula for illustration purposes, but everything up to now carries over in obvious ways to TMD pdfs and processes like Drell-Yan scattering. 
The steps bridge standard approaches to Type I and Type II phenomenology. 

Before continuing on to examples, notice 
that we may rewrite \eref{Wtermev0_finalversion} as 
\begin{widetext}
\begin{align}
W^{(n)}(\Tsc{q}{},Q)
&{}=
   H^{(n)}(\alpha_s(\mu_Q);C_2) 
    \int \frac{\diff[2]{\T{b}{}}}{(2 \pi)^2}
    ~ e^{-i\T{q}{}\cdot \T{b}{} }
    ~ \tilde{\param{D}}_{A}^{(n,d_r)}(z_A,\T{b}{};\mu_{\overline{Q}_0},\overline{Q}_0^2) 
    ~ \tilde{\param{D}}_{B}^{(n,d_r)}(z_B,\T{b}{};\mu_{\overline{Q}_0},\overline{Q}_0^2) \nonumber\\&
  \,\times
  \exp\left\{  
        \tilde{\param{K}}^{(n)}(\Tsc{b}{};\mu_{\overline{Q}_0}) \ln \parz{\frac{ Q^2 }{ \overline{Q}_0^2}}
           +\int_{\mu_{\overline{Q}_0}}^{\mu_Q}  \frac{ \diff{\mu'} }{ \mu' }
           \biggl[ 2 \gamma^{(n)}(\alpha_s(\mu'); 1) 
                 - \ln\frac{Q^2}{ {\mu'}^2 } \gamma_K^{(n)}(\alpha_s(\mu'))
           \biggr]
  \right\} \, 
\label{e.Wtermev0_finalversion_eliminate}
\end{align}
\end{widetext}
by directly substituting \eref{evol_paramb} and \eref{evolvedd3p} into \eref{Wtermev0_finalversion}. Which form to use is a matter of convenience, but each highlights different aspects of evolution at moderate and/or large $Q$. Equation~ \eqref{e.Wtermev0_finalversion} is written in a way that makes it obvious that TMD factorization exactly follows a TMD parton model description at $Q = Q_0$. That form of $W^{(n)}(\Tsc{q}{},Q)$ also makes it clear that if the TMD ffs and $\tilde{K}(\Tsc{b}{};\mu_{Q_0})$ are known exactly for all $\Tsc{b}{}$ at a fixed input scale, the evolution to larger $Q$ becomes technically trivial. Conversely, in \eref{Wtermev0_finalversion_eliminate} we have been able to eliminate $Q_0$ entirely from the expression, and the RG improved treatment of the $Q \gg Q_0$ limit is automatic in the $\sim 1/\Tsc{b}{}$ scale dependence at small $\Tsc{b}{}$. 
Indeed, this form is very similar to the standard 
presentation of the $W^{(n)}(\Tsc{q}{},Q)$, as we will see in \sref{css}.

\section{Low order examples}
\label{s.low_order}

We mean for the steps in \sref{summary} to apply to any model or nonperturbative method of parametrizing the intrinsic transverse momentum, so we have left the nonperturbative parts in \eref{K_conds} and \eref{param_conds} completely general. But specific examples make the steps much clearer, so we will illustrate them in this section with very minimalistic models of the nonperturbative intrinsic transverse momentum dependence. 

We caution that the nonperturbative parametrizations that we will use below should be regarded only as toy examples at this stage, only to be used for illustration purposes. In future work, we hope to implement the steps in \sref{summary} with input from at least some of the more sophisticated models or nonperturbative techniques referenced in the introduction. 

Since this section makes much use of the notation we have introduced in earlier sections, we remind the reader once more of our notation glossary \aref{notglossary}. We will assume that $\msbar$ renormalization is used everywhere, and we will always use $Q_0 = 2$~GeV for making example plots. 

\subsection{CS kernel example}
\label{s.low_order_cs}
 
To motivate a simple model for the nonperturbative part of the CS kernel, let us recall the physical effect that a nonzero $\tilde{K}(\Tsc{b}{};\mu_{Q_0})$ has on 
momentum-space cross sections at small transverse momentum. Consider the CS evolution
equation for $W(Q,\Tsc{q}{})$ in transverse momentum space. Fourier-Bessel transforms become transverse momentum convolutions so
\begin{align}
&{}\frac{\partial}{\partial \ln Q^2} W(Q,\T{q}{}) = \no 
&{} \int \diff{^2\T{k}{}} K\parz{\T{k}{};\mu} W(Q,\T{q}{}-\T{k}{}) + \text{Const.} \times W(Q,\T{q}{})  \, .
\label{e.momspaceevol}
\end{align}
(See, for example, Eq.~(25) of \cite{Collins:2014jpa}.) The partial derivative here indicates that $z_A$ and $z_B$ should be fixed. 
The ``$\text{Const.}$'' is independent of transverse momentum and is related to the anomalous dimension $\gamma$. It only contributes to a $Q$-dependent normalization.
Our task is to find a reasonable input  $K\parz{\T{k}{};\mu_{Q_0}}$ parametrization that satisfies \eref{K_conds} and also gives a good phenomenological description of cross sections in the region of $Q \approx Q_0$ close to the input scale. 
One way that $K\parz{\T{k}{};\mu}$ can contribute to the evolution of $W(Q \approx Q_0,\T{q}{})$ is simply through a normalization. To capture that behavior in \eref{momspaceevol}, the model parametrization should include a term proportional to a $\delta$-function, 
\begin{equation}
\label{e.K_zeroth}
\inpt{K}(\Tsc{k}{};\mu_{Q_0}) \propto  \delta^{(2)}(\T{k}{}) \, .
\end{equation}
This is an example of step \mref{choice_model} applied to $K(\Tsc{k}{};\mu_{Q_0})$. 
But as step \mref{model_mods} prescribes, $\inpt{K}(\Tsc{k}{};\mu_{Q_0})$ also needs to match the large $\Tsc{k}{}$ perturbative description,
\begin{equation}
\label{e.CS_first_match}
\inpt{K}(\Tsc{k}{};\mu_{Q_0}) \stackrel{\Tsc{k}{} \approx Q_0}{=} K^{(1)}(\Tsc{k}{};\mu_{Q_0})\, ,
\end{equation}
where for now we work at order $n = 1$. 
We also know that
\begin{equation}
K^{(1)}(\Tsc{k}{};\mu_{Q_0}) = \frac{\alpha_s(\mu_{Q_0}) C_F}{\pi^2}  \frac{1}{\Tscsq{k}{}} \, . \label{e.lowest_K}
\end{equation} 
(For a textbook derivation of \eref{lowest_K}, see section 13.10.2 of ~\cite{Collins:2011qcdbook}. Also see Ref.~\cite{Davies:1984hs} for earlier calculations.  For higher order $\tilde{K}$ expressions, see also~\cite{Li:2016ctv,Li:2016axz,Echevarria:2016scs,Lubbert:2016rku}, and see Ref.~\cite{Collins:2017oxh} 
for translating between different notations. See also \cite{Vladimirov:2020umg} for more discussion of the operator definition.)

As $\Tsc{k}{}$ decreases below $Q_0$, \eref{CS_first_match} needs to transition into a nonperturbative eparametrization in a way that is still phenemonologically successful at describing $Q \approx Q_0$ behavior. Existing evidence, both theoretical and phenomenological~\cite{Sun:2013dya,Sun:2013hua,Aidala:2014hva} and from lattice calculations~\cite{Shanahan:2020zxr}, points toward a shape for TMD pdfs and ffs that varies only very weakly with scale in the $Q \approx Q_0$ region. Our trial parametrization will reproduce this behavior if it is fairly sharply peaked around $\Tsc{k}{} \ll Q_0$ and then falls off rapidly for larger $\Tsc{k}{}$. Equation~\eqref{e.momspaceevol} with $K^{(1)}(\Tsc{k}{};\mu_{Q_0})$ captures that general behavior if we make the replacement $\Tscsq{k}{} \to \Tscsq{k}{} + m_K^2$ and keep the nonperturbative parameter $m_K$ small relative to $Q_0$.
Thus, we obtain a reasonable candidate for a $\inpt{K}^{(1)}(\Tsc{k}{};\mu_{Q_0})$ parametrization that satisfies~\eref{K_conds} if we combine the $\Tscsq{k}{} \to \Tscsq{k}{} + m_K^2$ modification of \eref{lowest_K} with \eref{K_zeroth}:
\begin{align}
&{}\inpt{K}^{(1)}(\Tsc{k}{};\mu_{Q_0}) = \no &{}\qquad \frac{\alpha_s(\mu_{Q_0}) C_F}{\pi^2}  \frac{1}{\Tscsq{k}{}+m_K^2} + C_K \delta^{(2)}(\T{k}{}) \, .
\label{e.Kparam_kt}
\end{align} 
The transformation into coordinate space is
\begin{equation}
\inpt{\tilde{K}}^{(1)}(\Tsc{b}{};\mu_{Q_0})= 
\frac{2\alpha_s(\mu_{Q_0}) C_F}{\pi} K_0(\Tsc {b}{} m_K) + C_K \, .
\end{equation}
Satisfying both \eref{param_K_RG} and \eref{K_closure} with the $\msbar$ expression for $\gamma_K^{(1)}$ requires 
\begin{equation}
C_K = \frac{2 \alpha_s(\mu_{Q_0}) C_F}{\pi} \ln\parz{\frac{m_K}{\mu_{Q_0}}} \  \label{e.C_K}
\end{equation}
So the input CS kernel is just the single parameter function
\begin{align}
&{} \inpt{\tilde{K}}^{(1)}(\Tsc{b}{};\mu_{Q_0})
= \no
&{}\qquad \frac{2\alpha_s(\mu_{Q_0}) C_F}{\pi} 
\left[ K_0(\Tsc {b}{} m_K) + 
 \ln\parz{\frac{m_K}{\mu_{Q_0}}} \right] \, .
\label{e.K_param}
\end{align}
Note that the same mass $m_K$ appears in \eref{C_K} and the first term of \eref{K_param} reproduces the known lowest order coordinate space $\tilde{K}^{(1)}(\Tsc{b}{};\mu_{Q_0})$ in $\msbar$ at small $\Tsc{b}{}$:
\begin{align}
&{}\lim_{\Tsc{b}{} \to 0} \inpt{\tilde{K}}^{(1)}(\Tsc{b}{};\mu_{Q_0}) = \no
&{}-\frac{2\alpha_s(\mu_{Q_0}) C_F}{\pi} \left[ \ln\parz{\frac{\Tsc{b}{} \mu_{Q_0} e^{\gamma_E}}{2}} + \ln\parz{\frac{m_K}{\mu_{Q_0}}} \right] + C_K \, \no
&{} = -\frac{2\alpha_s(\mu_{Q_0}) C_F}{\pi} \ln\parz{\frac{\Tsc{b}{} \mu_{Q_0} e^{\gamma_E}}{2}} \, . \label{e.small_bt_K}
\end{align}
At large $\Tsc{b}{}$, we get the expected (see~\cite[Sec. VII-A]{Collins:2014jpa}) constant negative behavior,
\begin{equation}
\lim_{\Tsc{b}{} \to \infty}  \inpt{\tilde{K}}^{(1)}(\Tsc{b}{};\mu_{Q_0}) =
\frac{2\alpha_s(\mu_{Q_0}) C_F}{\pi} 
 \ln\parz{\frac{m_K}{\mu_{Q_0}}} \, .
\end{equation}
\begin{figure}
\centering
\input{ktilde_underline_1_v2_inline.tex}
\caption{
The example parametrization for $\param{\tilde{K}}^{(1)}(\Tsc{b}{};\mu_{Q_0})$ from \eref{K_param_final}, obtained after performing the steps \mref{choice_model}, \mref{model_mods} and \mref{tilde} of \sref{summary}.
The top panel shows two scale transformation functions $\overline{Q}_0(\Tsc{b}{})$ that satisfy \eref{bdef}. The choice of functional form is \eref{qbar_param_a} from \aref{interp}, shown for two choices of $a$ (solid black and dashed red curves). For comparison, lines for the scales $\mu_{Q_0} = 2$~GeV (dash-dotted violet) and $C_1/\Tsc{b}{}$ (dashed violet) are also shown. The central panel is the  percent difference between the $\overline{Q}_0(\Tsc{b}{})$ obtained from the two values $a = 2$~GeV and $a = 4$~GeV, calculated as the difference divided by the average. The bottom panel is a plot of the actual $\tilde{\param{K}}^{(1)}(\Tsc{b}{};\mu_{Q_0})$ parametrization in \eref{K_param_final}. The results are shown for both $a = 2$~GeV and $a = 4$~GeV (black solid and red dashed curves), but the difference between the curves is not visible on the graph. The violet dashed curve in the lower plot is a $\Tsc{b}{} \to 0$ purely perturbative calculation, \eref{pure_pert}, shown for comparison. See text for details.}
\label{f.modelk1}
\end{figure}
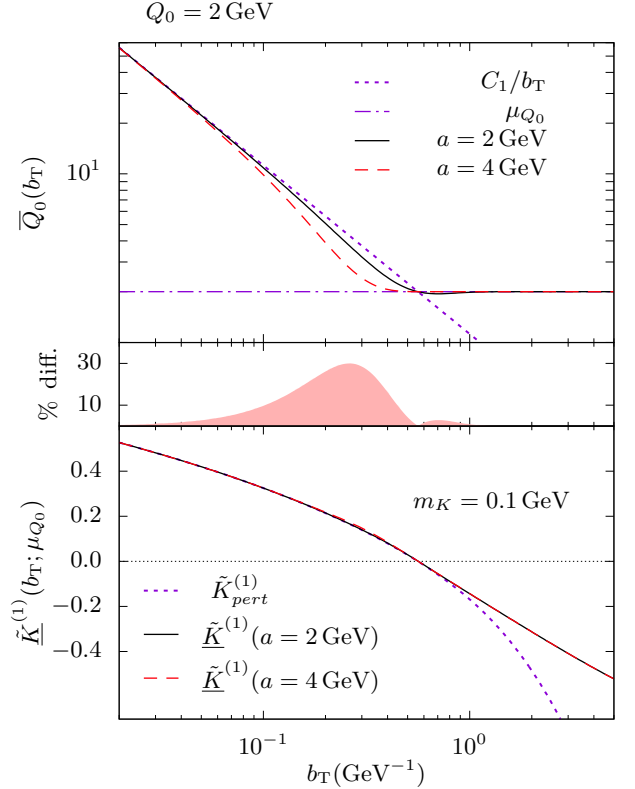
This completes steps \mref{choice_model} and \mref{model_mods}
insofar as they pertain to the CS kernel. 

To get a $\tilde{\param{K}}^{(1)}(\Tsc{b}{};\mu_{Q_0})$ that can be extended to calculations of $\tilde{K}(\Tsc{b}{};\mu_{Q_0})$ at $\Tsc{b}{} \ll 1/Q_0$, we need to proceed with step \mref{tilde} and choose a form for the scale transition function $\overline{Q}_0(\Tsc{b}{})$. For now we will use the form in \eref{qbar_param} from \aref{interp} for any numerical calculations and plots. Later, we will demonstrate that the details of this choice do not significantly affect calculations.

Finally, we get $\tilde{\param{K}}^{(1)}(\Tsc{b}{};\mu_{Q_0})$ by substituting the trial $\inpt{\tilde{K}}^{(1)}(\Tsc{b}{};\mu_{Q_0})$ from \eref{K_param} into \eref{evol_paramb}, 
\begin{align}
&{} \tilde{\param{K}}^{(1)}(\Tsc{b}{};\mu_{Q_0}) 
=
\inpt{\tilde{K}}^{(1)}(\Tsc{b}{};\mu_{\overline{Q}_0}) -
\int_{\mu_{\overline{Q}_0}}^{\mu_{Q_0}} \frac{\diff{\mu'}}{\mu'}  \gamma_K^{(1)}(\alpha_s(\mu')) \, \no
&{}= 
\frac{2\alpha_s(\mu_{\overline{Q}_0}) C_F}{\pi} 
\left[ K_0(\Tsc {b}{} m_K) + 
 \ln\parz{\frac{m_K}{\mu_{\overline{Q}_0}}} \right] \no
 &{} \qquad -
\int_{\mu_{\overline{Q}_0}}^{\mu_{Q_0}} \frac{\diff{\mu'}}{\mu'}  \gamma_K^{(1)}(\alpha_s(\mu')) \, ,
\label{e.K_param_final}
\end{align} 
Repeating \eref{adiff} confirms that variations in the form of $\overline{Q}_0(\Tsc{b}{})$ only enter at order
$\alpha_s(\mu_{Q_0})^2$,
\begin{align}
\label{e.adiffmodel1}
&{} \frac{\diff{}}{\diff{a}} \tilde{\param{K}}^{(1)}(\Tsc{b}{};\mu_{Q_0}) \no
&{} = \frac{1}{\mu_{\overline{Q}_0}} 
\frac{\diff{\mu_{\overline{Q}_0}}}{\diff{a}} 
\frac{2\beta(\mu_{\overline{Q}_0}) C_F}{\pi} 
\left[ K_0(\Tsc {b}{} m_K) + 
 \ln\parz{\frac{m_K}{\mu_{\overline{Q}_0}}} \right] \no
&{} \qquad = \order{\alpha_s(\mu_{Q_0})^2} \, 
\end{align}
for all $\Tsc{b}{}$.

This completes step \mref{tilde} from \sref{summary}, and it completes the model building 
for the CS kernel with $n=1$. To determine the phenomenological parameter $m_K$, one would need to then proceed to the next steps in \sref{summary}, in particular \phref{do_pheno_b}. But for that, one needs to first construct  a model for
the TMD ff itself, which we will do in the next subsection. 

Before proceeding to that step, it is instructive to examine the trial $\tilde{\param{K}}^{(1)}(\Tsc{b}{};\mu_{Q_0})$ graphically, and to verify that the properties described above hold. For example, we expect only a numerically mild, perturbatively suppressed, dependence in $\tilde{\param{K}}^{(1)}(\Tsc{b}{};\mu_{Q_0})$ on the 
scale transformation parameter $a$ in the region 
between small and large $b_{\text{T}}$, consistent with \eref{adiffmodel1}. The lower panel of~\fref{modelk1} shows plots of our trial 
$\tilde{\param{K}}^{(1)}(\Tsc{b}{};\mu_{Q_0})$ with $Q_0 = 2$~GeV and $m_K=0.1$~GeV, while the top panel shows two slightly different functional forms for $\overline{Q}_0(\Tsc{b}{})$ from \eref{bdef}, as given in \aref{interp} and \eref{qbar_param_a}. The two transition functions are generated by using two different values of $a$, $a = 2$~GeV and $a = 4$~GeV. The narrow central panel shows the $\%$ variation between the two $\overline{Q}_0(\Tsc{b}{})$, and it can be seen from the graph that they differ by a maximum of about $\approx 30\%$ for $\Tsc{b}{} \approx 0.3$ GeV$^{-1}$, but for $\Tsc{b}{}$ much larger or smaller than this the variation vanishes as per our requirements.

The $\tilde{\param{K}}^{(1)}(\Tsc{b}{};\mu_{Q_0})$ calculation in the lower panel of \fref{modelk1} was performed using both of the two $\overline{Q}_0(\Tsc{b}{})$ functions in the top panel. We have plotted them as solid black and  dashed-red curves, but the two are visually indistinguishable, confirming the good approximate scale insensitivity in \eref{adiffmodel1}. Despite the clearly 
visible difference between the two $\overline{Q}_0(\Tsc{b}{})$ functions in the upper panel in the region just below $\Tsc{b}{} \approx C_1/Q_0 \approx 0.5$~GeV$^{-1}$, the effect of switching between them essentially vanishes in the $\tilde{\param{K}}^{(1)}(\Tsc{b}{};\mu_{Q_0})$ calculations of the lower panel.

The normal way to calculate $\tilde{K}(\Tsc{b}{};\mu_{Q_0})$ in the small-$\Tsc{b}{}$ limit is to transform to an RG scale, $\mu_{Q_0} \to C_1/\Tsc{b}{}$, and use low order perturbative calculations. We should thus expect to recover this in the small-$\Tsc{b}{}$ limit of the $\tilde{\param{K}}^{(1)}(\Tsc{b}{};\mu_{Q_0})$ from \fref{modelk1}. To check this, let us 
define a ``purely perturbative'' calculation of the small-$\Tsc{b}{}$ $\tilde{K}(\Tsc{b}{};\mu_{Q_0})$ called $\tilde{K}_{\text{pert}}^{(1)}(\Tsc{b}{};\mu_{Q_0})$: 
\begin{align}
&{} \tilde{K}_{\text{pert}}^{(1)}(\Tsc{b}{};\mu_{Q_0}) \equiv \no 
&{} \tilde{K}^{(1)}(\Tsc{b}{};C_1/\Tsc{b}{}) - \int_{C_1/\Tsc{b}{}}^{\mu_{Q_0}} \frac{\diff{\mu'}{}}{\mu'} \gamma^{(1)}_K(\alpha_s(\mu'))\,. \label{e.pure_pert}
\end{align} 
We have shown $\tilde{K}_{\text{pert}}^{(1)}(\Tsc{b}{};\mu_{Q_0})$ as the violet dashed curve in the lower panel of \fref{modelk1}. 
As expected, the purely perturbative curve diverges at large $\Tsc{b}{}$, while it merges with the $\tilde{\param{K}}^{(1)}(\Tsc{b}{};\mu_{Q_0})$ curves for $\Tsc{b}{} \ll 1/\Lambda_\text{QCD}$.

Care is needed when interpreting the numerically small $\overline{Q}_0(\Tsc{b}{})$ dependence observed in plots like  \fref{modelk1}. In cross section calculations like \eref{Wtermev0}, $\tilde{\param{K}}^{(1)}(\Tsc{b}{};\mu_{Q_0})$ multiplies a $\ln(Q^2/Q_0^2)$, so small errors will be amplified as $Q \gg Q_0$. 
The effect of $\tilde{\param{K}}^{(1)}(\Tsc{b}{};\mu_{Q_0})$ on the $\Tsc{b}{}$-space cross section is entirely through the overall factor of $\exp \left\{ \tilde{\param{K}}^{(1)}(\Tsc{b}{};\mu_{Q_0})  \ln(Q^2/Q_0^2) \right\}$, as seen in \eref{Wtermev0}. So, we get a better sense of the size of errors after evolution by plotting this exponent for a wide range of $Q > Q_0$. 
For example, we may calculate $\tilde{\param{K}}^{(1)}(\Tsc{b}{};\mu_{Q_0})$ using different $\overline{Q}_0(\Tsc{b}{})$ and compare the resulting 
$\exp \left\{ \tilde{\param{K}}^{(1)}(\Tsc{b}{};\mu_{Q_0})  \ln(Q^2/Q_0^2) \right\}$ for several $Q > Q_0$. 
We have shown an example of this in~\fref{NPevoratio} for the two different choices of $\overline{Q}_0(\Tsc{b}{})$ in the top panel of \fref{modelk1}. In the top panel of~\fref{NPevoratio}, we have replotted the two $\tilde{\param{K}}^{(1)}(\Tsc{b}{};\mu_{Q_0})$ curves from the lower pane in \fref{modelk1}, but now with axis ranges shifted until the variation between the two calculations starts to become visible on the plot. It is evident from the curves that we should expect any effect from varying $a$ to be significant only in a band around $\Tsc{b}{} \approx 0.3$~GeV$^{-1}$. In the lower panel we have plotted the ratio of $\exp \left\{ \tilde{\param{K}}^{(1)}(\Tsc{b}{};\mu_{Q_0})  \ln(Q^2/Q_0^2) \right\}$ for the two different values of $a$ from the top panel. That is, we plot
\begin{equation}
r(a_1,a_2) \equiv  \frac{\left. \exp \left\{ \tilde{\param{K}}^{(1)}(\Tsc{b}{};\mu_{Q_0})  \ln(Q^2/Q_0^2) \right\} \right|_{a = a_2}}{\left. \exp \left\{ \tilde{\param{K}}^{(1)}(\Tsc{b}{};\mu_{Q_0})  \ln(Q^2/Q_0^2) \right\} \right|_{a = a_1}} \, 
\label{e.exp_ratio}
\end{equation}
for $Q = 4$~GeV and $Q = 100$~GeV. After evolving from $Q = 2$~GeV to $Q = 100$~GeV, the maximum effect is just under $6\%$. 
We will see in more detail how the scale sensitivity propagates to the cross section in \sref{cs_example}.

So far, our example plots have only used $m_K = 0.1$~GeV, but nonperurbative parameters like these will in general need to be adjusted in fits. In our setup, adjustments to parameters like $m_K$ will have a negligible effect on the $\Tsc{b}{} \lesssim 1/Q_0$ region of $\tilde{\param{K}}^{(1)}(\Tsc{b}{};\mu_{Q_0})$ so long as $m_K$ is kept small relative to $Q_0$. To illustrate this, we have plotted $\tilde{\param{K}}^{(1)}(\Tsc{b}{};\mu_{Q_0})$ once again in \fref{ktildemKdep} for several values of $m_K$, now on a linear horizontal axis to magnify the effect on the large $\Tsc{b}{}$ region. The plot confirms that the region of $\Tsc{b}{} \lesssim 0.5$~GeV$^{-1}$ is essentially unaffected by the values of the $m_K$ parameter between $\approx 0.1$~GeV and $\approx 0.5$~GeV, so long as those values are kept reasonably small relative to $Q_0$. A vertical line indicates the  $\Tsc{b}{}=0.3$~GeV$^{-1}$ position where we previously found the greatest scale sensitivity in the perturbative part of the calculation -- the peak of the bump in \fref{NPevoratio}.  
 In contrast to the small scale sensitivity in the bottom panel of \fref{modelk1}, sensitivity to changes in the value of $m_K$ is large and clearly visible, but only in the region of large $\Tsc{b}{}$. The step of fitting the purely nonperturbative parameter has been sequestered from the treatment of the transition into the perturbative regime.
\begin{figure}
    \centering
    \input{NPevoratio_v2_inline.tex}
    
    \caption{
    Top panel: A blown-up version of the curves in the lower panel of \fref{modelk1}. The axes have been adjusted so that the deviation between the two 
    $\tilde{\param{K}}^{(1)}(\Tsc{b}{};\mu_{Q_0})$ calculations for different $\overline{Q}_0(\Tsc{b}{})$ are visible, and we can see that the most significant variation is in a narrow band around $\Tsc{b}{} \approx 0.3$~GeV$^{-1}$. 
    Bottom panel: The effect of the choice of the transformation function on the ratio in \eref{exp_ratio}. All curves are obtained from our trial $n=1$ parametrization for $Q = 4$~GeV and $Q = 100$~GeV. At $Q=100$~GeV, the ratio $r(a_1,a_2)$ deviates from unity by a maximum of about $6\%$ in a transition region around $\Tsc{b}{} \approx 0.3$~GeV$^{-1}$.
    }
    \label{f.NPevoratio}
\end{figure}
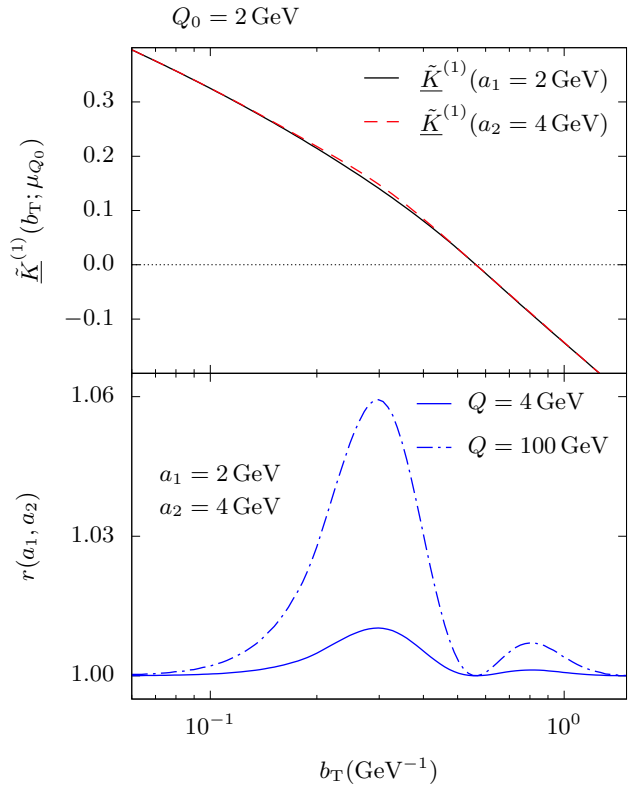
\begin{figure}
    \centering
    \input{ktildemK_dep_inline.tex}
    \vspace*{1mm}
    \caption{
    The trial $\tilde{\param{K}}^{(1)}(\Tsc{b}{};\mu_{Q_0})$ from \eref{K_param_final} calculated for three different values of the nonperturbative parameter $m_K$. The transformation function that was used is  \eref{qbar_param_a} for $a=2\,\text{GeV}$ as before. The black solid curve is identical to the black solid curve in the lower panel of \fref{modelk1}, but now we show two additional values of $m_K$, and with a linear horizontal axis. The sensitivity to $m_K$ is only visible for $\Tsc{b}{} \gtrsim C_1/Q_0\approx0.5\,\text{GeV}^{-1}$, as expected. The vertical line at $\Tsc{b}{} = 0.3$~GeV marks the point where we previously found significant scale sensitivity, as seen in \fref{NPevoratio}. }   
    \label{f.ktildemKdep}
\end{figure}
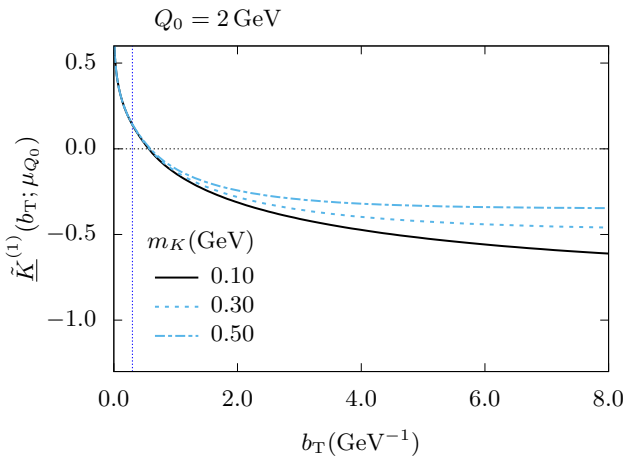

In phenomenological applications, one converges on an unambiguous $\tilde{\param{K}}$ as one repeats the steps above but with higher orders for the large $\Tsc{k}{}$ region. That amounts to constructing parametrizations for  $\tilde{\param{K}}^{(2)}(\Tsc{b}{};\mu_{Q_0})$, $\tilde{\param{K}}^{(3)}(\Tsc{b}{};\mu_{Q_0})$, etc. Going to larger $n$ reduces sensitivity to arbitrary choices like the functional form for $\bar{Q}_0(\Tsc{b}{})$. Quantities like $a$ and $m_K$ are also increasingly constrained as more data from larger $Q$ are included in fitting. 

Extending the above construction of $\tilde{\param{K}}^{(1)}(\Tsc{b}{};\mu_{Q_0})$ to the case of $\tilde{\param{K}}^{(2)}(\Tsc{b}{};\mu_{Q_0})$ is straightforward and instructive, but we leave it to future work. 

\subsection{TMD ff example}
\label{s.FF_example}

Next we need to repeat steps \mref{choice_model}-\mref{tilde} from \sref{summary} for the TMD ffs themselves. To keep the discussion here simple, we will assume that the TMD ffs are the same for hadrons A and B, and we will continue to focus only on the $n=1$ case. Fortunately, the steps are very analogous to the CS kernel, so much of the below will be repetition. 

A typical parametrization, common in TMD parton-model descriptions of Type I processes, is a Gaussian,
\begin{equation}
\inpt{D}^{(0,d_r)}(z,z \T{k}{};\mu_{Q_0},Q_0^2)  = \frac{C}{\pi M^2} e^{-z^2 \Tscsq{k}{} /M^2} \, . \label{e.candidate00}
\end{equation}
This fails to satisfy \eref{param_conds} when we try to extend it directly to  
$n=1$ because it does not have the right functional form to match to $D^{(n,d_r)}\parz{z,z \T{k}{};\mu_{Q_0},Q_0^2}$ when $\Tsc{k}{} \approx Q_0$. In order to construct a TMD ff for $n=1$, we need to describe the transition from a nonperturbative peak like \eref{candidate00} to a perturbative large $\Tsc{k}{}$ power-law tail. The simplest way to do this
is to just append $D^{(n,d_r)}\parz{z,z \T{k}{};\mu_{Q_0},Q_0^2}$ to \eref{candidate00} as an additive term. Inside $D^{(n,d_r)}\parz{z,z \T{k}{};\mu_{Q_0},Q_0^2}$, we can then make the replacement $\Tscsq{k}{} \to \Tscsq{k}{} + m_D^2$, where $m_D$ is a nonperturbative parameter, to smooth the $\Tsc{k}{} \to 0$ behavior into a nonperturbative peak, analogous to what we did in \eref{Kparam_kt}. Thus, our trial input parametrization is
\begin{align}
&{} \inpt{D}^{(1,d_r)}(z,z \T{k}{};\mu_{Q_0},Q_0^2) \no
&{} =  \frac{1}{2 \pi z^2} \frac{1}{\Tscsq{k}{} + m_D^2} \left[A^{(d_r)}(z;\mu_{Q_0}) \right. \no
&{} \left. + B^{(d_r)}(z;\mu_{Q_0}) \ln \frac{Q_0^2 }{\Tscsq{k}{}+m_D^2} \right] 
+ \frac{C^{(d_r)}}{\pi M^2} e^{-z^2 \Tscsq{k}{} /M^2} \, , \label{e.candidate}
\end{align}
where we have utilized the following abbreviations, 
\begin{align}
&{} A^{(d_r)}(z;\mu) \equiv \frac{\alpha_s(\mu)}{\pi}
\left\{ \left[(P_{qq}\otimes{d_r})(z;\mu)\right] \vphantom{\frac{3 C_F}{2} d_r(z;\mu)} \right. \no 
&{} \qquad  \left. -  \frac{3 C_F}{2} d_r(z;\mu)  \right\} \, , \label{e.A_def} \\
&{} B^{(d_r)}(z;\mu) \equiv \frac{\alpha_s(\mu) C_F}{\pi}d_r(z;\mu) \, . \label{e.AB_def}
\end{align}
For a textbook derivation of the large $\Tsc{k}{}$ perturbative behavior see, for example, Eq.(13.101) and Eq.(13.66) of Ref.~\cite{Collins:2011qcdbook}. At the order-$\alpha_s$ we are working in, the expressions are independent of the exact renormzalization scheme $r$ used for the collinear ffs, so we have left it general in \eref{candidate} for now. We will specialize to $r = \msbar$ later.     

The TMD ff should at least approximately 
match its collinear perturbative expansion for $\Tsc{k}{} \approx Q_0$, so $m_D$ and $M$ should be kept small relative to $Q_0$ in any fits.

One of our requirements from step \mref{model_mods} is that parametrizations of the TMD ffs must satisfy the integral relation in 
\eref{cutff1} with \eref{d_little_underline} satisfied. We will use this constraint to fix $C^{(d_r)}$, which so far is just another nonperturbative parameter. Evaluating the transverse momentum integral of \eref{candidate}, expanding in small 
$m/\mu_{Q_0},$ and solving for $C^{(d_r)}$ in terms of $\param{d}_c^{(1,d_r)}(z;\mu_{Q_0})$ gives
\begin{align}
 C^{(d_r)} &{}= \param{d}_c^{(1,d_r)}(z;\mu_{Q_0}) - A^{(d_r)}(z;\mu_{Q_0}) \ln \parz{\frac{\mu_{Q_0}}{m_D}} \no 
&{} -  B^{(d_r)}(z;\mu_{Q_0}) \ln \parz{\frac{\mu_{Q_0}}{m_D}} \ln \parz{\frac{Q_0^2}{\mu_{Q_0} m_D} } \no
&{} + \order{\frac{m}{\mu_{Q_0}}} \, . \label{e.C_val_err}
\end{align}
We still need to choose a form for $\param{d}_c^{(1,d_c)}(z;\mu_{Q_0})$, but our requirement is that it must satisfy \eref{d_little_underline}, where we allow the $m/\mu_{Q_0}$-suppressed contributions to be chosen to give an optimal parametrization. Thus, let us define the power-suppressed terms in \eref{d_little_underline} (again, for $n=1$) to exactly equal those in \eref{C_val_err}.
Then,
\begin{align}
 C^{(d_r)} &{}= d_c^{(1,d_r)}(z;\mu_{Q_0}) - A^{(d_r)}(z;\mu_{Q_0}) \ln \parz{\frac{\mu_{Q_0}}{m_D}} \no 
&{} \qquad -  B^{(d_r)}(z;\mu_{Q_0}) \ln \parz{\frac{\mu_{Q_0}}{m_D}} \ln \parz{\frac{Q_0^2}{\mu_{Q_0} m_D} } \no 
&{}= d_r(z;\mu_{Q_0}) - A^{(d_r)}(z;\mu_{Q_0}) \ln \parz{\frac{\mu_{Q_0}}{m_D}} \no 
&{} \qquad -  B^{(d_r)}(z;\mu_{Q_0}) \ln \parz{\frac{\mu_{Q_0}}{m_D}} \ln \parz{\frac{Q_0^2}{\mu_{Q_0} m_D} } \no
&{} \qquad + \Delta^{(n,d_r)}(\alpha_s(\mu_Q))
\, , \label{e.C_val} 
\end{align}
where in the last line we have used \eref{dnr_rel}. These last few steps are necessary if we wish to relate $C^{(d_r)}$ to known collinear ffs in standard schemes. 

Equation~\eqref{e.candidate}, with \eref{C_val} now for $C^{(d_r)}$, is the parametrization of the TMD ff that is to be substituted into \eref{tmd_factorization_input2} and, in accordance with step \phref{do_pheno_a}, used phenomenologically for describing Type I processes with standard TMD parton model techniques near the input scale $Q \approx Q_0$. 

To get a general sense of what the \eref{candidate} parametrization of the TMD ff looks like at $Q = Q_0$, we have plotted it in \fref{modelD1} using $\msbar$ collinear ffs and 
reasonable values of the mass parameters 
$m_D$, and $M$. (The nonperturbative mass parameters are kept small relative to $Q_0 = 2.0$~GeV.) For this case, $\Delta^{(n,d_{\msbar})}(\alpha_s(\mu_Q))$ is given by \eref{msbar_trans}. The plot is for a sample value of $z = 0.3$, but other values of $z$ produce qualitatively similar curves, as can be easily checked. Since there is a general expectation from existing Type I TMD phenomenology that the small transverse momentum region in moderate $Q$ processes is well-described by Gaussian TMDs, we have overlaid a Gaussian curve on top of \eref{candidate}, confirming that the $\Tsc{k}{} \ll Q_0$ region retains a generally Gaussian shape.

As per step \phref{do_pheno_a}, the small $\Tsc{k}{}$ region is to be described by fitting the $m_D$ and $M$ parameters to measurements. So long as these mass parameters are reasonably small compared to $Q_0$, the parametrization at least approximately recovers the lowest order perturbative description in collinear factorization around $\Tsc{k}{} \approx Q_0$. 
In principle $M$ and $m_D$ can both have $z$-dependence:
\begin{equation}
M \to M(z)\, , \qquad m_D \to m_D(z) \, ,
\end{equation}
but we will not include this in any of our example plots.

While we intend for the above parametrization to be only a toy example to illustrate broader procedural points, it is worth noting that at least some phenomenological support for an additive two-component model (like \eref{candidate}) exists in the observation from \cite{COMPASS:2017mvk} that a sum of two peaked nonperturbative functions provides a good fit to data in the moderate $Q$ region. In that case, the two peaked functions were both Gaussians, but the trend is nevertheless suggestive of a two component form more generally.  Reference~\cite{Anselmino:2006rv} has also confirmed that a combination of a Gaussian peak at small transverse momentum and a power-law tail at large transverse momentum provides a reasonable description of moderate $Q$ data.

The two model parameters in our construction have natural interpretations: The lower-case $m_D$ describes exactly how the TMD ff transitions from the Gaussian-like behavior typically ascribed to nonperturbative dependence to the power-law behavior more typical of a perturbative tail, while the capital $M$ parameters controls the precise shape of the Gaussian peak at small $\Tsc{k}{}$.

The inverse Fourier-Bessel transform into transverse coordinate space is
\begin{align}
&{} z^2 \inpt{\tilde{D}}^{(1,d_r)}(z,\T{b}{};\mu_{Q_0},Q_0^2) =  A^{(d_r)}(z;\mu_{Q_0})  K_0(\Tsc{b}{} m_D) \no 
&{} \qquad +  B^{(d_r)}(z;\mu_{Q_0})   K_0(\Tsc{b}{} m_D) \ln\parz{\frac{\Tsc{b}{}}{2 m_D} Q_0^2 e^{\gamma_E}}  
\no 
&{} \qquad + C^{(d_r)} \exp\parz{-\frac{\Tscsq{b}{} M^2}{4z^2}} \, , \label{e.btdparam2}
\end{align}
where $C^{(d_c)}$ is now defined as in \eref{C_val}.
For checking various properties of the input TMD ff, it will also be useful to have the $m \Tsc{b}{} \to 0$ limit,
\begin{widetext}
\begin{align}
& z^2 \inpt{\tilde{D}}^{(1,d_r)}(z,\T{b}{};\mu_{Q_0},Q_0^2) = \no
&{}  \qquad -A^{(d_r)}(z;\mu_{Q_0})   \ln \parz{\frac{\Tsc{b}{} \mu_{Q_0} e^{\gamma_E}}{2}}   -  
B^{(d_r)}(z;\mu_{Q_0})  \left[      \ln^2 \parz{\frac{\Tsc{b}{} \mu_{Q_0} e^{\gamma_E}}{2}} + \ln \parz{\frac{\Tsc{b}{} \mu_{Q_0} e^{\gamma_E}}{2}} \ln \parz{\frac{Q_0^2}{\mu_{Q_0}^2}} \right] \no
&{} \qquad + d_r(z;\mu_{Q_0}) + \Delta^{(1,d_r)}(\alpha_s(\mu_Q)) + \order{m \Tsc{b}{}} \, ,
\label{e.btdparam2lim}
\end{align}
Or, specializing to $r = \msbar$,
\begin{align}
& z^2 \inpt{\tilde{D}}^{(1,d_{\msbar})}(z,\T{b}{};\mu_{Q_0},Q_0^2) =  
\no
&{}  
\qquad -A^{(d_{\msbar})}(z;\mu_{Q_0})   \ln \parz{\frac{\Tsc{b}{} \mu_{Q_0} e^{\gamma_E}}{2}}   -  
B^{(d_{\msbar})}(z;\mu_{Q_0})  \left[      \ln^2 \parz{\frac{\Tsc{b}{} \mu_{Q_0} e^{\gamma_E}}{2}} + \ln \parz{\frac{\Tsc{b}{} \mu_{Q_0} e^{\gamma_E}}{2}} \ln \parz{\frac{Q_0^2}{\mu_{Q_0}^2}} \right] \no
&{} \qquad +\frac{\alpha_s(\mu_Q)}{2 \pi} \int_{z}^1 \frac{\diff{z'}{}}{z^{'}}
d_{\msbar}(z/z',\mu_Q) \left[ 2 P_{qq}(z') \ln z'  + C_F (1 - z')  \right]  + d_{\msbar}(z;\mu_{Q_0}) + \order{m \Tsc{b}{}} \, ,
\label{e.btdparam2limb2}
\end{align}
\end{widetext}
Equation~\eqref{e.btdparam2limb2} is the usual small-$\Tsc{b}{}$ OPE applied to $\tilde{D}(z,\T{b}{};\mu_{Q_0},Q_0^2)$ in the $\msbar$ scheme through order $\alpha_s(\mu_{Q_0})$. Using \eref{btdparam2lim}, it is straightforward to verify \erefs{CSPDFinpt}{RGinpt} by directly applying to \eref{btdparam2lim} a partial derivative with respect to $Q_0$ and a total derivative with respect to $\mu_{Q_0}$. 

This completes our steps \mref{choice_model} and \mref{model_mods} from \sref{summary}. Now we have a $\inpt{\tilde{D}}^{(1,d_c)}(z,\T{b}{};\mu_{Q_0},Q_0^2)$ parametrization that is suitable for Type I phenomenological applications. 


Finally, for step \mref{tilde} we need to construct a parametrization  $\tilde{\param{D}}^{(n,d_r)}(z,\T{b}{};\mu_{Q_0},Q_0^2)$
that applies not only near the 
input scale $Q \approx Q_0$ but also to $Q \gg Q_0$. To switch to the final underlined parametrization, we can continue to use the same $\overline{Q}_0(\Tsc{b}{})$
from \sref{cskernel} and \aref{interp} that we used for the CS kernel.
Implementing step \mref{tilde} amounts to simply substituting our $\inpt{\tilde{D}}^{(1,d_c)}(z,\T{b}{};\mu_{Q_0},Q_0^2)$ from \eref{btdparam2} into the right-hand side of \eref{evolvedd3p} along with the $\inpt{\tilde{K}}^{(1)}(\Tsc{b}{};\mu_{\overline{Q}_0})$ that we already constructed in \sref{low_order_cs}, \eref{K_param}. The final expression is,
\begin{widetext}
\begin{align}
&\tilde{\param{D}}^{(1,d_{\msbar})}(z,\T{b}{};\mu_{Q_0},Q_0^2) \no
&= \inpt{\tilde{D}}^{(1,d_{\msbar})}(z,\T{b}{};\mu_{\overline{Q}_0},\overline{Q}_0^2)
\exp \left\{
\int_{\mu_{\overline{Q}_0}}^{\mu_{Q_0}} \frac{d \mu^\prime}{\mu^\prime} \left[\gamma^{(1)}(\alpha_s(\mu^\prime);1) 
- \ln \frac{Q_0}{\mu^\prime} \gamma_K^{(1)}(\alpha_s(\mu^\prime))
  \right] +\ln \frac{Q_0}{\overline{Q}_0} \inpt{\tilde{K}}^{(1)}(\Tsc{b}{};\mu_{\overline{Q}_0}) \right\} \, . \label{e.evolvedd4}
\end{align} 
\end{widetext}
This is simply \eref{evolvedd3p} again but now we mean it to be implied that it is being used with the specific models from 
\eref{Kparam_kt} and \eref{candidate}. 
From \eref{btdparam2}, 
\begin{align}
&{} z^2 \inpt{\tilde{D}}^{(1,d_{\msbar})}(z,\T{b}{};\mu_{\overline{Q}_0},\overline{Q}_0^2) = \no
&{} \qquad A^{(d_{\msbar})}(z;\mu_{\overline{Q}_0})  K_0(\Tsc{b}{} m_D)  \no
&{} \qquad +  B^{(d_{\msbar})}(z;\mu_{\overline{Q}_0})   K_0(\Tsc{b}{} m_D) \ln\parz{\frac{\Tsc{b}{}}{2 m_D} \overline{Q}_0^2 e^{\gamma_E}} \no
&{} \qquad + C^{(d_{\msbar})} \exp\parz{-\frac{\Tscsq{b}{} M^2}{4 z^2}} \, , \label{e.btdparam3}
\end{align}
and $\param{\tilde{K}}^{(1)}(\Tsc{b}{};\mu_{\overline{Q}_0})$ is the same $n=1$ result already 
written in \eref{K_param_final}. $C^{(d_{\msbar})}$ is given by \eref{C_val}.
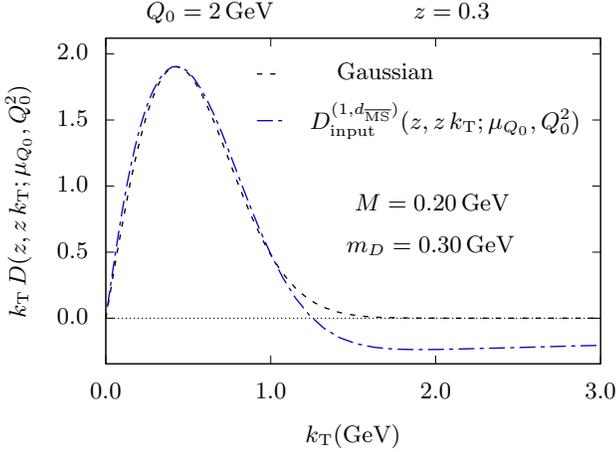
\begin{figure}
\centering
\input{Dinput_inline.tex}
\vspace*{1mm}
\caption{
The $n=1$ input TMD ff from \eref{candidate}. The function is shown for $M=0.2\,\text{GeV}$ and $m_D=0.3\,\text{GeV}$ at a fixed value of $z=0.3$ (blue dot-dashed). For comparison, we have also overlaid a Gaussian (black dashed) curve. Up to $\Tsc{k}{}\approx1.0\,\text{GeV}$, both lines exhibit similar profiles. The change in sign at larger $\Tsc{k}{}$  is due to matching to the perturbative collinear factorizaton expression using $\msbar$ collinear ffs.}
\label{f.modelD1}
\end{figure}

For illustration,~\fref{Dtrans} is a plot of our trial 
$\param{\tilde{D}}^{(1,d_{\msbar})}(z,\T{b}{};\mu_{Q_0},Q_0^2)$ from \eref{evolvedd4}, plotted in coordinate space, where as before we have used an input scale $Q_0=2\,\text{GeV}$ and $z=0.3$. We have chosen typical sizes for the nonperturbative mass parameters: $M=0.2\,\text{GeV}$, $m_D=0.3\,\text{GeV}$ and $m_K=0.1\,\text{GeV}$. As in the case of $\tilde{\param{K}}^{(1)}(\Tsc{b}{};\mu_{Q_0})$, we are able to test scale sensitivity in the intermediate $\Tsc{b}{}$ region by varying the transition function $\overline{Q}_0(\Tsc{b}{})$. In~\fref{Dtrans}, we do this by again switching between the two $\overline{Q}_0(\Tsc{b}{})$ functions from the upper panel of \fref{modelk1}; the solid black and dashed red curves are for $a = 2$~GeV and $a = 4$~GeV respectively. The weakness of the observed variation confirms that the setup is behaving as intended (recall \eref{adiff_Dunderline}). As with $\tilde{\param{K}}^{(1)}(\Tsc{b}{};\mu_{Q_0})$, sensitivity to parameters like $a$ can in principle be reduced still further by including higher orders and fitting at larger $Q$.  
This requires matching to a higher order treatment of the large $\Tsc{q}{}$ tail -- see, for example, Refs.~\cite{Echevarria:2016scs,Luo:2019hmp,Luo:2020epw}.
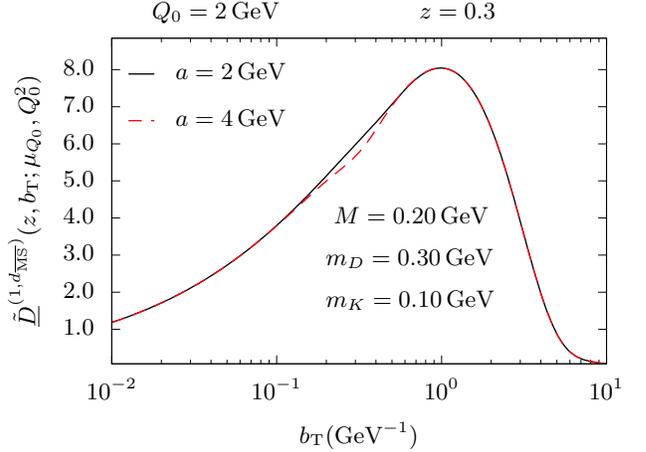
\begin{figure}
\centering
\input{Dtilde_underline_1_inline.tex}
\vspace*{1mm}
\caption{
A plot of the example 
$\param{\tilde{D}}^{(1,d_{\msbar})}(z,\T{b}{};\mu_{Q_0},Q_0^2)$ from \eref{evolvedd4}, with typical nonperturbative mass parameters chosen for illustration purposes; $m_D$ and $M$ have the same values as in \fref{modelD1} while $m_K$ has the same value as in the lower panel of \fref{modelk1}. 
As in all previous plots, we have also fixed $Q_0=2\,\text{GeV}$.
The difference between the two curves corresponds to switching between the two transition functions $\overline{Q}_0(\Tsc{b}{})$ in the upper panel of \fref{modelk1}, as indicated by the two values of $a$.
This figure is the culmination of steps \mref{choice_model}, \mref{model_mods} and \mref{tilde} from \sref{summary}.
}
\label{f.Dtrans}
\end{figure}

\subsection{Cross section examples}
\label{s.cs_example}

\begin{figure}
\centering
\input{WqT_step_B3_2_inline.tex}
\vspace*{1mm}
\caption{
$W^{(1)}(\Tsc{q}{},Q)$ from \eref{Wtermev0_finalversion} calculated using  $\tilde{\param{D}}^{(1,d_{\msbar})}(z,\T{b}{};\mu_{Q_0},Q_0^2)$ and $\tilde{\param{K}}^{(1)}(\Tsc{b}{};\mu_{Q_0})$ from \eref{K_param_final} and \eref{evolvedd4} (solid black curves) for $Q = Q_0$, $2 Q_0$ and $5 Q_0$. Here, as in all remaining plots that we will show, we use an input scale of $Q_0=2\,\text{GeV}$ and $z=0.3$. The scale transformation function $\overline{Q}_0(\Tsc{b}{})$ used to produce these curves is the solid black function in the upper panel of \fref{modelk1} corresponding to $a = 2$~GeV. The model parameters are the same $M=0.2\,\text{GeV}$, $m_D=0.3\,\text{GeV}$ and $m_K=0.1\,\text{GeV}$ that we used in the plots of the previous subsection.
For comparison, $W^{(1)}(\Tsc{q}{},Q)$ calculations are also shown wherein $\inpt{\tilde{K}}^{(1)}(\Tsc{b}{};\mu_{\overline{Q}_0})$ and 
$\inpt{\tilde{D}}^{(1,d_{\msbar})}(z,\T{b}{};\mu_{Q_0},Q_0^2)$ from \eref{K_param} and \eref{btdparam2limb2} are used in \eref{Wtermev0_finalversion} (blue dot-dashed curves), in place of the underlined parametrizations. Note the negligible different for $Q \approx Q_0$ and small $\Tsc{q}{}$. See text for further discussion. 
}
\label{f.WclosetoQ0}
\end{figure}
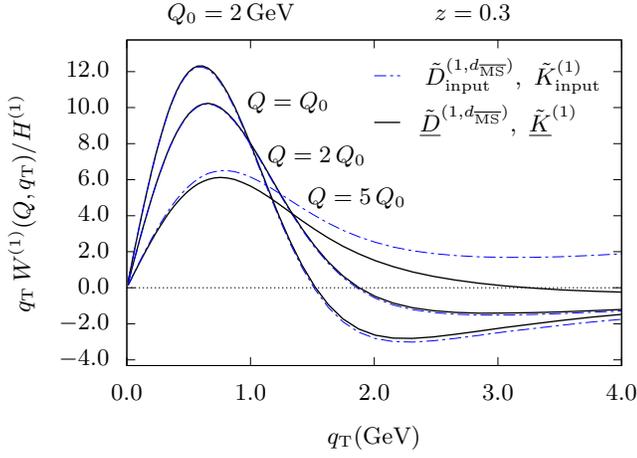
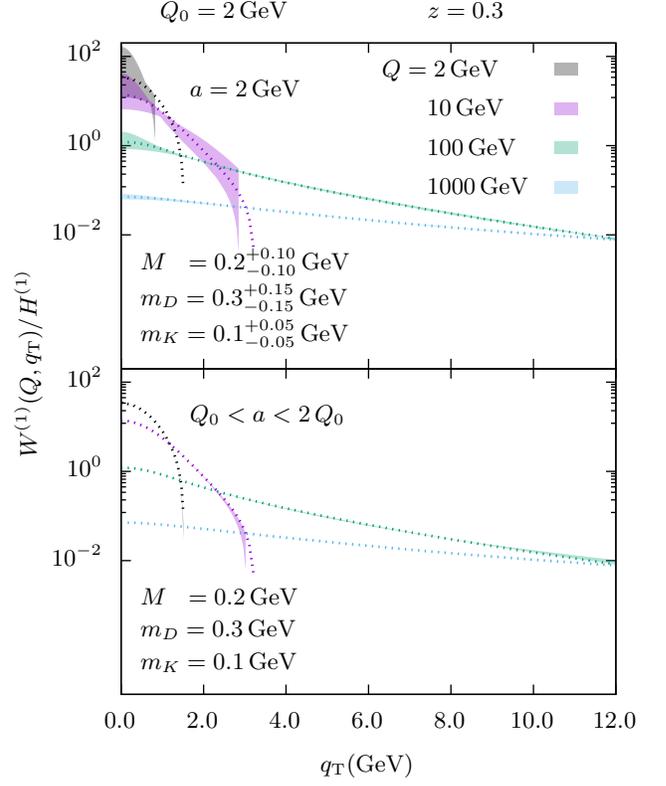
\begin{figure}
\centering
\input{WqT_bands_inline.tex}
\caption{
$W^{(1)}(\Tsc{q}{},Q)$, calculated using the trial underlined parametrizations from this section and evolved to scales much larger than $Q_0$. The dotted lines are calculated using fixed intrinsic scales $M=0.2\,\text{GeV}$, $m_D=0.3\,\text{GeV}$ and $m_K=0.1\,\text{GeV}$ and the usual scale transformation function with  $a=2\,\text{GeV}$.
The colored bands in the top panel show the effect of varying nonperturbative mass parameters by $\pm50\%$ with respect to the values used for dotted lines. The bands in the bottom panel show the perturbative scale sensitivity from varying $a$ between $2\,\text{GeV}$ and $4\,\text{GeV}$, with masses fixed to those of the dotted lines.
For reference, we also show the $Q=Q_0$ case as the black dotted line.
}
\label{f.WlargeQ}
\end{figure}

\begin{figure}
\centering
\input{WqT_bands_linear_inline.tex}
\caption{ 
Same as the top panel of \fref{WlargeQ}, but with the horizontal axes now restricted to $\Tsc{q}{} < 4$~GeV, linear vertical axes, and with each $W^{(1)}(\Tsc{q}{},Q)$ curve normalized to its value at $\Tsc{q}{} = 0$. The upper panel shows the lower $Q$ values, $Q = Q_0 = 2$~GeV and $Q = 10$~GeV while the lower panel shows the large $Q$ values $Q = 100$, $1000$~GeV.
}
\label{f.WlargeQ_linear}
\end{figure}
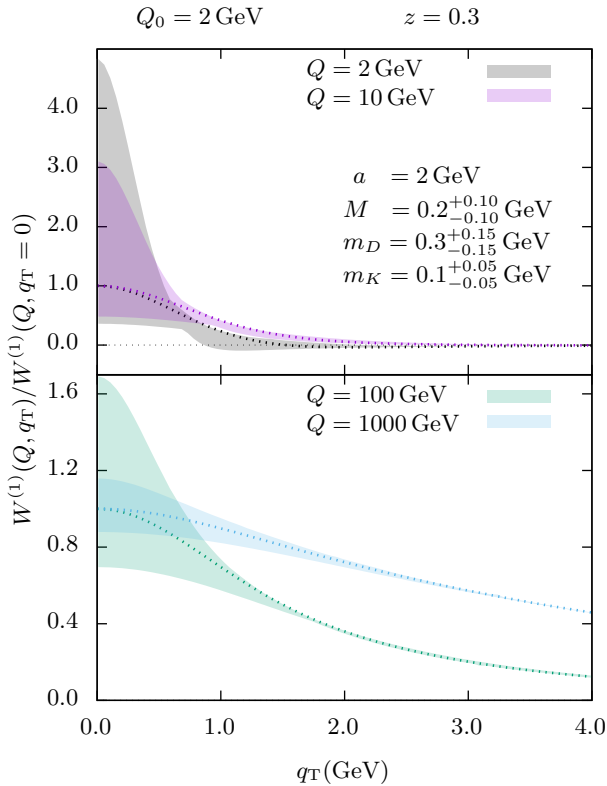


With \eref{evolvedd4} now completely set up, all that is needed to get the cross section is to substitute it, along with \eref{K_param_final}, into  \eref{Wtermev0_finalversion} to obtain a calculation of $W^{(1)}(\Tsc{q}{},Q)$ for any $Q \geq Q_0$. To illustrate how the features of the 
$\tilde{\param{D}}^{(1,d_{\msbar})}(z,\T{b}{};\mu_{Q_0},Q_0^2)$ and $\tilde{\param{K}}^{(1)}(\Tsc{b}{};\mu_{Q_0})$ parametrizations from the previous subsections influence $W^{(1)}(\Tsc{q}{},Q)$, and to finish reviewing the steps of \sref{summary}, we will end this section below by examining several example plots of $W^{(1)}(\Tsc{q}{},Q)$.

First, \fref{WclosetoQ0} shows $\Tsc{q}{} W^{(1)}(\Tsc{q}{},Q)$ (divided by an uninteresting normalization $H^{(1)}$), plotted versus $\Tsc{q}{}$ and with a selection of $Q$ values just above the input scale $Q_0 = 2$~GeV. The nonperturbative parameters are the same $M=0.2\,\text{GeV}$, $m_D=0.3\,\text{GeV}$ and $m_K=0.1\,\text{GeV}$ values that we used in our illustrations from the previous subsection, and the $\overline{Q}_0(\Tsc{b}{})$ is the same transition function from the upper panel of \fref{modelk1} with $a = 2$~GeV. 
The solid black curves are what are obtained if we substitute the underlined functions of \eref{K_param_final} and \eref{evolvedd4} into \eref{Wtermev0_finalversion}. For comparison, the blue dot-dashed curves are what are obtained when we simply substitute the input parametrizations, \eref{K_param} and \eref{btdparam2limb2} into
\eref{Wtermev0_finalversion}, instead of the final underlined parametrizations optimized for the small $\Tsc{b}{}$ limit. Showing both curves confirms that switching between ``$\text{input}$'' and underlined parametrizations results in a negligible difference in the cross section calculation at $\Tsc{q}{} \ll Q$ when $Q$ is only slightly larger than $Q_0$. In \fref{WclosetoQ0}, the difference between the solid black and dot-dashed blue curves is nearly invisible for $Q$ between $Q_0$ and $2 Q_0$, and only becomes significant for $Q \approx 5 Q_0$ and larger $\Tsc{q}{}$. Making these observations corresponds to step \phref{check_tilde} from \sref{summary}. They confirm that the $\text{input}$ or the underlined parametrizations are both valid and interchangeable for phenomenological Type I  applications near $Q \approx Q_0$.

Next, \fref{WlargeQ} shows the $W^{(1)}(\Tsc{q}{},Q)$ calculation (from now on, we will always use the underlined parametrizations), plotted against $\Tsc{q}{}$ on a logarithmic scale for a selection of $Q$ covering a large range between $Q_0 = 2$~GeV and $Q = 1000$~GeV. The top panel is calculated with a fixed scale transformation function $\overline{Q}_0(\Tsc{b}{})$, specifically the $a = 2$~GeV curve in \fref{modelk1}. The bands in the top panel were generated by varying the parameters $M$, $m_D$, and $m_K$ associated with intrinsic transverse momentum by $50\%$ around the values we used in the previous subsections, $M=0.2\,\text{GeV}$, $m_D=0.3\,\text{GeV}$ and $m_K=0.1\,\text{GeV}$. 
The lower panel in \fref{WlargeQ} shows the same $W^{(1)}(\Tsc{q}{},Q)$ calculations, but now with the intrinsic nonperturbative scales fixed to their previous values. Instead, the bands are generated by varying the scale transformation function $\overline{Q}_0(\Tsc{b}{})$ between the two curves in the upper panel of \fref{modelk1} corresponding to $a = 2$~GeV and $a = 4$~GeV. Comparing the upper and lower panels in \fref{WlargeQ} shows that, within the range of parameters considered here, scale sensitivity is far weaker than the sensitivity to intrinsic transverse momentum parameters. On the logarithmic scale, the scale sensitivity is only visible, even at large $Q$, around the node where $W^{(1)}(\Tsc{q}{},Q)$ crosses zero. 

Nevertheless, sensitivity to intrinsic transverse momentum parameters does clearly diminish with increasing $Q$, especially for $\Tsc{q}{} \gtrsim Q_0$. To get a sense of how rapidly it decreases with our parametrizations, we have plotted the top curves from \fref{WlargeQ} again in \fref{WlargeQ_linear}, but now with linear axes and only for the region of $\Tsc{q}{} < 4.0$~GeV in order to magnify sensitivity to variations in $M$, $m_D$, and $m_K$. We have also normalized the curves by their values at $\Tsc{q}{} = 0$. The upper panel in \fref{WlargeQ_linear} shows the bands for smaller $Q = 2$~GeV and $Q = 10$~GeV scales, and it shows that the small transverse momentum region $\Tsc{q}{} \ll Q$ is very sensitive to nonperturbative intrinsic mass scales. At $\Tsc{q}{} = 0$, the width of the band is about an order of magnitude for $Q \approx Q_0$. 
The lower panel of \fref{WlargeQ_linear} shows the bands for the larger $Q = 100$~GeV and $Q = 1000$~GeV scales. There, the sensitivity to intrinsic mass parameters at $\Tsc{q}{} = 0$ is much weaker than for the smaller $Q$, and it becomes essentially invisible above about $\Tsc{q}{} \gtrsim 0.02 Q$. 
This weak sensitivity to intrinsic nonperturbative transverse momentum parameters at $\Tsc{q}{} \gg m$ will be especially important if we need to consistently match to fixed order asymptotic calculations~\cite{Collins:2016hqq}.

In phenomenological applications such as fitting, nonperturbative parameters like $M$, $m_D$, and $m_K$, along with scale setting choices like the value of $a$, become better constrained each time data from somewhat higher $Q$ are included in the fitting. As higher $Q$ are incorporated into fits, and input parameters become better constrained, it eventually becomes unambiguous how to evolve to still higher $Q$. The steps that we have described in this subsection, along with the plots used to illustrate them, 
correspond to steps \evref{evolve_W} and \evref{evolve_W_largerQ} in \sref{summary}.

The illustrative examples in this subsection are to confirm that the setup in \sref{summary} reproduces general expectations. We emphasize once again, before closing the discussion of examples, that our purpose here is not to advocate for a particular choice of a model parametrization for small $\Tsc{k}{}$ dependence, but rather to illustrate the general steps from \sref{summary} in  concrete situations. Ultimately, it is up to phenomenological tests to assess the success of any particular model or calculation of nonperturbative transverse momentum dependence.  

\section{Integral Relations II}
\label{s.int_rels_2}

It is worthwhile to return again to the integral 
relations discussed in \sref{tmd_integrals} in light of parametrizations like the one we constructed above. 
Now recall how integral relations often appear 
in phenomenological extractions of TMD functions near the input scale.  For $Q \approx 1-2$~GeV and $\Tsc{k}{} \ll Q_0$, it is well-known that the shapes of transverse momentum distributions are generally well approximated by Gaussians, so one might reasonably adopt a parametrization of the form
\begin{equation}
D(z,z \T{k}{};\mu_{Q_0},Q_0^2) \stackrel{??}{=}\frac{d(z;\mu_{Q_0})}{\pi M^2} e^{-z^2 \Tscsq{k}{} /M^2} \, , \label{e.gauss_mod}
\end{equation}
where we have dropped renormalization subscripts etc to simplify expressions. The TMD and collinear ffs parametrized in this way automatically satisfy the parton model integral relation
\begin{align}
2 \pi z^2 \int \diff{\Tsc{k}{}}{} \Tsc{k}{} D(z,z \T{k}{};\mu_{Q_0},Q_0^2) = 
d(z;\mu_{Q_0}) \, . 
\label{e.cutff1p}
\end{align} 
A parametrization like \eref{gauss_mod} imposes a strong suppression on large $\Tsc{k}{}$, cutting off the large $\Tsc{k}{}$ tail. Because large $\Tsc{k}{}$-dependence is generally regarded as perturbatively calculable, it is tempting to assume that it is possible to fix 
\eref{gauss_mod} in later steps simply by appending a purely perturbative tail for the region of $\Tsc{k}{} \gtrsim Q_0$ while leaving the initial Gaussian of \eref{gauss_mod} completely unchanged. Taking that view literally would mean that, as regards the integral relation, one does not need to modify either \eref{gauss_mod} or \eref{cutff1p} other than to explicitly indicate an upper cutoff on the integral on the right-hand side of \eref{cutff1p} and to specify that higher orders in $\alpha_s$ and/or $1/Q_0$ 
are neglected:
\begin{align}
&{} 2 \pi z^2 \int_0^{\mu_{Q_0}} \diff{\Tsc{k}{}}{} \Tsc{k}{} D(z,z \T{k}{};\mu_{Q_0},Q_0^2) \no
&{} \qquad \stackrel{??}{=} d(z;\mu_{Q_0}) + \text{small corrections}\, . 
\label{e.cutff1p_cut}
\end{align} 
A problem with this approach can be seen, however, in our examples from the previous section, where \eref{cutff1p_cut} would amount to  applying the transverse momentum integral in \eref{cutff1p_cut} to \eref{candidate} with $C^{(d_r)} = d_r(z;\mu_{Q_0})$ instead of \eref{C_val} so that
\begin{align}
&{} z^2 \inpt{\tilde{D}}^{(1,d_r)}(z,\T{b}{};\mu_{Q_0},Q_0^2) =  A^{(d_r)}(z;\mu_{Q_0})  K_0(\Tsc{b}{} m_D) \no 
&{} \qquad +  B^{(d_r)}(z;\mu_{Q_0})   K_0(\Tsc{b}{} m_D) \ln\parz{\frac{\Tsc{b}{}}{2 m_D} Q_0^2 e^{\gamma_E}}  
\no 
&{} \qquad + d_r(z;\mu_{Q_0}) \exp\parz{-\frac{\Tscsq{b}{} M^2}{4z^2}} \, . \label{e.btdparam2_cut}
\end{align} 
But then the integral on the left side of \eref{cutff1p_cut} becomes
\begin{align}
&{} 2 \pi z^2 \int_0^{\mu_{Q_0}} \diff{\Tsc{k}{}}{} \Tsc{k}{} \inpt{D}(z,z \T{k}{};\mu_{Q_0},Q_0^2) = \no
&{} \; d_r(z;\mu_{Q_0}) + A^{(d_{r})}(z;\mu_{Q_0})  \ln \parz{\frac{\mu_{Q_0}}{m_D}} \no
&{} \; + B^{(d_{r})}(z;\mu_{Q_0}) \ln^2 \parz{\frac{\mu_{Q_0}^2}{m_D^2}} \, . 
\label{e.cutff1pp}
\end{align} 
The last two terms involving $A^{(d_{r})}$ and $B^{(d_{r})}$ would thus need to be identified with the ``small corrections'' of \eref{cutff1p_cut}.
However, the factorization theorem applies to the limit that 
$m/\mu_{Q_0} \to 0$, so the last two terms are not the purely perturbative corrections implied by an expression like \eref{cutff1p_cut}. 
While $A^{(d_{r})}(z;\mu_{Q_0})$ and $B^{(d_{r})}(z;\mu_{Q_0})$ involve a coupling $\alpha_s(\mu_{Q_0})$ that vanishes asymptotically like
$$
\alpha_s(\mu_{Q_0}) \sim \frac{1}{\ln(\mu_{Q_0}/\Lambda_\text{QCD})} \, ,
$$
at large $\mu_{Q_0}$,
the logarithms in \eref{cutff1pp} more than compensate for this in the limit of large $\mu_{Q_0}$. Indeed, the term in \eref{cutff1pp} involving $B^{(d_{r})}(z;\mu_{Q_0})$ blows up as $\mu_{Q_0}/m \to \infty$. To keep a consistent integral relation that matches a parton model interpretation while also accounting for tails, evolution, etc, the coefficient of the Gaussian in \eref{gauss_mod} needs to be 
an expression like \eref{C_val} rather than a simple collinear ff. 

To write a version of \eref{cutff1p_cut}, that is consistent with the presence of a large $\Tsc{k}{}$ tail region, as with our example in \sref{FF_example} with \eref{btdparam2}, it was necessary to interpolate between the nonperturbative and tail regions first. 

\section{Comparison with the standard presentation}
\label{s.css}

After step \evref{evolve_W_largerQ} of \sref{summary}, we noted that it is possible to recast final results into a form more familiar from past applications of TMD evolution in Type II contexts. We will show how to perform that translation in this section. We emphasize that the steps below are not necessary for implementing the approach above, so this section may be skipped without missing the main points of this article. 
Most of the steps below amount to reshuffling factors in the cross section expression. 
Ultimately, however, the translation is important for comparing approaches. 

\subsection{The $\bstarsc$ method}

\begin{figure}
\centering
\input{gk_inline.tex}
\caption{
The example $\tilde{\param{K}}^{(1)}(\Tsc{b}{};\mu_{Q_0})$ from \eref{K_param_final} and the corresponding ${}\param{g}_K^{(1)}(\Tsc{b}{})$ (\eref{gKdef_mod}) and $\tilde{\param{K}}^{(1)}(\bstarsc;\mu_{Q_0})$ calculated in the $\bstarsc$-prescription with \eref{bstar}. Results are obtained using the same $m_K=0.1\,\text{GeV}$ and $a=2\,\text{GeV}$ as in \fref{modelk1}. 
The top panel is the case of 
$\bmax=0.1\,\text{GeV}^{-1}$ and the bottom panel is the case of $b_{\text{max}}=1.0\,\text{GeV}^{-1}$. The $\bmax$-dependence in $\tilde{\param{K}}^{(1)}(b_{*};\mu_{Q_0})$ (violet dot-dashed) cancels that of $-{}\param{g}_K^{(1)}(\Tsc{b}{})$ (blue-dashed). The solid black curve showing $\tilde{\param{K}}^{(1)}(\Tsc{b}{};\mu_{Q_0})$ is identical in the top and bottom.
}
\label{f.gk_underline}
\end{figure}
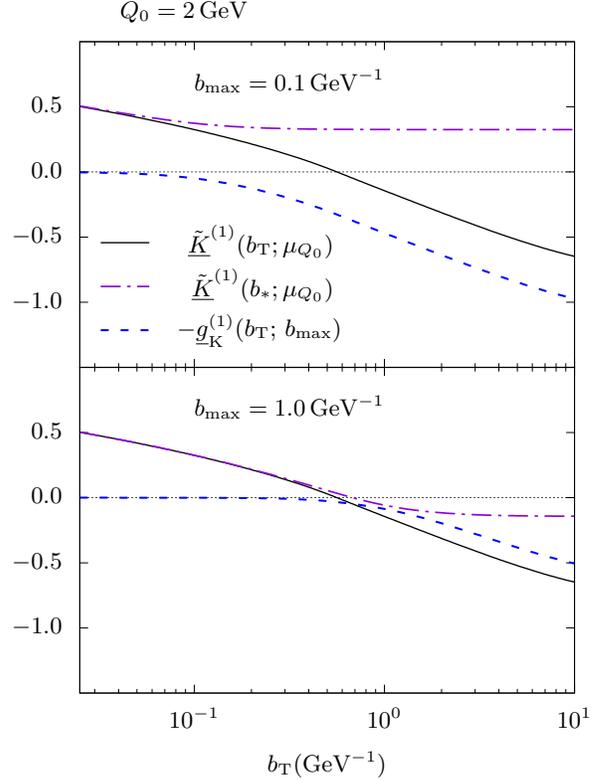
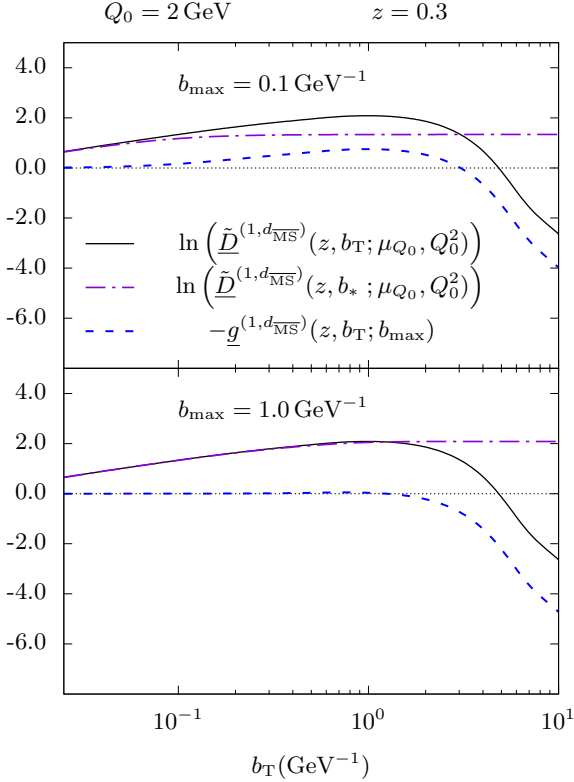
\begin{figure}
\centering
\input{gD_inline.tex}
\caption{
The example $\tilde{\param{D}}^{(1,d_{\msbar})}(z,\Tsc{b}{};\mu_{Q_0},Q_0^2)$ from \eref{evolvedd4}, and the corresponding ${}\param{g}^{(1,d_{\msbar})}(z,\Tsc{b}{})$ (\eref{gdef_mod}) and $\tilde{\param{D}}^{(1,d_{\msbar})}(z,\bstarsc;\mu_{Q_0},Q_0^2)$ from the $\bstarsc$-prescription (\eref{bstar}). The curves are generated using the same $M=0.2\,\text{GeV}$, $m_D=0.3\,\text{GeV}$, and $a=2\,\text{GeV}$ as in \fref{WclosetoQ0}. 
The top panel is the case of 
$\bmax=0.1\,\text{GeV}^{-1}$ and the bottom panel is the case of  $\bmax=1.0\,\text{GeV}^{-1}$. The $\bmax$-dependence in $\ln(\tilde{\param{D}}^{(1,d_{\msbar})}(z,b_{*};\mu_{Q_0},Q_0^2))$ (violet dot-dashed) cancels that of $-\param{g}^{(1,d_{\msbar})}(z,\Tsc{b}{})$ (blue-dashed). The solid black curves for $\tilde{\param{D}}^{(1,d_{\msbar})}(z,\Tsc{b}{};\mu_{Q_0},Q_0^2)$ are identical in the top and bottom panels.
}
\label{f.gd_underline}
\end{figure}
Readers who are familiar with standard implementations of the CSS formalism might find the surface appearance of our expressions for the evolved $W^{(n)}(\Tsc{q}{},Q)$ in sections \sref{cskernel}-\sref{low_order} somewhat odd. Normally, the cross section is written with nonperturbative TMD effects contained in separate exponential factors usually notated 
\begin{equation}
\label{e.g_funcs}
e^{-g_A(z_A,\Tsc{b}{})}, \;\; e^{-g_B(z_B,\Tsc{b}{})},  \;\; \text{and} \;\; e^{-g_K(\Tsc{b}{}) \ln \frac{Q^2}{Q_0^2}} \, ,
\end{equation}
where the nonperturbative transverse momentum dependence is encoded in the (coordinate dependence of) the lower-case $g$-functions in the exponents. (The $g_A$ and $g_B$ functions are placed inside exponents so that they retain the appearance of Sudakov form factor contributions.)
Then, in the usual presentation, the rest of the factors in the cross section automatically get expressed in terms of collinear functions by using the OPE to approximate the small $\Tsc{b}{}$ behavior. For a specific example of what we mean here, consider Eq.(13.81) of \cite{Collins:2011qcdbook}. 

In this subsection, we will review the steps for transforming the low-$\Tsc{q}{}$ cross section (or rather $W(\Tsc{q}{},Q)$) in \eref{Wtermev0} into the form that involves \eref{g_funcs} $g$-functions.  

Setting up the usual presentation of the cross section begins with a partition of the coordinate space $\tilde{W}(\Tsc{b}{},Q)$ into regions considered large and small $\Tsc{b}{}$. One does this by defining an arbitrary function of $\Tsc{b}{}$, traditionally called $\bstar(\Tsc{b}{})$. The 
function should smoothly interpolate between $\T{b}{}$ at small values of $\Tsc{b}{}$ and a maximum transverse size $\vect{b}_{\rm max}$ as $\Tsc{b}{}$ grows to $\Tsc{b}{} \gg \bmax$. It is otherwise arbitrary. In other words, 
\begin{equation}
\bstar(\Tsc{b}{}) = 
\begin{dcases}
\T{b}{} & b_T \ll b_{\rm max} \\
\vect{b}_{\rm max} & b_T \gg b_{\rm max} \,  \label{e.bdefold}
\end{dcases}\, .
\end{equation}
The value of $\bmax$ is also arbitrary, but it is usually interpreted roughly as a value somewhere near the boundary between nonperturbatively large and perturbatively small regions of $\Tsc{b}{}$. 
The purpose of the ``$\bstar$ method''~
\cite{Collins:1981va} is to sequester a purely perturbative 
calculation of transverse coordinate dependence away 
from a part that involves nonperturbative 
modeling or fitting.
While any reasonably well-behaved, smooth function of $\Tsc{b}{}$ that obeys the right side of~\eref{bdefold} is a valid $\bstar(\Tsc{b}{})$, the most often used choice is
\begin{align}
\label{e.bstar}
  \bstar(\Tsc{b}{}) = \frac{ \T{b}{} }{ \sqrt{ 1 + \Tscsq{b}{}/\bmax^2} } \, . 
\end{align}
Later on, we will also need to define another hard scale that approaches the RG improved value of $\mu = C_1/\Tsc{b}{}$ appropriate to the $\Tsc{b}{} \to 0$ limit but that levels off at a fixed scale at large $\Tsc{b}{}$.  The simplest (and standard) way to do this is to just use the inverse of $\bstarsc$ and define
\begin{equation}
\label{e.mubst}
\mu_{\bstarsc} \equiv C_1/\bstarsc \, .
\end{equation}
The behavior of $\mu_{\bstarsc}$ is similar to that of our $\overline{Q}_0(\Tsc{b}{})$, but $\overline{Q}_0(\Tsc{b}{})$ approaches $Q_0$ at large $\Tsc{b}{}$ while $\mu_{\bstarsc}$ approaches $C_1/\bmax$. (Indeed, in our treatment below we could opt to use $\overline{Q}_0(\Tsc{b}{})$ instead of $\mu_{\bstarsc}$, but we will continue with $\mu_{\bstarsc}$ to make the comparison with standard expressions clear.)  

Next, one solves \eref{RG} to relate a TMD ff (for hadron A, for example) at an input scale $Q_0$ to the TMD ff at 
any other scale $\sqrt{\zeta}$ by
\begin{align}
&{}\tilde{D}_A(z,\T{b}{};\mu,\zeta) \no
&{} \qquad = \tilde{D}_A(z,\T{b}{};\mu,Q_0^2) \exp \left\{ \tilde{K}(\Tsc{b}{};\mu) \ln \parz{\frac{\sqrt{\zeta}}{Q_0}} \right\} \, .
\label{e.Dzeta}
\end{align}
Exactly the same equation applies independently of the  
transverse coordinate $\T{b}{}$, so we also have
\begin{align}
&{} \tilde{D}_A(z,\bstar;\mu,\zeta) \no
&{} \qquad = 
\tilde{D}_A(z,\bstar;\mu,Q_0^2) \exp \left\{ \tilde{K}(\bstarsc;\mu) \ln \parz{\frac{\sqrt{\zeta}}{Q_0}} \right\} \, .
\label{e.Dzeta_star}
\end{align}
Then the ratio of \eref{Dzeta} and \eref{Dzeta_star} is
\begin{align}
&{} \frac{\tilde{D}_A(z,\T{b}{};\mu,\zeta)}{\tilde{D}_A(z,\bstar;\mu,\zeta)} = \frac{\tilde{D}_A(z,\T{b}{};\mu,Q_0^2)}{\tilde{D}_A(z,\bstar;\mu,Q_0^2)} \times \no
&{} \qquad \times \exp \left\{ -\left[ \tilde{K}(\bstarsc,\mu) -\tilde{K}(\Tsc{b}{},\mu) \right] \ln \parz{\frac{\sqrt{\zeta}}{Q_0}}  \right\} \no
&{} =
\frac{\tilde{D}_A(z,\T{b}{};\mu,Q_0^2)}{\tilde{D}_A(z,\bstar;\mu,Q_0^2)} \exp \left\{ -g_K(\Tsc{b}{}) \ln \parz{\frac{\sqrt{\zeta}}{Q_0}}  \right\} \no
&{} =
\frac{\tilde{D}_A(z,\T{b}{};\mu_{Q_0},Q_0^2)}{\tilde{D}_A(z,\bstar;\mu_{Q_0},Q_0^2)} \exp \left\{ -g_K(\Tsc{b}{}) \ln \parz{\frac{\sqrt{\zeta}}{Q_0}}  \right\} \, , \label{e.ratio_one}
\end{align}
where on the second line we have \emph{defined}
\begin{equation}
\label{e.gKdef}
  g_K(\Tsc{b}{}) 
  \equiv \tilde{K}(\bstarsc,\mu) -\tilde{K}(\Tsc{b}{},\mu) \, .
\end{equation}
Since the $\mu$-dependence 
of $\tilde{K}(\Tsc{b}{},\mu)$ is also  $\Tsc{b}{}$-independent, $g_K(\Tsc{b}{})$ is 
$\mu$-independent. That is, $\mu$-dependence cancels between the two terms, so $g_K(\Tsc{b}{})$ is $\mu$-independent by definition. 
Also, on the last line of \eref{ratio_one} we have used that the $\mu$-dependence of $\tilde{D}_A(z,\T{b}{};\mu,Q_0^2)$ is a $\Tsc{b}{}$-independent overall factor -- recall the evolution equation in \eref{RGgamma} -- to specialize to the case of $\mu = \mu_{Q_0}$. 

Next, one defines the logarithm of the ratio on the last line of \eref{ratio_one} by the symbol $-g_A(z,\T{b}{})$:
\begin{align}
-g_A(z,\T{b}{}) &{} \equiv \ln \parz{ \frac{\tilde{D}_A(z,\T{b}{};\mu_{Q_0},Q_0^2)}{\tilde{D}_A(z,\bstar;\mu_{Q_0},Q_0^2)}}\, , \label{e.gdef}
\end{align}
with the $A$ subscript reminding of potential sensitivity to the identity of the final state hadron. 
Combining \eref{ratio_one} and \eref{gdef} gives
\begin{align}
&{} \tilde{D}_A(z,\T{b}{};\mu,\zeta) 
 = 
\tilde{D}_A(z,\bstar;\mu,\zeta) \times \no \qquad 
&{} \times \exp\left\{ -g_A(z,\Tsc{b}{}) - g_K(\Tsc{b}{}) \ln \parz{\frac{\sqrt{\zeta}}{Q_0}} \right\} \, .
\label{e.pre_evol}
\end{align}
The $\tilde{D}_A(z,\bstar;\mu,\zeta)$ on the 
right-hand side is still the exact operator definition, but it is only ever evaluated at $\Tsc{b}{} \leq \bmax$. 
The remaining exponential factor is sensitive to
the large $\Tsc{b}{}$ region. 
As of yet, there are no approximations. 
In particular, any sensitivity to 
$\bmax$ or the choice of the $\bstar$ parametrization in \eref{bdefold} cancels exactly between the factors on the right-hand side of \eref{pre_evol}. We have simply taken the original definition 
of $\tilde{D}_A(z,\bstar;\mu,\zeta)$ and partitioned it into two factors.

The logarithm on the right side of the definition in \eref{gdef} is cosmetic; expressing the  nonperturbative ratio as the exponential of a function $-g_A(z,\Tsc{b}{})$ gives it the appearance of a type of contribution to a Sudakov exponent. 

Despite the apparent arbitrariness of the above steps, one can aniticipate the motivation for writing the TMD ff as in \eref{pre_evol} by looking ahead. We obtain the full cross section by substituting \eref{pre_evol} into the evolved $W(\Tsc{q}{},Q)$ in \eref{Wtermev0} with $\mu = \mu_{Q_0}$ and $\sqrt{\zeta} = Q_0$. In the resulting cross section expression,  $\tilde{D}_A(z,\bstar;\mu_{Q_0},Q_0^2)$ will be well-approximated by collinear factorization at $\Tsc{b}{} \approx 1/Q_0$ so long as $\bmax \approx 1/Q_0$ and $Q_0$ is reasonably large compared to nonperturbative scales. This would be sufficient for calculations with $Q \approx Q_0$, where the $1/Q_0 \lesssim \Tsc{b}{} < \infty$ region is the only relevant contribution. However, if we plan to evolve to very large $Q$, then we also need an accurate treatment of $\tilde{D}_A(z,\bstar;\mu_{Q_0},Q_0^2)$ in the $\Tsc{b}{} \ll 1/Q_0$ limit.  But the fixed order calculations of $\tilde{D}_A(z,\bstar;\mu_{Q_0},Q_0^2)$ in collinear factorization are poorly behaved as $\Tsc{b}{} \mu_{Q_0} \to 0$, even though this is the limit where perturbative QCD should be most reliable. 

As usual, therefore, we need to apply the evolution equations (\eref{evolvedd}) once again in order to evolve $\tilde{D}_A(z,\bstar;\mu,\zeta)$ from $\mu, \zeta$ to the RG-improved $\mu_{\bstarsc}, \mu_{\bstarsc}^2$. The evolution equations allow us to rewrite \eref{pre_evol} as
\begin{widetext}
\begin{align}
\tilde{D}_A(z,\T{b}{};\mu,\zeta) &{}= 
\tilde{D}_A(z,\bstar;\mu,\zeta) \,\exp\left\{ -g_A(z,\Tsc{b}{}) - g_K(\Tsc{b}{}) \ln \parz{\frac{\sqrt{\zeta}}{Q_0}} \right\} \no
&{}=
\tilde{D}_A(z,\bstar;\mubstar,\mubstar^2)
\exp\left\{ \int_{\mubstar}^{\mu} \frac{d \mu^\prime}{\mu^\prime} \left[\gamma(\alpha_s(\mu^\prime);1) 
- \ln \frac{\sqrt{\zeta}}{\mu^\prime} \gamma_K(\alpha_s(\mu^\prime))
  \right] +\ln \frac{\sqrt{\zeta}}{\mubstar} \tilde{K}(\bstarsc;\mubstar)
  \right\} \no
&{} \qquad \times \exp\left\{
-g_A(z,\Tsc{b}{}) - g_K(\Tsc{b}{}) \ln \parz{\frac{\sqrt{\zeta}}{Q_0}} \right\} \, .
\label{e.Dnoapp}
\end{align}
Recall that $\mu_{\bstarsc}$ is defined in \eref{mubst}.
There is of course an exactly analogous equation for $\tilde{D}_B(z,\bstar;\mubstar,\mubstar^2)$.   Substituting the evolved versions of $\tilde{D}_A(z,\bstar;\mubstar,\mubstar^2)$ and 
$\tilde{D}_B(z,\bstar;\mubstar,\mubstar^2)$ into the $W$-term factorization formula \eref{tmd_factorization} and setting the final scales equal to $\mu = \mu_Q$ and $\zeta = Q^2$ in \eref{tmd_factorization} gives
\begin{align}
W(\Tsc{q}{},Q) &{}=  H(\mu_{Q};C_2)
    \int \frac{\diff[2]{\T{b}{}}}{(2 \pi)^2}
    ~ e^{-i\T{q}{}\cdot \T{b}{} } 
    \tilde{D}_{A}(z_A,\bstar;\mubstar,\mubstar^2) 
    \tilde{D}_{B}(z_B,\bstar;\mubstar,\mubstar^2) \no
    &{} \times \exp
    \left\{ 2 \int_{\mubstar}^{\mu_{Q}} \frac{d \mu^\prime}{\mu^\prime} \left[\gamma(\alpha_s(\mu^\prime);1) 
- \ln \frac{Q}{\mu^\prime} \gamma_K(\alpha_s(\mu^\prime))
  \right] +\ln \frac{Q^2}{\mubstar^2} \tilde{K}(\bstarsc;\mubstar)
  \right\}
  \no
&{} \times \exp\left\{
-g_A(z_A,\Tsc{b}{}) - g_B(z_B,\Tsc{b}{}) - g_K(\Tsc{b}{}) \ln \parz{\frac{Q^2}{Q_0^2}} \right\} \, . \label{e.before_app}
\end{align}
\end{widetext}
Equation~\eqref{e.before_app} is very close to the standard way of expressing the CSS-evolved $W$-term.\footnote{There are, however, a large number of minor but not always obvious variations in the form of the expression in the literature. There are also many different systems of notation. See \cite{Collins:2017oxh} for some translation.} As we have written 
it, there are still no approximations; the solutions to the evolution equations are exact and the steps above simply reorganize the original factorization formula in \eref{tmd_factorization}. However, by writing $W(\Tsc{q}{},Q)$ as in \eref{before_app}, we have isolated on the first two lines those factors that can be confidently approximated in perturbation theory using collinear factorization. The value of $\Tsc{b}{}$ never rises above $\bmax$ and the scale $\mu_{\bstarsc}$ never drops below $C_1/\bmax$. Therefore, one obtains well-behaved perturbative calculations by replacing $H(\mu_{Q};C_2)$, $\gamma(\alpha_s(\mu^\prime);1)$, $\gamma_K(\alpha_s(\mu^\prime))$ and $\tilde{K}(\bstarsc;\mubstar)$ by their $n^\text{th}$-order perturbative calculations. 

For the TMD ffs themselves on the first line, the choice of $\mu = \sqrt{\zeta} = \mubstar$ implements RG improvement for the limit of small $\Tsc{b}{}$. As long as $\bmax$ is small enough, $\tilde{D}_{A,B}(z_{A,B},\bstar;\mubstar,\mubstar^2)$ can be expanded in an OPE:
\begin{align}
&\tilde{D}^{(n,d_r)}(z,\bstar;\mubstar,\mubstar^2) \no
&{} = \int_z^1 \frac{\diff{\hat{z}}{}}{\hat{z}^{3 - 2 \epsilon}} d_r(\hat{z};\mu_{\bstarsc}) \tilde{C}^{(n)}_D(z/\hat{z},\Tsc{b}{};\mu_{\bstarsc}^2,\mu_{\bstarsc},\alpha_s(\mu_{\bstarsc})) \no
&{} + \order{m \Tsc{b}{}} \, , \label{e.OPE_for_ff}
\end{align}
which is a more explicit version of \eref{D_fact} but in $\Tsc{b}{}$-space. (Here, as usual, $m$ represents any of the small intrinsic mass scales, including now $1/\bmax$.) Substituting \eref{OPE_for_ff} for both 
$\tilde{D}_{A}(z_{A},\bstar;\mubstar,\mubstar^2)$ and 
$\tilde{D}_{B}(z_{B},\bstar;\mubstar,\mubstar^2)$, along with the other perturbative approximations mentioned above, recovers the standard CSS expression -- compare, for example, with the Drell-Yan version of TMD factorization in Eq.~(22) of \cite{Collins:2014jpa}.

The $\bstarsc$ method, as it is explained here, has 
several desirable properties. There is the elegant feature that, in dealing with the nonperturbative region of large $\Tsc{b}{}$, one never modifies or approximates the operator definitions of the TMD ffs themselves. 
Rather, on the first line of \eref{before_app} we have simply changed their arguments from $\Tsc{b}{}$ to $\bstarsc$. Along the same lines, the $g$-functions on the last line have explicit definitions in terms of the underlying QCD operators. The final result for the cross section, \eref{before_app}, is exactly independent of the choice of the $\bstar(\Tsc{b}{})$ function in
\eref{bdefold} or of the value of parameters like $\bmax$. Since changing them simply amounts to reshuffling contributions between the perturbative and non-perturbative factors, the $\bstarsc$-independence is a version of RG invariance that we can express as
\begin{equation}
\frac{\diff{}}{\diff{\bmax}} W(\Tsc{q}{},Q) = 0 \, .
\label{e.bmax_rg}
\end{equation}
Or, if we consider other more general $\bstarsc(\Tsc{b}{})$ functions determined by a collection of possibly many parameters $\{ \text{b-params} \}$, we can express the same relation schematically as 
\begin{equation}
\frac{\diff{}}{\diff{\{ \text{b-params} \}}} W(\Tsc{q}{},Q) = 0 \, .
\label{e.params_rg}
\end{equation}
These relations are exact for \eref{before_app}.
Therefore, it is legitimate to say that perturbative calculations of the first two lines in \eref{before_app} is completely perturbative in that they have no dependence on nonperturbative parameters beyond whatever power suppressed errors are introduced when we substitute \eref{OPE_for_ff}. Thus, the $\bstarsc$-method is not itself a model when used as originally set up. It is only a method for reorganizing contributions between the various factors in \eref{before_app}.

For all of this to be preserved in practice when performing calculations and applying them to phenomenology, the \eref{g_funcs} $g$-functions need to be parametrized nonperturbatively but in a way that preserves \eref{params_rg}. If a change of $\bstarsc$ or $\bmax$ induces a change in the perturbative parts, a compensating change is needed in the \eref{g_funcs} functions. (For more on this, see for example the discussion in section~13.13.2 of \cite{Collins:2011qcdbook}.) When specific paramtrizations and approximations are substituted for the symbols in \eref{before_app}, the statements above that are exact become approximate. Ideally, however, one hopes to do this in such a way that the errors are small and controlled. 

But implementing this style of approach in practice is complicated by the fact that it leaves very little said about the details of the \eref{g_funcs} $g$-functions other than that they must vanish like a power at small $\Tsc{b}{}$. In particular, it leaves unaddressed the treatment of the transition between the purely perturbative and purely nonperturbative regions in the parametrizations, along with the role that integral relations like \eref{cutff1p_cut} play in that transition. For \emph{very} large $Q$, where the $g$-functions are anyway expected to give only small corrections, these details might not have an important effect on calculations. But they become important once one enters regimes where one is sensitive to the details of hadron structure.
In the earliest implementations, $\Tscsq{b}{}$ power-laws~\cite{Davies:1984sp} and/or $\Tsc{b}{}$  power-laws~\cite{Ladinsky:1993zn} were proposed, and these work reasonably well for many applications. But if a $g$-function from \eref{g_funcs} is modeled with a simple ansatz, relations like \eref{params_rg} tend to be violated in ways that can significantly affect  results~\cite{Qiu:2000hf,Qiu:2000ga}. In some applications, the $\bstarsc$-method and the choice of functions like~\eref{bdefold} get reinterpreted as being part of the nonperturbative model itself, due to the large $\bstarsc$ sensitivity that is introduced by most simple ansatzes as discussed above. But then the perturbative calculation of the small $\Tsc{b}{}$ behavior become intertwined with the nonperturbative modeling, which is a situation that the $\bstarsc$ method of organization in \eref{before_app} is meant to avoid.  

These and related complications are all very well known, and efforts have been made to overcome them. One essentially needs to reverse engineer the ansatzes for the $g$-functions to recover some approximate consistency with relations like \eref{params_rg}.
For example, Refs.~\cite{Qiu:2000hf,Grewal:2020hoc} impose continuity for both the functions and their derivatives at a point $\bmax$ in $\Tsc{b}{}$-space that they designate as a boundary between perturbative and nonperturbative regions. Similarly, in treating  $g_K(\Tsc{b}{})$ Ref.~\cite{Collins:2014jpa} proposed to expand the nonperturbative parametrization in a power series of $\bmax^2$ at small $\Tsc{b}{}$ and match to the corresponding powers in the perturbative part of $\tilde{K}{}(\bstar;\mu)$.

\subsection{Bottom-up in the $\bstarsc$ method}

Since the bottom-up procedure that we described in Secs.~\ref{s.cskernel}-\ref{s.summary} explicitly interpolates between nonperturbative and perturbative transverse momentum dependence, there was no need for a $\bstarsc$ method. The $\overline{Q}_0(\Tsc{b}{})$ that we introduced in \eref{bdef} plays a role that is in some sense analogous to the $\mubstar$ from \eref{mubst} in that both impose an RG improved $\mu \sim 1/\Tsc{b}{}$ scale in the small $\Tsc{b}{}$ limit. However, in the earlier sections of this paper we never attempted to completely sequester purely perturbative and purely nonperturbative parts. 

Nevertheless, the steps for the $\bstarsc$-method carry over directly to our final expressions from the bottom-up procedure and \erefs{evolvedd3p}{Wtermev0_finalversion}, and the problems summarized at the end of the last subsection are automatically avoided. One simply arrives at expressions for \eref{g_funcs} that are connected to specific nonperturbative TMD models or calculations, constrained from the outset to satisfy all important properties.  In this subsection, we will show the details of how to rewrite the bottom-up expression in \eref{Wtermev0_finalversion} using the $\bstarsc$-method. Most steps are simply a repetition of the normal way of writing the CSS evolution in terms of $\bstar$, $\mubstar$, etc, that we reviewed in the last subsection but now using the underlined functions and their evolution equations.

By construction, all of the underlined 
objects satisfy the specific evolution equations in \eref{param_K_RG2}, \eref{CSPDF_underline}, and \eref{RGgamma_underline} exactly.
Therefore, steps identical to those leading from~\eref{Dzeta} to \eref{before_app} apply also to 
\eref{Wtermev0_finalversion} as long as we maintain all the appropriate underlines and ``$(n)$'' superscripts everywhere. They thus allow us to rewrite $W^{(n)}(\Tsc{q}{},Q)$ as 
\begin{widetext}
\begin{align}
W^{(n)}(\Tsc{q}{},Q) &{}=  H^{(n)}(\mu_{Q};C_2)
    \int \frac{\diff[2]{\T{b}{}}}{(2 \pi)^2}
    ~ e^{-i\T{q}{}\cdot \T{b}{} } 
    \tilde{\param{D}}_{A}^{(n,d_{\msbar})}(z_A,\bstar;\mubstar,\mubstar^2) 
    \tilde{\param{D}}_{B}^{(n,d_{\msbar})}(z_B,\bstar;\mubstar,\mubstar^2) \no
    &{} \times \exp
    \left\{ 2 \int_{\mubstar}^{\mu_{Q}} \frac{d \mu^\prime}{\mu^\prime} \left[\gamma^{(n)}(\alpha_s(\mu^\prime);1) 
- \ln \frac{Q}{\mu^\prime} \gamma_K^{(n)}(\alpha_s(\mu^\prime))
  \right] +\ln \frac{Q^2}{\mubstar^2} \tilde{\param{K}}^{(n)}(\bstarsc;\mubstar)
  \right\}
  \no
&{} \times \exp\left\{
-\param{g}_A^{(n,d_{\msbar})}(z_A,\Tsc{b}{}) - \param{g}_B^{(n,d_{\msbar})}(z_B,\Tsc{b}{}) - \param{g}^{(n)}_K(\Tsc{b}{}) \ln \parz{\frac{Q^2}{Q_0^2}} \right\} \, , \label{e.before_app_bstar}
\end{align}
\end{widetext}
but now with
\begin{align}
 &{}\param{g}_K^{(n)}(\Tsc{b}{}) 
 \equiv \tilde{\param{K}}^{(n)}(\bstarsc,\mu_{Q_0}) -\tilde{\param{K}}^{(n)}(\Tsc{b}{},\mu_{Q_0}) \, , \label{e.gKdef_mod} \\
-&{}\param{g}^{(n,d_{\msbar})}(z,\T{b}{}) \no
&{} \qquad \equiv \ln \parz{ \frac{\tilde{\param{D}}^{(n,d_{\msbar})}(z,\T{b}{};\mu_{Q_0},Q_0^2)}{\tilde{\param{D}}^{(n,d_{\msbar})}(z,\bstar;\mu_{Q_0},Q_0^2)}}\,.  \label{e.gdef_mod}
\end{align}
The functions defined in \eref{gKdef_mod} and \eref{gdef_mod} are now written
in terms of whatever model or nonperturbative calculation is being used at $\mu=\mu_{Q_0}$, but
just as with \eref{gKdef} and \eref{gdef}, they are $\mu$-independent. The underlines and superscripts indicate that these $g$-functions are defined with specific model parametrization and perturbative calculations in mind, rather than the purely abstract $g$-functions of \erefs{gKdef}{gdef}. 

It is instructive to substitute the example
parametrizations, \eref{K_param_final} and \eref{evolvedd4}, that we constructed in \sref{low_order} into \erefs{before_app_bstar}{gdef_mod} and confirm the $\bmax$-independence and other expected properties. In \fref{gk_underline}, the solid black curves are the same $\tilde{\param{K}}^{(1)}(\Tsc{b}{};\mu_{Q_0})$ as in \fref{modelk1} and \fref{NPevoratio}, but for comparison we have also shown the separate $\tilde{\param{K}}^{(1)}(\bstarsc;\mu_{Q_0})$ and $-\param{g}_K^{(1)}(\Tsc{b}{})$ curves. The two panels compare the graphs for different values of $\bmax$: the top panel is for $\bmax = 0.1$~GeV$^{-1}$ while the bottom panel is for 
$\bmax = 1.0$~GeV$^{-1}$. In both cases, $-\param{g}_K^{(1)}(\Tsc{b}{})$ approaches zero as $\Tsc{b}{} m \to 0$, and $\tilde{\param{K}}^{(1)}(\bstarsc;\mu_{Q_0})$ approaches the unapproximated $\tilde{\param{K}}^{(1)}(\Tsc{b}{};\mu_{Q_0})$ as $\Tsc{b}{} m \to 0$.
Ultimately, it is only 
\begin{equation}
\label{e.K_separated}
\tilde{\param{K}}^{(1)}(\Tsc{b}{};\mu_{Q_0}) = 
\tilde{\param{K}}^{(1)}(\bstarsc;\mu_{Q_0}) - \param{g}_K^{(1)}(\Tsc{b}{})
\end{equation}
that appears in \eref{Wtermev0_finalversion}, and indeed the sum of the dashed and dot-dashed curves in \fref{gk_underline} always reproduces exactly the solid black $\tilde{\param{K}}^{(1)}(\Tsc{b}{};\mu_{Q_0})$ curve. The separate terms in \eref{K_separated} are different for different $\bmax$ but of course $\tilde{\param{K}}^{(1)}(\Tsc{b}{};\mu_{Q_0})$ is not. 

A similar comparison for $-\param{g}_A^{(1,d_{\msbar})}(z,\T{b}{})$ is shown in \fref{gd_underline}. Now from \eref{gdef_mod} the combination of terms that is independent of $\bmax$ is
\begin{align}
&{} \ln \parz{ \tilde{\param{D}}^{(1,d_{\msbar})}(z,\T{b}{};\mu_{Q_0},Q_0^2)} \no
&{}= \ln \parz{ \tilde{\param{D}}^{(1,d_{\msbar})}(z,\bstar;\mu_{Q_0},Q_0^2)} -\param{g}^{(1,d_{\msbar})}(z,\T{b}{}) \, .
\end{align}
The solid black curve in \fref{gd_underline} is 
$\ln \parz{ \tilde{\param{D}}^{(1,d_{\msbar})}(z,\T{b}{};\mu_{Q_0},Q_0^2)}$ and it is the same for the upper ($\bmax = 0.1$~GeV$^{-1}$) and lower ($\bmax = 1.0$~GeV$^{-1}$) panels. The dashed and dot-dashed curves are for $\ln \parz{ \tilde{\param{D}}^{(1,d_{\msbar})}(z,\bstar;\mu_{Q_0},Q_0^2)}$ and $-\param{g}^{(1,d_{\msbar})}(z,\T{b}{})$ respectively. The sum of the latter two always reproduces $\ln \parz{ \tilde{\param{D}}^{(1,d_{\msbar})}(z,\T{b}{};\mu_{Q_0},Q_0^2)}$ while $-\param{g}^{(1,d_{\msbar})}(z,\T{b}{})$ goes to zero like a power at small $\Tsc{b}{}$.

Note carefully that there are no error terms in going from \eref{Wtermev0_finalversion} to \eref{before_app_bstar}. So far, the expression is just another way to write \eref{Wtermev0_finalversion}. Plots of \eref{before_app_bstar} that use our parametrization from \sref{low_order} produce figures identical to
\fref{WclosetoQ0}, \fref{WlargeQ}, and \fref{WlargeQ_linear}.

There are actually two ways that we can recast the methods of \sref{cskernel}-\sref{summary} in the $\bstarsc$-method. In the first, we can simply notice that \eref{before_app_bstar} is already very close to the standard form. This version of the $\bstarsc$-expression is literally identical to the bottom-up expression in \eref{Wtermev0_finalversion}. However, it requires that we use the exact $\tilde{\param{K}}^{(1)}(\Tsc{b}{};\mu_{Q_0})$ and $\tilde{\param{D}}_{A,B}^{(n,d_{\msbar})}(z_A,\bstar;\mubstar,\mubstar^2)$. In our specific examples from earlier, this would be \eref{K_param_final} and \eref{evolvedd4}. Of course, this defeats the purpose of the $\bstarsc$ approach.

However, now we can also use that
\begin{equation}
\tilde{\param{K}}^{(n)}(\bstarsc;\mubstar) = 
\tilde{K}^{(n)}(\bstarsc;\mubstar) + \order{m \bmax} \, \label{e.final_k_approx}
\end{equation}
and 
\begin{align}
&{} \tilde{\param{D}}^{(n,d_{\msbar})}(z,\bstar;\mubstar,\mubstar^2) \no &{} \; = 
\tilde{D}^{(n,d_{\msbar})}_{\text{OPE}}(z,\bstar;\mubstar,\mubstar^2) + \order{m \bmax} \, , 
\label{e.final_f_approx}
\end{align}
where $\tilde{D}^{(n,d_{\msbar})}_{\text{OPE}}(z,\bstar;\mubstar,\mubstar^2)$ is the first term on the right side of \eref{OPE_for_ff}. In other words, we can use purely perturbative expressions for these functions since they are constrained to arguments $\Tsc{b}{} < \bmax$. Dropping the order $m \bmax$ terms and substituting \erefs{final_k_approx}{final_f_approx} gives us our 
last expression for $W^{(n)}(\Tsc{q}{},Q)$,
\begin{widetext}
\begin{align}
W^{(n)}_{\bstarsc}(\Tsc{q}{},Q) &{}=  H^{(n)}(\mu_{Q};C_2)
    \int \frac{\diff[2]{\T{b}{}}}{(2 \pi)^2}
    ~ e^{-i\T{q}{}\cdot \T{b}{} } 
    \tilde{D}_{A,\text{OPE}}^{(n,d_{\msbar})}(z_A,\bstar;\mubstar,\mubstar^2) 
    \tilde{D}_{B, \text{OPE}}^{(n,d_{\msbar})}(z_B,\bstar;\mubstar,\mubstar^2) \no
    &{} \times \exp
    \left\{ 2 \int_{\mubstar}^{\mu_{Q}} \frac{d \mu^\prime}{\mu^\prime} \left[\gamma^{(n)}(\alpha_s(\mu^\prime);1) 
- \ln \frac{Q}{\mu^\prime} \gamma_K^{(n)}(\alpha_s(\mu^\prime))
  \right] +\ln \frac{Q^2}{\mubstar^2} \tilde{K}^{(n)}(\bstarsc;\mubstar)
  \right\}
  \no
&{} \times \exp\left\{
-\param{g}_A^{(n,d_{\msbar})}(z_A,\Tsc{b}{}) - \param{g}_B^{(n,d_{\msbar})}(z_B,\Tsc{b}{}) - \param{g}^{(n)}_K(\Tsc{b}{}) \ln \parz{\frac{Q^2}{Q_0^2}} \right\} \, . \label{e.before_app_bstar_css}
\end{align}
\end{widetext}
The $\bstarsc$ subscript on the left is to indicate that we have dropped the underlines on $\tilde{\param{K}}^{(n)}(\bstarsc;\mubstar)$ and 
$\tilde{\param{D}}^{(n,d_{\msbar})}(z,\bstar;\mubstar,\mubstar^2)$ and we have neglected the $\order{m \bmax}$ terms in \erefs{final_k_approx}{final_f_approx}. These functions are just calculated in collinear perturbation theory now.  
Thus, 
\begin{equation}
W^{(n)}(\Tsc{q}{},Q_0) - W^{(n)}_{\bstarsc}(\Tsc{q}{},Q_0) = \order{m \bmax} \, . \label{e.Werror}
\end{equation}
Equation~\eqref{e.before_app_bstar_css} is the evolved $W(\Tsc{q}{},Q)$ as it is normally presented in the CSS formalism. (For comparison, consider Eq.~(22) from Ref.~\cite{Collins:2014jpa}.) 
However, now there is no ambiguity about what the functions $g_A^{(n,d_{\msbar})}(z_A,\Tsc{b}{})$, $g_B^{(n,d_{\msbar})}(z_B,\Tsc{b}{})$ and $g^{(n)}_K(\Tsc{b}{})$ are, assuming that one has already followed the steps in \sref{summary}. 
Whatever model, parametrization, or calculational technique we have used to describe the nonperturbative small transverse momentum dependence near the input scale (steps~\mref{choice_model},\mref{model_mods}) simply gets substituted into the right side of \erefs{gKdef_mod}{gdef_mod}.
The integrand in~\eref{before_app_bstar_css} equals that of \eref{Wtermev0_finalversion} up to corrections that vanish like a power of $\Tsc{b}{} \leq \bmax$. 
As long as $\bmax$ is small relative to the intrinsic mass parameters in the model, 
\eref{params_rg} is satisfied automatically.

Now in the bottom-up approach here there are no drawbacks to choosing $\bmax$ small enough to make the right side of \eref{Werror} as small as desired. It amounts simply to including more of the perturbative part of the calculation inside the $g$-functions. 

Of course, the above applies to any other ff renormalization scheme, we have used $\msbar$ because it is the most common.

One may check the above directly by using the explicit $n=1$ example from \sref{low_order}. Equation~\eqref{e.before_app_bstar} and ~\eref{before_app_bstar_css} are both plotted in \fref{ope_underline} for two values of $Q$ and with a range of $\bmax$. The solid black curves are the same $W^{(1)}(\Tsc{q}{},Q)$ from \sref{low_order} with the same sample parameters, while the other curves are generated by \eref{before_app_bstar_css} for a sequence of decreasing $\bmax$. As expected, \eref{before_app_bstar_css} converges to \eref{before_app_bstar}/\eref{Wtermev0_finalversion}, and $\bmax$ dependence vanishes, for small enough $\bmax$. This happens more rapidly at the larger $Q$ where there is less sensitivity to large $\Tsc{b}{}$ region. The size of the $\order{m \Tsc{b}{}}$ errors are of course dependent upon the size of the parameters $M$, $m_K$, $m_D$, and $a$ in the parametrization of the $g$-functions. In practice, we may of course simply choose $\bmax$ small enough that this error is negligible. 

\begin{figure}
\centering
\input{WqT_ope_inline.tex}
\caption{
A comparison of the cross section $W^{(1)}(\Tsc{q}{},Q)$ calculated using \eref{before_app_bstar} and \eref{before_app_bstar_css}, 
with the example models from \sref{low_order}, at scales of $Q=2\,\text{GeV}$ (top panel) and $Q=10\,\text{GeV}$ (bottom panel), and with fixed values of intrinsic masses $M$, $m_D$ and $m_K$ as indicated in the labels. In each case, the solid lines implement the cross section in the usual notation of the CSS formalism, i.e. through the use of \eref{before_app_bstar} and $\param{g}$ functions defined in terms of our model examples, as in \eref{gKdef_mod} and \eref{gdef_mod}. The dashed lines are obtained by use of \eref{before_app_bstar_css}, with the same $\param{g}$ as solid lines but with \eref{before_app_bstar_css}. Note that \eref{before_app_bstar} is identical to \eref{Wtermev0_finalversion} and is thus independent of $b_{\text{max}}$ by construction. 
}
\label{f.ope_underline}
\end{figure}
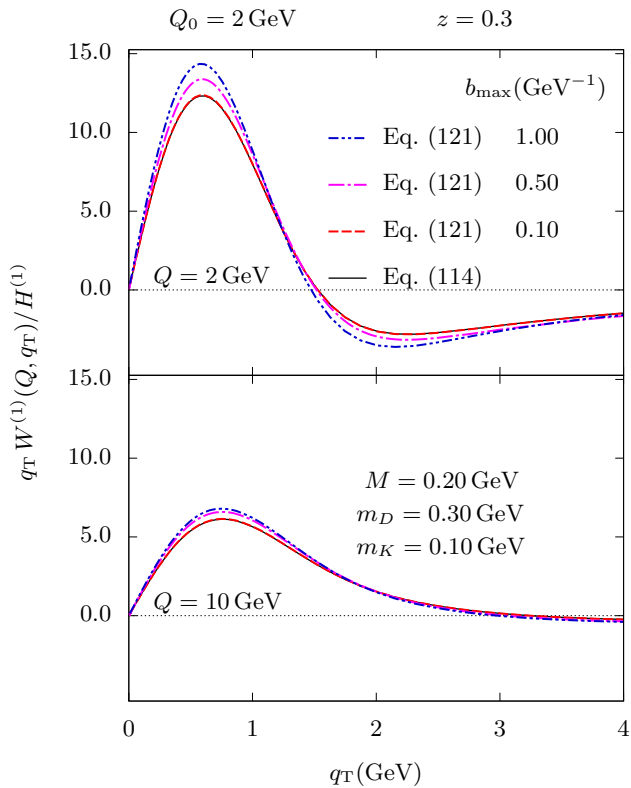

Compare what we have done above with the way that $g_A^{(n,d_{\msbar})}(z_A,\Tsc{b}{})$ (and the analogous function for the $B$ hadron) is often treated in top-down styles of approaches, where a $g$ functions is usually modeled by an ansatz:
\begin{equation}
g_A^{(n,d_{\msbar})}(z_A,\Tsc{b}{}) \longrightarrow g_A^{(n,d_{\msbar})}(z_A,\Tsc{b}{};\{c_1,c_2,\dots\}) \, .
\end{equation}
Here, the $\{c_1,c_2,\dots\}$ is a list of ansatz parameters. But notice that the integral 
\begin{equation*}
\int_0^{\mu_{Q_0}} \diff{\Tsc{k}{}}{} \Tsc{k}{} e^{-g_A^{(n,d_{\msbar})}(z_A,\Tsc{b}{};\{c_1,c_2,\dots\})}
\end{equation*}
is constrained by \erefs{cutff1}{d_little_underline}. Therefore, the initially postulated  $\{c_1,c_2,\dots\}$ parameters are not independent if a separate set of collinear ffs is known and available. At least one of the parameters is fixed in terms of the other parameters and the collinear functions. 
In the example from \sref{low_order}, this corresponds to the step where we replaced the $C^{(d_c)}$ in \eref{candidate} by the right side of \eref{C_val}. 
Without that step, the parameters of $g_A^{(n,d_{\msbar})}(z_A,\Tsc{b}{})$ are underconstrained, and the model of the nonperturbative transverse momentum will generally be inconsistent with the collinear correlation functions.\footnote{Alternatively, one could in principle leave the TMD ff model unconstrained by~\eref{cutff1} and instead calculate all collinear ffs directly in terms of the parametrization for the TMD ff. This may be impractical, however, given the existing extensive phenomenology that constrains collinear functions as compared to the TMD ones.} To our knowledge, such a step is not explicitly performed in the top-down implementations of TMD factorization and evolution. 

These constraints also mean that the nonperturbative $g$-functions are strongly correlated with the collinear functions, consistent with observed phenomenology~\cite{Bury:2022czx}.

Before closing this section, we emphasize once more that the steps above involving $\bstarsc$ are not strictly necessary if one has already adopted the bottom-up steps of \sref{summary} and implemented \eref{Wtermev0_finalversion}, but they may be helpful in comparing with past results.

\section{Conclusion}
\label{s.conclusion}

In this paper, we have explained the advantages of a  bottom-up/forward-evolution style of treating TMD correlation functions modeled or calculated nonperturbatively at a moderate input scale. Before concluding, let us reemphasize that we have not made any fundamental modifications to standard TMD factorization and evolution -- indeed, see \eref{before_app_bstar_css}. Rather, our purpose has been to make explicit the steps for combining any choice of nonperturbative model or parametrization of transverse momentum dependence with full TMD evolution in a reasonably automated way. This did lead to a significant amount of reorganization of previously known results, but we hope that the final prescription is fairly straightforward. Overall, our discussion in this paper might be summarized as a prescription for constructing the ``$g$-functions'' of typical CSS implementations in a way that conforms simultaneously with the TMD parton model viewpoint and TMD evolution.

The phenomenological strategy is to take TMDs generated from models, nonperturbative calculations, and/or fits near an input scale $Q_0$, and use these to make predictions for higher $Q$ measurements. Based on how successful those predictions are, models and/or fit parameters can be updated. The expected trend is that the amount of parameter adjustment diminishes as larger $Q$ measurements are included. 

Future steps involve constructing nonperturbative parametrizations analogous to our example functions in 
\sref{low_order}, but with more sophisticated input from nonperturbative QCD, including the models discussed in the introduction. 
It is likely (as was our intent) that many existing ``Type I'' phenomenological results can be combined with our approach with minimal modifications. Using the steps from \sref{summary} to implement evolution with specific models like the spectator or bag models is a natural next task. Of course, all expressions need to be extended to all flavors and channels and to TMD pdfs. Other minor tasks still to be completed include extending to higher $n$ (for $\tilde{K}$ this is simple), including the singlet gluon channel, and extending it to a treatment of gluon TMDs~\cite{Scarpa:2019fol}.

Over the past several years, there has been significant progress also in lattice-based methods for calculating TMD functions~\cite{Chu:2022mxh,Ji:2020ect,Constantinou:2020hdm,Cichy:2018mum,Ebert:2022fmh,Ebert:2019tvc,Ebert:2019okf,Ebert:2018gzl}. The techniques discussed in this paper will also be important for connecting those developments to experimental data and phenomenological extractions. 

We have focused on discussing unpolarized cross sections, but the same steps carry over in a straightforward way to 
polarized processes. A point of caution is that the integral relations analogous to \eref{dc_deff} become more subtle in some spin dependent TMD functions, and the translation between correlation functions defined with cutoffs and in other schemes is not as straightforward~\cite{Qiu:2020oqr,Rogers:2020tfs}. 

Nonperturbative $g$-functions exactly analogous to \eref{g_funcs} appear in the TMD factorization formula for \emph{double} scattering with double parton TMDs -- see Eq.(6.31) of 
\cite{Buffing:2017mqm}. Therefore, techniques analogous to what we have described for single scattering should in principle apply there as well.

An application where our procedure also likely helps, but which we have not yet discussed in detail, is in the matching to large transverse momentum. The so-called ``asymptotic'' term discussed in, for example, \cite{Collins:2016hqq} is the large $\Tsc{k}{}$ asymptote of \eref{tmd_factorization_input}, and it is an important ingredient for matching to the fixed order collinear calculation at large $\Tsc{q}{} \approx Q$. In top-down/backward evolution approaches, errors that grow larger as $Q$ decreases tend to spoil reasonable agreement between direct calculations of \eref{tmd_factorization_input} and the asymptotic term~\cite{Boglione:2014oea}. By contrast, in the bottom-up approach we have laid out in this paper, the matching to the asymptotic term at $Q \approx Q_0$ is automatic by construction. A separate but related issue is the difficulty observed in some processes, especially at moderate $Q$, of explaining $\Tsc{q}{} \approx Q$ using standard fixed order collinear factorization and existing collinear pdfs and ffs~\cite{Daleo:2004pn,Gonzalez-Hernandez:2018ipj,Bacchetta:2019tcu,Wang:2019bvb}.

In large $Q$ measurements, there is reason to expect that nonperturbative transverse momentum eventually becomes phenomenologically irrelevant. 
It can remain important, however, when very high precision is a goal, such as measurements of vector boson masses~\cite{Banfi:2011dx,Bacchetta:2018lna,Guzzi:2013aja,Nadolsky:2004vt,CDF:2022hxs}.

We leave all of these considerations to future work.

\vskip 0.3in
\acknowledgments
Fatma Aslan and Abha Rajan were involved in early versions of this project and we thank her for useful discussions. We also thank Mariaelena Boglione for useful comments. 
T.R. was supported by the U.S. Department of Energy, Office of Science, Office of Nuclear Physics, under Award Number DE-SC0018106. 
This work 
was also supported by the DOE Contract No. DE- AC05-06OR23177, under which 
Jefferson Science Associates, LLC operates Jefferson Lab.
The work of N.S. was supported by the DOE, Office of Science, Office of Nuclear Physics in the Early Career Program.


\begin{appendix}

\section{Notation Glossary}
\label{a.notglossary}

\begin{itemize}
\item $X^{(n)}$ 
\begin{quote}
An object $X$ calculated \emph{through} order $\alpha_s^n$.
\end{quote}

\item $W(\Tsc{q}{},Q)$ 
\begin{quote}
The low-$\Tsc{q}{}$ contribution to the cross section TMD factorization, the ``$W$-term.'' See \eref{tmd_factorization_momspace}.
\end{quote}

\item $m$ 
\begin{quote}
Any small mass scale of hadronic size or smaller, e.g., $\Lambda_\text{QCD}$, $m_\pi$, $m_q$ etc. 
\end{quote}

\item $m_K$, $m_D$, $M$ 
\begin{quote}
The nonperturbative mass scales in the parametrizations from \sref{low_order}. See \eref{Kparam_kt} and \eref{candidate}. 
\end{quote}

\item Renormalization group scales:
\begin{quote}
$\mu$: A generic scale. \\
$\mu_Q = C_2 Q$, \\ 
$\mu_{Q_0} = C_2 Q_0$, \\
$\mu_{\overline{Q}_0} = C_2 \overline{Q}_0(\Tsc{b}{})$. 
\\ See below for $\overline{Q}_0(\Tsc{b}{})$, and see also \aref{interp}.
\end{quote}

\item Hard factors:
\begin{enumerate}
\item $H(\alpha_s(\mu_Q);C_2)$: The exact overall hard factor for the $W$-term in \eref{Wtermev0}. 
\item $H^{(n)}(\alpha_s(\mu_Q);C_2)$: The hard factor in perturbation theory truncated beyond order $\alpha_s(\mu_Q)^n$.
\item $\mathcal{C}^{(n)}_D(z\Tsc{k}{})$: The hard coefficient in the factorization formula relating the TMD ff to the collinear ff at $\Tsc{k}{} \approx \mu_{Q}$, calculated in perturbation theory and truncated past order $\alpha_s(\mu_Q)^n$. See \eref{D_fact}. 
\item $\mathcal{C}^{(n)}_\Delta$: The hard coefficient in the correction term relating the cutoff integrated TMD ff to the collinear ff at $\Tsc{k}{} \approx \mu_{Q}$, calculated in perturbation theory and truncated past order $\alpha_s(\mu_Q)^n$. See \eref{delta_fact}. 
\end{enumerate}

\item $d_{A}(z_A;\mu)$ 
\begin{quote}
\underline{Definition}: collinear fragmentation function for a quark with momentum fraction $z_A$ fragmenting to hadron $A$.
\end{quote}

\item $d_{A,r}(z_A;\mu_Q)$
\begin{quote}
\underline{Definition}: A more precise notation for $d_{A}(z_A;\mu_Q)$. Collinear fragmentation function with UV divergences handled in the $r$ renormalization and/or regularization schemes. (e.g., $d_{\msbar}$ is the standard collinear fragmentation function defined in the $\msbar$ scheme.)
The auxiliary scale $\mu$ is the renormalization scale and in the above we have already replaced it with $\mu = \mu_Q = C_2 Q$. Throughout this paper, we always assume $C_2 = 1$. 
\end{quote}

\item $D_{A}\parz{z_A,z_A \T{k}{A};\mu_Q,Q^2}$
\begin{quote}
\underline{Definition}: The TMD quark ff for a quark with transverse momentum $\T{k}{A}$ fragmenting to hadron $A$. The auxiliary parameters associated with evolution are $\mu$ and $\zeta$, and in the above they have already been replaced by $\mu = \mu_Q$ and $\zeta = \mu_Q^2 = Q^2$. We use different symbols for $\mu$ and $\zeta$ to keep derivatives clear. 
\end{quote}

\item $D_{A}^{(n,d_r)}\parz{z_A,z_A \T{k}{A};\mu_Q,Q^2}$ 
\begin{quote}
A \emph{calculation} of a TMD quark ff in perturbative $n^\text{th}$-order collinear factorization optimized for $\Tsc{k}{} \approx Q$ and using collinear ffs $d_r$ defined in the $r$ scheme.
\end{quote}

\item $d_{A,c}(z_A;\mu_Q)$
\begin{quote}
\underline{Definition}: A special case of $d_{A,r}(z_A;\mu_Q)$ where the UV divergences are handled using the cutoff scheme. See~\eref{dc_deff}.
\end{quote}

\item $\Delta^{(n,d_r)}(\alpha_s(\mu_Q))$
\begin{quote}
The correction term for relating $d_{A,c}(z_A;\mu_Q)$ to 
the collinear fragmentation function $d_{A,r}(z_A;\mu_Q)$ in another scheme $r$. See~\eref{cutff}.
\end{quote}

\item $d_{A,c}^{(n,d_r)}(z_A;\mu_Q)$
\begin{quote}
The \emph{approximate} expression for $d_{A,c}(z_A;\mu_Q)$ in terms of $d_{A,r}(z_A;\mu_Q)$, accurate up to order $\alpha_s^{n+1}$ and power suppressed errors.  See~\erefs{dnr_rel}{dc_err}. 
\end{quote}

\item $\underline{d}_{A,c}^{(n,d_r)}(z_A;\mu_Q)$
\begin{quote}
A parametrization of $d_{A,c}(z_A;\mu_Q)$ obtained from a the cutoff integral of the underlined TMD ff, $\tilde{\param{D}}_A^{(n,d_r)}(z_A,\T{b}{};\mu_{Q_0},Q_0^2)$.  See~\erefs{cutff1}{d_little_underline} and the definition of $\tilde{\param{D}}_A^{(n,d_r)}(z_A,\T{b}{};\mu_{Q_0},Q_0^2)$ below. 
\end{quote}

\item $\tilde{K}(\Tsc{b}{};\mu)$
\begin{quote}
\underline{Definition}: The coordinate space CS kernel. See \erefs{CSPDF}{RG}.  Related to its momentum space version via
\begin{equation*}
\label{e.coord_2_mom_K}
\tilde{K}(\Tsc{b}{};\mu) \equiv
\int \diff{^2\T{k}{}}{} e^{i \T{k}{} \T{b}{}}
K(\Tsc{k}{};\mu) \, .
\end{equation*}
\end{quote}

\item $K^{(n)}(\Tsc{k}{};\mu)$
\begin{quote}
A fixed $n^\text{th}$-order, fixed scale perturbative calculation of $K(\Tsc{k}{};\mu)$. See \erefs{Kexp}{kappa_def}. 
\end{quote}

\item $\inpt{K}^{(n)}\parz{\Tsc{k}{};\mu_{Q_0}}$
\begin{quote}
The \emph{parametrization} of the momentum space CS kernal that interpolates between a nonperterbative parametrization of $K(\Tsc{k}{};\mu_{Q_0})$ at small $\Tsc{k}{}$ and $K^{(n)}(\Tsc{k}{};\mu_{Q_0})$ at large $\Tsc{k}{}$. See \eref{K_conds}. Optimized for applications in the $Q \approx Q_0$, $\Tsc{k}{} \lesssim Q_0$ region of \eref{Wtermev0}. The coordinate space version $\inpt{\tilde{K}}^{(n)}\parz{\Tsc{b}{};\mu_{Q_0}}$ is found from \eref{Kinput_def}. 
\end{quote}

\item $\overline{Q}_0(\Tsc{b}{})$
\begin{quote}
Transformation function for switching from $\mu = Q_0$ to $\mu = C_1/\Tsc{b}{}$ RG scales. See \eref{bdef}. See also \aref{interp}.
\end{quote} 

\item $\inpt{\tilde{K}}^{(n)}(\Tsc{b}{};\mu_{\overline{Q}_0})$
\begin{quote}
Same as $\inpt{\tilde{K}}^{(n)}\parz{\Tsc{b}{};\mu_{Q_0}}$ but with $\mu_{\overline{Q}_0} = C_2 \overline{Q}_0(\Tsc{b}{})$ as the scale.
\end{quote} 

\item $\tilde{\param{K}}^{(n)}(\Tsc{b}{};\mu_{Q_0})$
\begin{quote}
Final parametrization of $\tilde{K}(\Tsc{b}{};\mu_{Q_0})$ optimized for all $\Tsc{b}{}$, including the RG improved treatment of the $\Tsc{b}{} \ll 1/Q_0$ region. See \eref{evol_paramb}.
\end{quote} 

\item $\inpt{D}^{(n,d_r)}\parz{z,z \T{k}{};\mu_{Q_0},Q_0^2}$
\begin{quote}
The input \emph{parametrization} of the TMD ff with a nonperturbative parametrization for small $\Tsc{k}{}$ and interpolating to an $n^\text{th}$-order perturbative calculation at $\Tsc{k}{} \sim Q_0$. Applicable to phenomenology at $Q \approx Q_0$ where $0 \leq \Tsc{k}{} \lesssim Q_0$ is the relevant kinematical region. The perturbative part uses $d_r$ collinear ffs in the $r$ renormalization/regularization scheme. See \eref{param_conds}.
\end{quote} 

\item $\tilde{\param{D}}_A^{(n,d_r)}(z_A,\T{b}{};\mu_{Q_0},Q_0^2)$
\begin{quote}
The final \emph{parametrization} of the coordinate space TMD ff at the input scale and optimized for all $\Tsc{b}{}$, including the RG improved treatment of the $\Tsc{b}{} \ll 1/Q_0$ region. The perturbative part uses $d_r$ collinear ffs in the $r$ renormalization/regularization scheme. See \eref{evolvedd3p}.
\end{quote} 

\item $A^{(d_c)}(z;\mu)$, $B^{(d_c)}(z;\mu)$, $C^{(d_c)}$
\begin{quote}
Abbreviations for factors that involve collinear factorizaton. See \eref{A_def}, \eref{AB_def}, and \eref{C_val}.
\end{quote} 

\item $W^{(n)}(\Tsc{q}{},Q)$
\begin{quote}
A calculation of $W(\Tsc{q}{},Q)$ in \eref{Wtermev0} using $\tilde{\param{D}}_{A,B}^{(n,d_r)}(z_{A,B},\T{b}{};\mu_{Q_0},Q_0^2)$, $H^{(n)}(\alpha_s(\mu_Q);C_2)$, $\tilde{\param{K}}^{(n)}(\Tsc{b}{};\mu_{Q_0})$, $\gamma_K^{(n)}(\alpha_s(\mu))$, and $\gamma^{(n)}(\alpha_s(\mu);1)$.
\end{quote} 

\end{itemize}

\section{Scale setting}
\label{a.scale_setting}

The expansion in $m/Q$ that gives the TMD factorization in \eref{tmd_factorization_momspace} as its leading power must eventually break down as $Q$ becomes small. In practice, this means one must choose a minimum $Q_0$ below which one stops trusting factorization in phenomenological applications, and this is what we have called the input scale $Q_0$. Values of $\mu$ near $Q_0$ are understood to be close to the boundary where perturbative expansions in $\alpha_s{}(\mu)$ and power expansions in $m/Q_0$ cease to be valid even approximately. For the purposes of the present paper, $Q_0$ may be treated as a phenomenologically determined number. In practice it is usually taken to be somewhere in the range of $1$-$3$~GeV. 

The value of $Q_0$ fixes the overall RG ($\mu$) and rapidity ($\zeta$) scales in the input TMD ffs in \eref{tmd_factorization_input}. 
 There is nothing incorrect in principle with working only with TMD factorization in the form that is written in \eref{tmd_factorization_momspace} and \eref{Wtermev0}. One may regard $\tilde{D}_{A}(z_A,\T{b}{};\mu_{Q_0},Q_0^2)$, $\tilde{D}_{B}(z_B,\T{b}{};\mu_{Q_0},Q_0^2)$, and $\tilde{K}(\Tsc{b}{};\mu_{Q_0})$ as entirely unknown functions to be determined phenomenologically or by other nonperturbative means. This is essentially the TMD parton model and is a standard approach to much phenomenology applied to nucleon structure studies near $Q \approx Q_0$. However, it leaves the TMD ffs very badly constrained at large $\Tsc{k}{}$ where perturbative calculations should be possible. When $\Tsc{k}{A,B} \approx Q_0$ in \eref{tmd_factorization_momspace}, or equivalently when 
$\Tsc{b}{} \approx 1/Q_0$ in \eref{Wtermev0}, it is possible to expand the TMD functions perturbatively in collinear factorization and thereby increase predictive power. 
When $\Tsc{k}{A,B}$ is much larger than $Q_0$, one can further improve perturbative calculations by choosing $\Tsc{k}{A,B}$ itself to set the hard scale in the individual TMD ffs rather than $Q_0$. Of course, when $\Tsc{k}{A,B}$ is small, it is not appropriate as a hard scale and one should instead continue to use $Q_0$. Indeed, it is the nonperturbative $\Tsc{k}{}$-dependence that is often of primary interest in TMD hadron structure studies.

In this appendix, we elaborate on the issue of the transition between the choices of $Q_0$ and $\Tsc{k}{}$ as scales for the TMD ffs, keeping in mind that the ultimate goal is to smoothly transition between traditional Type I and type II styles of approach. The discussion will be somewhat schematic and is meant to further motivate the main body of the text. Also, we will work in transverse momentum space where our main points will be somewhat more intuitive, but the discussion applies equally in transverse coordinate space, which is more common.

To see the issues clearly, 
recall that when $\Tsc{k}{}$ is large we may express the $\Tsc{k}{}$-dependence in a single input TMD ff in the form 
\begin{align}
&{}D(z,z\T{k}{};\mu_{Q_0},Q_0^2) \stackrel{\Tsc{k}{} \gg m}{=} \no &{} \qquad \frac{1}{\Tscsq{k}{}} \left[ \delta \parz{\alpha_s(\mu_{Q_0}),\frac{\Tsc{k}{}}{\mu_{Q_0}},\frac{Q_0^2}{\mu_{Q_0}^2}} + \order{\frac{m}{\mu_{Q_0}},\frac{m}{\Tsc{k}{}}} \right] \, 
\label{e.Dkt_exp}
\end{align}
where $\delta$ is some function of the coupling and the ratios of the scales that generally appear in logarithms. 

When $\Tsc{k}{} \gg Q_0$, the $\Tsc{k}{}/\mu_{Q_0}$ ratio on the right side of \eref{Dkt_exp} diverges so that perturbative calculations of $\delta$ with fixed $\mu_{Q_0}$ degrade in accuracy. Optimizing $\delta$ in perturbation theory requires another application of evolution from $\mu_{Q_0}$ to $\mu_{\Tsc{k}{}} = C_k \Tsc{k}{}$ and $Q_0^2$ to $\mu_{\Tsc{k}{}}^2 = C_k^2 \Tscsq{k}{}$ where $C_k$ is an order unity numerical constant analogous to $C_1$ and $C_2$ in the main body of the text.
After evolution, the TMD ff that we work with instead has the general behavior
\begin{align}
&{}D(z,z\T{k}{};\mu_{\Tsc{k}{}},\mu_{\Tsc{k}{}}^2) = \no &{} \qquad \frac{1}{\Tscsq{k}{}} \left[ \delta \parz{\alpha_s(C_k \Tsc{k}{} ),\order{1},\order{1}} + \order{\frac{m}{\Tsc{k}{}}} \right] \, . 
\label{e.Dkt_exp1}
\end{align}
Now as $\Tsc{k}{} / Q_0 \to \infty$ the convergence properties of a perturbative expansion of $\delta$ only improves as the coupling and power suppressed terms vanish. Most of the standard Type II/top-down styles of TMD factorization implementations thus only use a scale choice analogous to \eref{Dkt_exp1} (or rather its coordinate space analog) and rarely the fixed scale version in \eref{Dkt_exp}. 

However, small-$\alpha_s$ calculations of $\delta$ in \eref{Dkt_exp1} come with numerically unstable truncation errors in the region around $\Tsc{k}{} \approx Q_0$ simply because $Q_0$ is at the border between perturbative and nonperturbative scales. Small variations in, for example, the value of $C_k$ can  have a big effect on calculations. The problem is exacerbated by the fact that the transformation from $\mu_{Q_0}, Q_0^2$ to $\mu_{\Tsc{k}{}}, \mu_{\Tsc{k}{}}^2$ effectively involves a resummation of many higher order logarithms in region of $\Tsc{k}{A,B}$. That is, what appear in the factorization after the scale transformation is not \eref{Dkt_exp1} alone but 
\begin{align}
&{}\sim D(z,z\T{k}{};\mu_{\Tsc{k}{}},\mu_{\Tsc{k}{}}^2) \no
& \; \times \left[ 1 + \sum_\text{terms like} \; \alpha_s(\mu_k)^m \ln^n \frac{\mu_{\Tsc{k}{}}}{\mu_{Q_0}}  \right] \, .
\label{e.Dkt_exp_logs}
\end{align}
However, when $\Tsc{k}{} \approx \mu_{Q_0}$ the logarithms on 
the second line are anyway no larger than any other contributions that are higher order in $\alpha_s(\Tsc{k}{})$.\footnote{It is also worth recalling here that QCD perturbation series are only asymptotic rather than convergent.} The advantage of the scale transformation is lost here. More problematically, because of the rapidly rising coupling $\alpha_s(\mu_k)$ just below $\Tsc{k}{} \approx Q_0$, the higher order logarithmic terms in \eref{Dkt_exp_logs} are numerically unstable over the range of $\Tsc{k}{}$ extending from just below to just above $Q_0$.
Thus, calculations that use \eref{Dkt_exp1} are rather poorly behaved in the $\Tsc{k}{} \approx Q_0$ borderline region where we should expect a reasonably smooth transition to a nonperturbative region.    

But calculations with a fixed scale can still be at least approximately valid around $\Tsc{k}{} \approx Q_0$ where \eref{Dkt_exp} is  
\begin{align}
&{}D(z,z\T{k}{};\mu_{Q_0},Q_0^2) \stackrel{\Tsc{k}{} \approx Q_0}{=} \no &{} \qquad \frac{1}{\Tscsq{k}{}} \left[ \delta \parz{\alpha_s(\mu_{Q_0}),\order{1},\order{1}} + \order{\frac{m}{Q_0}} \right] \, .
\label{e.Dkt_exp2}
\end{align}
As long as $Q_0$ is not very small, one can use low order perturbative calculations to approximate $D$. And there are no large logarithms in \eref{Dkt_exp} if $\Tsc{k}{} \sim Q_0$, so the advantages of changing scales to \eref{Dkt_exp1} are absent. 

Therefore, our preferred choice of scale for the transition region around $\Tsc{k}{} \approx Q_0$ is actually the fixed scale in \eref{Dkt_exp}, not the $\Tsc{k}{} \gg Q_0$ RG improved scale of \eref{Dkt_exp1}. Another advantage of using the fixed scale is that it automatically gives the fixed order asymptotic term in momentum space at $Q = Q_0$ and $\Tsc{q}{} \approx Q_0$, which is also a fixed scale calculation in momentum space, when the TMD ff is substituted into \eref{tmd_factorization_momspace}. 

The above suggests that it is best to categorize the $\Tsc{k}{}$-dependence not into just two regions of small $\Tsc{k}{}$ and large $\Tsc{k}{}$, but into three regions: small $\Tsc{k}{}$ ($\Tsc{k}{} \ll Q_0$), large $\Tsc{k}{}$ ($\Tsc{k}{} \approx Q_0$), and very large $\Tsc{k}{}$ ($\Tsc{k}{} \gg Q_0$).
The large $\Tsc{k}{}$ region is where $\Tsc{k}{}$ is just barely large enough for small coupling descriptions of transverse momentum dependence to be reasonable. 

In the discussion above we have worked in transverse momentum space because that better matches the intuition of models and phenomenology. However, the analogous observations apply straightforwardly to transverse coordinate space. In that case, there is a region of very large $\Tsc{b}{} \gg 1/Q_0$ that is entirely nonperturbative, a region of very small $\Tsc{b}{} \ll 1/Q_0$ that is purely perturbative as long as an RG improved $\mu \sim 1/\Tsc{b}{}$ is used, and an intermediate region of $\Tsc{b}{} \sim 1/Q_0$ where fixed order, fixed scale calculations are ideal.

One of our tasks in the main body of the paper is to interpolate between these three regions in our parameterizations. For the intermediate region of transverse momentum, transitioning between scales only introduces higher order errors. 

Notice that our discussion of large logarithms above is a mirror image of how large logarithms are often introduced in explanations of top down approaches. There, one starts with \eref{Dkt_exp} but using $\mu_Q$ and $Q^2$ instead of the input $\mu_{Q_0}$ and $Q_0^2$ and assuming $Q \gg Q_0$:
\begin{align}
&{}D(z,z\T{k}{};\mu_{Q},Q^2) \stackrel{\Tsc{k}{} \gg m}{=} \no &{} \qquad \frac{1}{\Tscsq{k}{}} \left[ \delta \parz{\alpha_s(\mu_{Q}),\frac{\Tsc{k}{}}{\mu_{Q}},\frac{Q^2}{\mu_{Q}^2}} + \order{\frac{m}{\mu_{Q}},\frac{m}{\Tsc{k}{}}} \right] \, 
\label{e.Dkt_exp_topdown}
\end{align}
Then the task is to determine how to resum large logarithms of $\ln(\Tsc{k}{}/\mu_{Q})$ as $\Tsc{k}{}$ gets \emph{small} relative to $Q$, rather than as $\Tsc{k}{}$ gets large relative to $Q_0$. 


\section{Scale transformation function}
\label{a.interp}

For the scale transition function in \eref{bdef}, we must arrange for the transition from $\sim 1/\Tsc{b}{}$ to $Q_0$ to occur at $\Tsc{b}{}$ somewhat smaller than $1/Q_0$ to avoid modifying the treatment of \eref{Wtermev0} in the $Q \approx Q_0$ region. 
One choice that satisfies this for a $Q_0 = 2$~GeV is 
\begin{align}
\label{e.qbar_param}
&{}\overline{Q}_0(\Tsc{b}{}) \no
&{}= 2.0~\text{GeV} \left[ 1 - \parz{1 - \frac{C_1}{(2.0~\text{GeV}) \Tsc{b}{}}}
e^{- (4~\text{GeV}^2) \Tscsq{b}{}}\right] \, .
\end{align}
If we wish to adjust the exact shape in the $\approx 1/Q_0$ transition region by adding a parameter as in \eref{Qbar_params}, we may modify \eref{qbar_param} by introducing a parameter $a$,
\begin{align}
\label{e.qbar_param_a}
&\overline{Q}_0(\Tsc{b}{},a) \no
&{}= 2.0~\text{GeV} \left[ 1 - \parz{1 - \frac{C_1}{(2.0~\text{GeV}) \Tsc{b}{}}}
e^{-\Tscsq{b}{} a^2} \right] \, .
\end{align}
Here the transition between the two RG scales takes place around $\Tsc{b}{} \sim 1/a$. We can use the \eref{qbar_param_a} form to check approximate scale independence in the transition region by varying $a$ slightly. $C_1$ is the usual numerical constant, $C_1 = 2 e^{-\gamma_E} \approx 1.123$.

\end{appendix}

\bibliography{bibliography}

\end{document}

%% file: ktilde_underline_1_v2_inline.tex
\begingroup
  \makeatletter
  \providecommand\color[2][]{%
    \GenericError{(gnuplot) \space\space\space\@spaces}{%
      Package color not loaded in conjunction with
      terminal option `colourtext'%
    }{See the gnuplot documentation for explanation.%
    }{Either use 'blacktext' in gnuplot or load the package
      color.sty in LaTeX.}%
    \renewcommand\color[2][]{}%
  }%
  \providecommand\includegraphics[2][]{%
    \GenericError{(gnuplot) \space\space\space\@spaces}{%
      Package graphicx or graphics not loaded%
    }{See the gnuplot documentation for explanation.%
    }{The gnuplot epslatex terminal needs graphicx.sty or graphics.sty.}%
    \renewcommand\includegraphics[2][]{}%
  }%
  \providecommand\rotatebox[2]{#2}%
  \@ifundefined{ifGPcolor}{%
    \newif\ifGPcolor
    \GPcolortrue
  }{}%
  \@ifundefined{ifGPblacktext}{%
    \newif\ifGPblacktext
    \GPblacktextfalse
  }{}%
  \let\gplgaddtomacro\g@addto@macro
  \gdef\gplbacktext{}%
  \gdef\gplfronttext{}%
  \makeatother
  \ifGPblacktext
    \def\colorrgb#1{}%
    \def\colorgray#1{}%
  \else
    \ifGPcolor
      \def\colorrgb#1{\color[rgb]{#1}}%
      \def\colorgray#1{\color[gray]{#1}}%
      \expandafter\def\csname LTw\endcsname{\color{white}}%
      \expandafter\def\csname LTb\endcsname{\color{black}}%
      \expandafter\def\csname LTa\endcsname{\color{black}}%
      \expandafter\def\csname LT0\endcsname{\color[rgb]{1,0,0}}%
      \expandafter\def\csname LT1\endcsname{\color[rgb]{0,1,0}}%
      \expandafter\def\csname LT2\endcsname{\color[rgb]{0,0,1}}%
      \expandafter\def\csname LT3\endcsname{\color[rgb]{1,0,1}}%
      \expandafter\def\csname LT4\endcsname{\color[rgb]{0,1,1}}%
      \expandafter\def\csname LT5\endcsname{\color[rgb]{1,1,0}}%
      \expandafter\def\csname LT6\endcsname{\color[rgb]{0,0,0}}%
      \expandafter\def\csname LT7\endcsname{\color[rgb]{1,0.3,0}}%
      \expandafter\def\csname LT8\endcsname{\color[rgb]{0.5,0.5,0.5}}%
    \else
      \def\colorrgb#1{\color{black}}%
      \def\colorgray#1{\color[gray]{#1}}%
      \expandafter\def\csname LTw\endcsname{\color{white}}%
      \expandafter\def\csname LTb\endcsname{\color{black}}%
      \expandafter\def\csname LTa\endcsname{\color{black}}%
      \expandafter\def\csname LT0\endcsname{\color{black}}%
      \expandafter\def\csname LT1\endcsname{\color{black}}%
      \expandafter\def\csname LT2\endcsname{\color{black}}%
      \expandafter\def\csname LT3\endcsname{\color{black}}%
      \expandafter\def\csname LT4\endcsname{\color{black}}%
      \expandafter\def\csname LT5\endcsname{\color{black}}%
      \expandafter\def\csname LT6\endcsname{\color{black}}%
      \expandafter\def\csname LT7\endcsname{\color{black}}%
      \expandafter\def\csname LT8\endcsname{\color{black}}%
    \fi
  \fi
    \setlength{\unitlength}{0.0500bp}%
    \ifx\gptboxheight\undefined%
      \newlength{\gptboxheight}%
      \newlength{\gptboxwidth}%
      \newsavebox{\gptboxtext}%
    \fi%
    \setlength{\fboxrule}{0.5pt}%
    \setlength{\fboxsep}{1pt}%
\begin{picture}(5040.00,6300.00)%
    \gplgaddtomacro\gplbacktext{%
      \csname LTb\endcsname
      \put(674,3408){\makebox(0,0)[r]{\strut{}}}%
      \put(674,3791){\makebox(0,0)[r]{\strut{}}}%
      \put(674,4015){\makebox(0,0)[r]{\strut{}}}%
      \put(674,4174){\makebox(0,0)[r]{\strut{}}}%
      \put(674,4297){\makebox(0,0)[r]{\strut{}}}%
      \put(674,4397){\makebox(0,0)[r]{\strut{}}}%
      \put(674,4483){\makebox(0,0)[r]{\strut{}}}%
      \put(674,4556){\makebox(0,0)[r]{\strut{}}}%
      \put(674,4621){\makebox(0,0)[r]{\strut{}}}%
      \put(674,4680){\makebox(0,0)[r]{\strut{}$10^{1}$}}%
      \put(674,5062){\makebox(0,0)[r]{\strut{}}}%
      \put(674,5286){\makebox(0,0)[r]{\strut{}}}%
      \put(674,5445){\makebox(0,0)[r]{\strut{}}}%
      \put(674,5568){\makebox(0,0)[r]{\strut{}}}%
      \put(674,5669){\makebox(0,0)[r]{\strut{}}}%
      \put(1893,3188){\makebox(0,0){\strut{}$$}}%
      \put(3448,3188){\makebox(0,0){\strut{}$$}}%
    }%
    \gplgaddtomacro\gplfronttext{%
      \csname LTb\endcsname
      \put(158,4538){\rotatebox{-270}{\makebox(0,0){\strut{}$\overline{Q}_0(\Tsc{b}{})$}}}%
      \csname LTb\endcsname
      \put(4031,5370){\makebox(0,0)[r]{\strut{}$C_1/b_{\text{T}}$}}%
      \csname LTb\endcsname
      \put(4031,5150){\makebox(0,0)[r]{\strut{}$\mu_{Q_0}$}}%
      \csname LTb\endcsname
      \put(4031,4930){\makebox(0,0)[r]{\strut{}$a=2\,\text{GeV}$}}%
      \csname LTb\endcsname
      \put(4031,4710){\makebox(0,0)[r]{\strut{}$a=4\,\text{GeV}$}}%
    }%
    \gplgaddtomacro\gplbacktext{%
      \csname LTb\endcsname
      \put(674,2935){\makebox(0,0)[r]{\strut{}$10$}}%
      \put(674,3250){\makebox(0,0)[r]{\strut{}$30$}}%
      \put(1893,2558){\makebox(0,0){\strut{}}}%
      \put(3448,2558){\makebox(0,0){\strut{}}}%
    }%
    \gplgaddtomacro\gplfronttext{%
      \csname LTb\endcsname
      \put(290,3092){\rotatebox{-270}{\makebox(0,0){\strut{}$\%\,$ diff.}}}%
    }%
    \gplgaddtomacro\gplbacktext{%
      \csname LTb\endcsname
      \put(674,1077){\makebox(0,0)[r]{\strut{}$-0.4$}}%
      \put(674,1417){\makebox(0,0)[r]{\strut{}$-0.2$}}%
      \put(674,1757){\makebox(0,0)[r]{\strut{}$0.0$}}%
      \put(674,2097){\makebox(0,0)[r]{\strut{}$0.2$}}%
      \put(674,2437){\makebox(0,0)[r]{\strut{}$0.4$}}%
      \put(1893,347){\makebox(0,0){\strut{}$10^{-1}$}}%
      \put(3448,347){\makebox(0,0){\strut{}$10^{0}$}}%
      \csname LTb\endcsname
      \put(1008,5890){\makebox(0,0)[l]{\strut{}$Q_0=2\,\text{GeV}$}}%
      \put(3023,2205){\makebox(0,0)[l]{\strut{}$m_K=0.1\,\text{GeV}$}}%
    }%
    \gplgaddtomacro\gplfronttext{%
      \csname LTb\endcsname
      \put(127,1672){\rotatebox{-270}{\makebox(0,0){\strut{}$\tilde{\param{K}}^{(1)}(\Tsc{b}{};\mu_{Q_0})$}}}%
      \put(2670,149){\makebox(0,0){\strut{}$b_{\text{T}}(\text{GeV}^{-1})$}}%
      \csname LTb\endcsname
      \put(2771,1536){\makebox(0,0)[r]{\strut{}$\tilde{K}^{(1)}_{pert}\qquad\qquad\,\,\,$}}%
      \csname LTb\endcsname
      \put(2771,1206){\makebox(0,0)[r]{\strut{}$\param{\tilde{K}}^{(1)}(a=2\,\text{GeV})$}}%
      \csname LTb\endcsname
      \put(2771,876){\makebox(0,0)[r]{\strut{}$\param{\tilde{K}}^{(1)}(a=4\,\text{GeV})$}}%
    }%
    \gplbacktext
    \put(0,0){\includegraphics{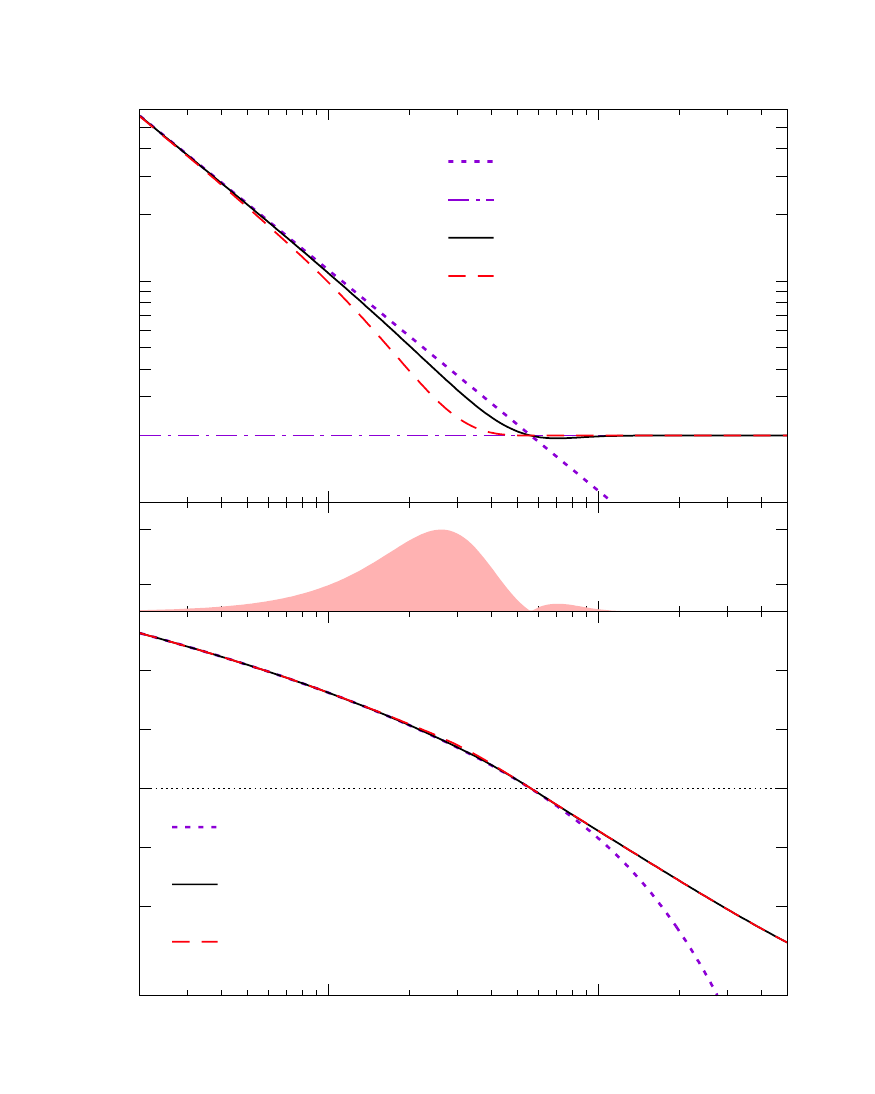}}%
    \gplfronttext
  \end{picture}%
\endgroup

%% file: NPevoratio_v2_inline.tex
\begingroup
  \makeatletter
  \providecommand\color[2][]{%
    \GenericError{(gnuplot) \space\space\space\@spaces}{%
      Package color not loaded in conjunction with
      terminal option `colourtext'%
    }{See the gnuplot documentation for explanation.%
    }{Either use 'blacktext' in gnuplot or load the package
      color.sty in LaTeX.}%
    \renewcommand\color[2][]{}%
  }%
  \providecommand\includegraphics[2][]{%
    \GenericError{(gnuplot) \space\space\space\@spaces}{%
      Package graphicx or graphics not loaded%
    }{See the gnuplot documentation for explanation.%
    }{The gnuplot epslatex terminal needs graphicx.sty or graphics.sty.}%
    \renewcommand\includegraphics[2][]{}%
  }%
  \providecommand\rotatebox[2]{#2}%
  \@ifundefined{ifGPcolor}{%
    \newif\ifGPcolor
    \GPcolortrue
  }{}%
  \@ifundefined{ifGPblacktext}{%
    \newif\ifGPblacktext
    \GPblacktextfalse
  }{}%
  \let\gplgaddtomacro\g@addto@macro
  \gdef\gplbacktext{}%
  \gdef\gplfronttext{}%
  \makeatother
  \ifGPblacktext
    \def\colorrgb#1{}%
    \def\colorgray#1{}%
  \else
    \ifGPcolor
      \def\colorrgb#1{\color[rgb]{#1}}%
      \def\colorgray#1{\color[gray]{#1}}%
      \expandafter\def\csname LTw\endcsname{\color{white}}%
      \expandafter\def\csname LTb\endcsname{\color{black}}%
      \expandafter\def\csname LTa\endcsname{\color{black}}%
      \expandafter\def\csname LT0\endcsname{\color[rgb]{1,0,0}}%
      \expandafter\def\csname LT1\endcsname{\color[rgb]{0,1,0}}%
      \expandafter\def\csname LT2\endcsname{\color[rgb]{0,0,1}}%
      \expandafter\def\csname LT3\endcsname{\color[rgb]{1,0,1}}%
      \expandafter\def\csname LT4\endcsname{\color[rgb]{0,1,1}}%
      \expandafter\def\csname LT5\endcsname{\color[rgb]{1,1,0}}%
      \expandafter\def\csname LT6\endcsname{\color[rgb]{0,0,0}}%
      \expandafter\def\csname LT7\endcsname{\color[rgb]{1,0.3,0}}%
      \expandafter\def\csname LT8\endcsname{\color[rgb]{0.5,0.5,0.5}}%
    \else
      \def\colorrgb#1{\color{black}}%
      \def\colorgray#1{\color[gray]{#1}}%
      \expandafter\def\csname LTw\endcsname{\color{white}}%
      \expandafter\def\csname LTb\endcsname{\color{black}}%
      \expandafter\def\csname LTa\endcsname{\color{black}}%
      \expandafter\def\csname LT0\endcsname{\color{black}}%
      \expandafter\def\csname LT1\endcsname{\color{black}}%
      \expandafter\def\csname LT2\endcsname{\color{black}}%
      \expandafter\def\csname LT3\endcsname{\color{black}}%
      \expandafter\def\csname LT4\endcsname{\color{black}}%
      \expandafter\def\csname LT5\endcsname{\color{black}}%
      \expandafter\def\csname LT6\endcsname{\color{black}}%
      \expandafter\def\csname LT7\endcsname{\color{black}}%
      \expandafter\def\csname LT8\endcsname{\color{black}}%
    \fi
  \fi
    \setlength{\unitlength}{0.0500bp}%
    \ifx\gptboxheight\undefined%
      \newlength{\gptboxheight}%
      \newlength{\gptboxwidth}%
      \newsavebox{\gptboxtext}%
    \fi%
    \setlength{\fboxrule}{0.5pt}%
    \setlength{\fboxsep}{1pt}%
\begin{picture}(5040.00,6300.00)%
    \gplgaddtomacro\gplbacktext{%
      \csname LTb\endcsname
      \put(573,931){\makebox(0,0)[r]{\strut{}$1.00$}}%
      \put(573,1984){\makebox(0,0)[r]{\strut{}$1.03$}}%
      \put(573,3037){\makebox(0,0)[r]{\strut{}$1.06$}}%
      \put(1297,536){\makebox(0,0){\strut{}$10^{-1}$}}%
      \put(3964,536){\makebox(0,0){\strut{}$10^{0}$}}%
      \put(1008,5890){\makebox(0,0)[l]{\strut{}$Q_0=2\,\text{GeV}$}}%
      \put(884,2440){\makebox(0,0)[l]{\strut{}$\,a_1=2\,\text{GeV}$}}%
      \put(884,2195){\makebox(0,0)[l]{\strut{}$\,a_2=4\,\text{GeV}$}}%
    }%
    \gplgaddtomacro\gplfronttext{%
      \csname LTb\endcsname
      \put(-75,1984){\rotatebox{-270}{\makebox(0,0){\strut{}$r(a_1,a_2)$}}}%
      \put(2569,206){\makebox(0,0){\strut{}$b_{\text{T}}(\text{GeV}^{-1})$}}%
      \csname LTb\endcsname
      \put(4302,2984){\makebox(0,0)[r]{\strut{}$Q=4\,\text{GeV}\hphantom{00}$}}%
      \csname LTb\endcsname
      \put(4302,2654){\makebox(0,0)[r]{\strut{}$Q=100\,\text{GeV}$}}%
    }%
    \gplgaddtomacro\gplbacktext{%
      \csname LTb\endcsname
      \put(573,3622){\makebox(0,0)[r]{\strut{}$-0.1$}}%
      \put(573,4032){\makebox(0,0)[r]{\strut{}$0.0$}}%
      \put(573,4441){\makebox(0,0)[r]{\strut{}$0.1$}}%
      \put(573,4850){\makebox(0,0)[r]{\strut{}$0.2$}}%
      \put(573,5260){\makebox(0,0)[r]{\strut{}$0.3$}}%
      \put(1297,2993){\makebox(0,0){\strut{}}}%
      \put(3964,2993){\makebox(0,0){\strut{}}}%
    }%
    \gplgaddtomacro\gplfronttext{%
      \csname LTb\endcsname
      \put(-75,4441){\rotatebox{-270}{\makebox(0,0){\strut{}$\tilde{\param{K}}^{(1)}(\Tsc{b}{};\mu_{Q_0})$}}}%
      \csname LTb\endcsname
      \put(4302,5441){\makebox(0,0)[r]{\strut{}$\param{\tilde{K}}^{(1)}(a_1=2\,\text{GeV})$}}%
      \csname LTb\endcsname
      \put(4302,5111){\makebox(0,0)[r]{\strut{}$\param{\tilde{K}}^{(1)}(a_2=4\,\text{GeV})$}}%
    }%
    \gplbacktext
    \put(0,0){\includegraphics{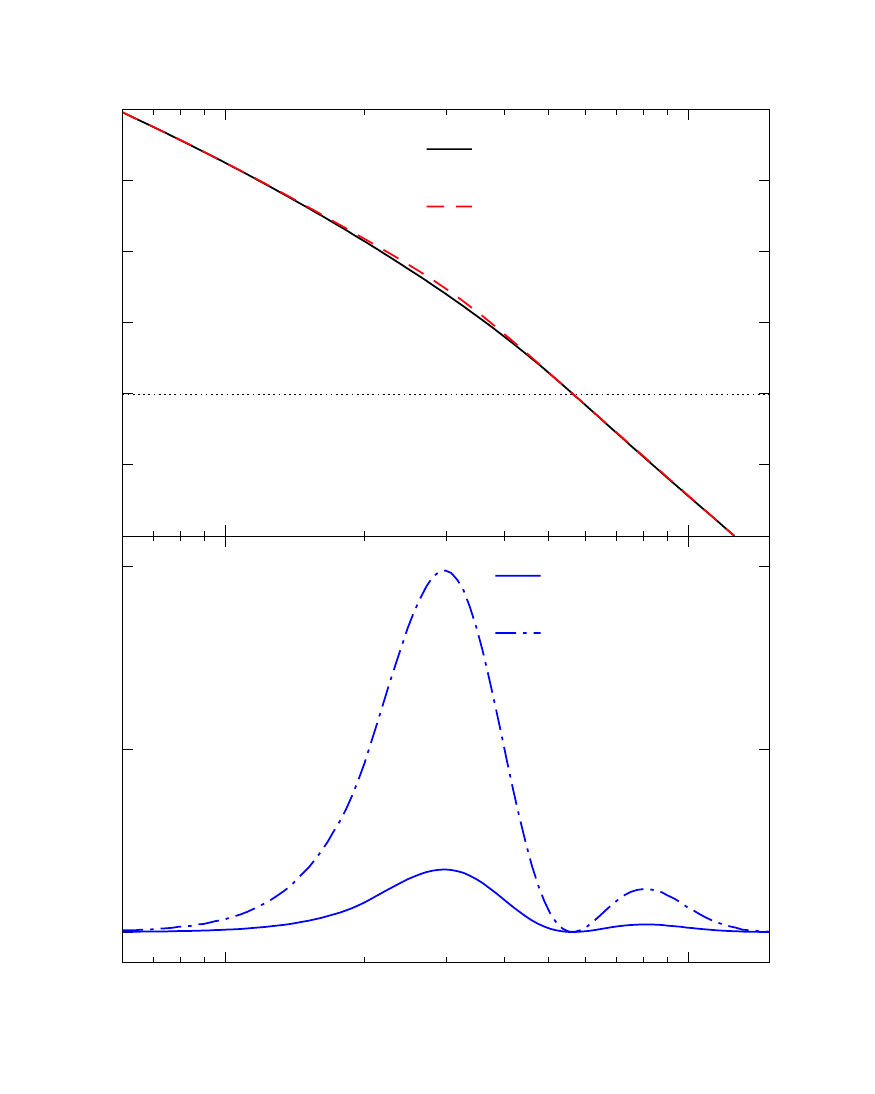}}%
    \gplfronttext
  \end{picture}%
\endgroup

%% file: ktildemK_dep_inline.tex
\begingroup
  \makeatletter
  \providecommand\color[2][]{%
    \GenericError{(gnuplot) \space\space\space\@spaces}{%
      Package color not loaded in conjunction with
      terminal option `colourtext'%
    }{See the gnuplot documentation for explanation.%
    }{Either use 'blacktext' in gnuplot or load the package
      color.sty in LaTeX.}%
    \renewcommand\color[2][]{}%
  }%
  \providecommand\includegraphics[2][]{%
    \GenericError{(gnuplot) \space\space\space\@spaces}{%
      Package graphicx or graphics not loaded%
    }{See the gnuplot documentation for explanation.%
    }{The gnuplot epslatex terminal needs graphicx.sty or graphics.sty.}%
    \renewcommand\includegraphics[2][]{}%
  }%
  \providecommand\rotatebox[2]{#2}%
  \@ifundefined{ifGPcolor}{%
    \newif\ifGPcolor
    \GPcolortrue
  }{}%
  \@ifundefined{ifGPblacktext}{%
    \newif\ifGPblacktext
    \GPblacktextfalse
  }{}%
  \let\gplgaddtomacro\g@addto@macro
  \gdef\gplbacktext{}%
  \gdef\gplfronttext{}%
  \makeatother
  \ifGPblacktext
    \def\colorrgb#1{}%
    \def\colorgray#1{}%
  \else
    \ifGPcolor
      \def\colorrgb#1{\color[rgb]{#1}}%
      \def\colorgray#1{\color[gray]{#1}}%
      \expandafter\def\csname LTw\endcsname{\color{white}}%
      \expandafter\def\csname LTb\endcsname{\color{black}}%
      \expandafter\def\csname LTa\endcsname{\color{black}}%
      \expandafter\def\csname LT0\endcsname{\color[rgb]{1,0,0}}%
      \expandafter\def\csname LT1\endcsname{\color[rgb]{0,1,0}}%
      \expandafter\def\csname LT2\endcsname{\color[rgb]{0,0,1}}%
      \expandafter\def\csname LT3\endcsname{\color[rgb]{1,0,1}}%
      \expandafter\def\csname LT4\endcsname{\color[rgb]{0,1,1}}%
      \expandafter\def\csname LT5\endcsname{\color[rgb]{1,1,0}}%
      \expandafter\def\csname LT6\endcsname{\color[rgb]{0,0,0}}%
      \expandafter\def\csname LT7\endcsname{\color[rgb]{1,0.3,0}}%
      \expandafter\def\csname LT8\endcsname{\color[rgb]{0.5,0.5,0.5}}%
    \else
      \def\colorrgb#1{\color{black}}%
      \def\colorgray#1{\color[gray]{#1}}%
      \expandafter\def\csname LTw\endcsname{\color{white}}%
      \expandafter\def\csname LTb\endcsname{\color{black}}%
      \expandafter\def\csname LTa\endcsname{\color{black}}%
      \expandafter\def\csname LT0\endcsname{\color{black}}%
      \expandafter\def\csname LT1\endcsname{\color{black}}%
      \expandafter\def\csname LT2\endcsname{\color{black}}%
      \expandafter\def\csname LT3\endcsname{\color{black}}%
      \expandafter\def\csname LT4\endcsname{\color{black}}%
      \expandafter\def\csname LT5\endcsname{\color{black}}%
      \expandafter\def\csname LT6\endcsname{\color{black}}%
      \expandafter\def\csname LT7\endcsname{\color{black}}%
      \expandafter\def\csname LT8\endcsname{\color{black}}%
    \fi
  \fi
    \setlength{\unitlength}{0.0500bp}%
    \ifx\gptboxheight\undefined%
      \newlength{\gptboxheight}%
      \newlength{\gptboxwidth}%
      \newsavebox{\gptboxtext}%
    \fi%
    \setlength{\fboxrule}{0.5pt}%
    \setlength{\fboxsep}{1pt}%
\begin{picture}(5040.00,3150.00)%
    \gplgaddtomacro\gplbacktext{%
      \csname LTb\endcsname
      \put(573,766){\makebox(0,0)[r]{\strut{}$-1.0$}}%
      \put(573,1412){\makebox(0,0)[r]{\strut{}$-0.5$}}%
      \put(573,2058){\makebox(0,0)[r]{\strut{}$0.0$}}%
      \put(573,2705){\makebox(0,0)[r]{\strut{}$0.5$}}%
      \put(705,158){\makebox(0,0){\strut{}$0.0$}}%
      \put(1637,158){\makebox(0,0){\strut{}$2.0$}}%
      \put(2570,158){\makebox(0,0){\strut{}$4.0$}}%
      \put(3502,158){\makebox(0,0){\strut{}$6.0$}}%
      \put(4434,158){\makebox(0,0){\strut{}$8.0$}}%
      \csname LTb\endcsname
      \put(1008,3023){\makebox(0,0)[l]{\strut{}$Q_0=2\,\text{GeV}$}}%
      \put(957,1354){\makebox(0,0)[l]{\strut{}$m_K(\text{GeV})$}}%
    }%
    \gplgaddtomacro\gplfronttext{%
      \csname LTb\endcsname
      \put(26,1606){\rotatebox{-270}{\makebox(0,0){\strut{}$\tilde{\param{K}}^{(1)}(\Tsc{b}{};\mu_{Q_0})$}}}%
      \put(2569,-172){\makebox(0,0){\strut{}$b_{\text{T}}(\text{GeV}^{-1})$}}%
      \csname LTb\endcsname
      \put(1764,1087){\makebox(0,0)[r]{\strut{}$0.10$}}%
      \csname LTb\endcsname
      \put(1764,867){\makebox(0,0)[r]{\strut{}$0.30$}}%
      \csname LTb\endcsname
      \put(1764,647){\makebox(0,0)[r]{\strut{}$0.50$}}%
    }%
    \gplbacktext
    \put(0,0){\includegraphics{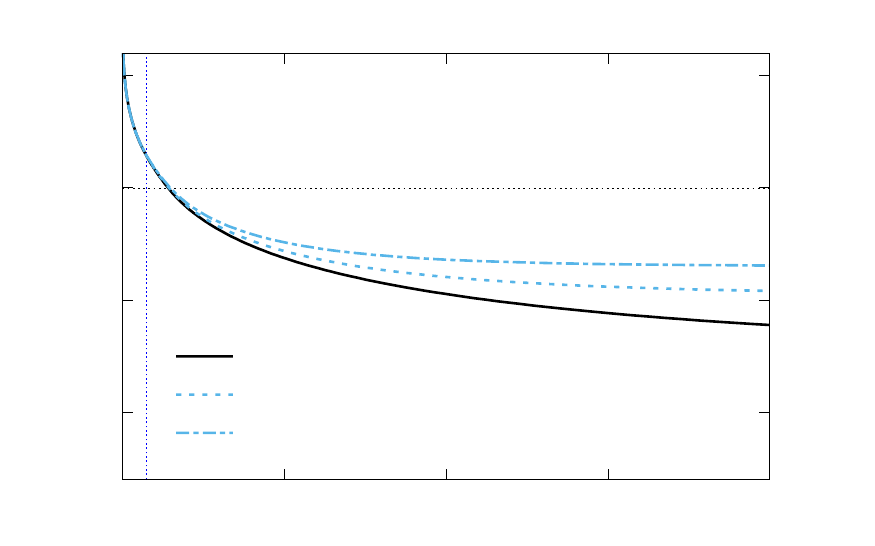}}%
    \gplfronttext
  \end{picture}%
\endgroup

%% file: Dinput_inline.tex
\begingroup
  \makeatletter
  \providecommand\color[2][]{%
    \GenericError{(gnuplot) \space\space\space\@spaces}{%
      Package color not loaded in conjunction with
      terminal option `colourtext'%
    }{See the gnuplot documentation for explanation.%
    }{Either use 'blacktext' in gnuplot or load the package
      color.sty in LaTeX.}%
    \renewcommand\color[2][]{}%
  }%
  \providecommand\includegraphics[2][]{%
    \GenericError{(gnuplot) \space\space\space\@spaces}{%
      Package graphicx or graphics not loaded%
    }{See the gnuplot documentation for explanation.%
    }{The gnuplot epslatex terminal needs graphicx.sty or graphics.sty.}%
    \renewcommand\includegraphics[2][]{}%
  }%
  \providecommand\rotatebox[2]{#2}%
  \@ifundefined{ifGPcolor}{%
    \newif\ifGPcolor
    \GPcolortrue
  }{}%
  \@ifundefined{ifGPblacktext}{%
    \newif\ifGPblacktext
    \GPblacktextfalse
  }{}%
  \let\gplgaddtomacro\g@addto@macro
  \gdef\gplbacktext{}%
  \gdef\gplfronttext{}%
  \makeatother
  \ifGPblacktext
    \def\colorrgb#1{}%
    \def\colorgray#1{}%
  \else
    \ifGPcolor
      \def\colorrgb#1{\color[rgb]{#1}}%
      \def\colorgray#1{\color[gray]{#1}}%
      \expandafter\def\csname LTw\endcsname{\color{white}}%
      \expandafter\def\csname LTb\endcsname{\color{black}}%
      \expandafter\def\csname LTa\endcsname{\color{black}}%
      \expandafter\def\csname LT0\endcsname{\color[rgb]{1,0,0}}%
      \expandafter\def\csname LT1\endcsname{\color[rgb]{0,1,0}}%
      \expandafter\def\csname LT2\endcsname{\color[rgb]{0,0,1}}%
      \expandafter\def\csname LT3\endcsname{\color[rgb]{1,0,1}}%
      \expandafter\def\csname LT4\endcsname{\color[rgb]{0,1,1}}%
      \expandafter\def\csname LT5\endcsname{\color[rgb]{1,1,0}}%
      \expandafter\def\csname LT6\endcsname{\color[rgb]{0,0,0}}%
      \expandafter\def\csname LT7\endcsname{\color[rgb]{1,0.3,0}}%
      \expandafter\def\csname LT8\endcsname{\color[rgb]{0.5,0.5,0.5}}%
    \else
      \def\colorrgb#1{\color{black}}%
      \def\colorgray#1{\color[gray]{#1}}%
      \expandafter\def\csname LTw\endcsname{\color{white}}%
      \expandafter\def\csname LTb\endcsname{\color{black}}%
      \expandafter\def\csname LTa\endcsname{\color{black}}%
      \expandafter\def\csname LT0\endcsname{\color{black}}%
      \expandafter\def\csname LT1\endcsname{\color{black}}%
      \expandafter\def\csname LT2\endcsname{\color{black}}%
      \expandafter\def\csname LT3\endcsname{\color{black}}%
      \expandafter\def\csname LT4\endcsname{\color{black}}%
      \expandafter\def\csname LT5\endcsname{\color{black}}%
      \expandafter\def\csname LT6\endcsname{\color{black}}%
      \expandafter\def\csname LT7\endcsname{\color{black}}%
      \expandafter\def\csname LT8\endcsname{\color{black}}%
    \fi
  \fi
    \setlength{\unitlength}{0.0500bp}%
    \ifx\gptboxheight\undefined%
      \newlength{\gptboxheight}%
      \newlength{\gptboxwidth}%
      \newsavebox{\gptboxtext}%
    \fi%
    \setlength{\fboxrule}{0.5pt}%
    \setlength{\fboxsep}{1pt}%
\begin{picture}(5040.00,3150.00)%
    \gplgaddtomacro\gplbacktext{%
      \csname LTb\endcsname
      \put(573,722){\makebox(0,0)[r]{\strut{}$0.0$}}%
      \put(573,1220){\makebox(0,0)[r]{\strut{}$0.5$}}%
      \put(573,1719){\makebox(0,0)[r]{\strut{}$1.0$}}%
      \put(573,2217){\makebox(0,0)[r]{\strut{}$1.5$}}%
      \put(573,2715){\makebox(0,0)[r]{\strut{}$2.0$}}%
      \put(705,158){\makebox(0,0){\strut{}$0.0$}}%
      \put(1948,158){\makebox(0,0){\strut{}$1.0$}}%
      \put(3191,158){\makebox(0,0){\strut{}$2.0$}}%
      \put(4434,158){\makebox(0,0){\strut{}$3.0$}}%
      \csname LTb\endcsname
      \put(1008,3023){\makebox(0,0)[l]{\strut{}$Q_0=2\,\text{GeV}$}}%
      \put(3023,3023){\makebox(0,0)[l]{\strut{}$z=0.3$}}%
      \put(2520,1575){\makebox(0,0)[l]{\strut{}$\,\,M=0.20\,\text{GeV}$}}%
      \put(2520,1260){\makebox(0,0)[l]{\strut{}$m_D=0.30\,\text{GeV}$}}%
    }%
    \gplgaddtomacro\gplfronttext{%
      \csname LTb\endcsname
      \put(57,1606){\rotatebox{-270}{\makebox(0,0){\strut{}$\Tsc{k}{}\,D(z,z \,\Tsc{k}{};\mu_{Q_0},Q_0^2)$}}}%
      \put(2569,-172){\makebox(0,0){\strut{}$k_{\text{T}}(\text{GeV})$}}%
      \csname LTb\endcsname
      \put(4283,2564){\makebox(0,0)[r]{\strut{}Gaussian$\qquad\qquad\qquad$}}%
      \csname LTb\endcsname
      \put(4283,2212){\makebox(0,0)[r]{\strut{}$\inpt{D}^{(1,d_{\msbar})}(z,z \,\Tsc{k}{};\mu_{Q_0},Q_0^2)$}}%
    }%
    \gplbacktext
    \put(0,0){\includegraphics{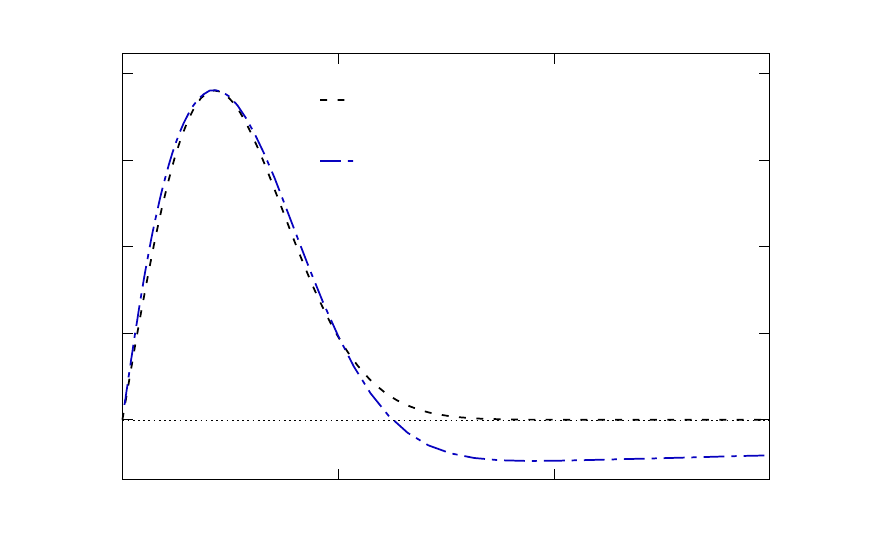}}%
    \gplfronttext
  \end{picture}%
\endgroup

%% file: Dtilde_underline_1_inline.tex
\begingroup
  \makeatletter
  \providecommand\color[2][]{%
    \GenericError{(gnuplot) \space\space\space\@spaces}{%
      Package color not loaded in conjunction with
      terminal option `colourtext'%
    }{See the gnuplot documentation for explanation.%
    }{Either use 'blacktext' in gnuplot or load the package
      color.sty in LaTeX.}%
    \renewcommand\color[2][]{}%
  }%
  \providecommand\includegraphics[2][]{%
    \GenericError{(gnuplot) \space\space\space\@spaces}{%
      Package graphicx or graphics not loaded%
    }{See the gnuplot documentation for explanation.%
    }{The gnuplot epslatex terminal needs graphicx.sty or graphics.sty.}%
    \renewcommand\includegraphics[2][]{}%
  }%
  \providecommand\rotatebox[2]{#2}%
  \@ifundefined{ifGPcolor}{%
    \newif\ifGPcolor
    \GPcolortrue
  }{}%
  \@ifundefined{ifGPblacktext}{%
    \newif\ifGPblacktext
    \GPblacktextfalse
  }{}%
  \let\gplgaddtomacro\g@addto@macro
  \gdef\gplbacktext{}%
  \gdef\gplfronttext{}%
  \makeatother
  \ifGPblacktext
    \def\colorrgb#1{}%
    \def\colorgray#1{}%
  \else
    \ifGPcolor
      \def\colorrgb#1{\color[rgb]{#1}}%
      \def\colorgray#1{\color[gray]{#1}}%
      \expandafter\def\csname LTw\endcsname{\color{white}}%
      \expandafter\def\csname LTb\endcsname{\color{black}}%
      \expandafter\def\csname LTa\endcsname{\color{black}}%
      \expandafter\def\csname LT0\endcsname{\color[rgb]{1,0,0}}%
      \expandafter\def\csname LT1\endcsname{\color[rgb]{0,1,0}}%
      \expandafter\def\csname LT2\endcsname{\color[rgb]{0,0,1}}%
      \expandafter\def\csname LT3\endcsname{\color[rgb]{1,0,1}}%
      \expandafter\def\csname LT4\endcsname{\color[rgb]{0,1,1}}%
      \expandafter\def\csname LT5\endcsname{\color[rgb]{1,1,0}}%
      \expandafter\def\csname LT6\endcsname{\color[rgb]{0,0,0}}%
      \expandafter\def\csname LT7\endcsname{\color[rgb]{1,0.3,0}}%
      \expandafter\def\csname LT8\endcsname{\color[rgb]{0.5,0.5,0.5}}%
    \else
      \def\colorrgb#1{\color{black}}%
      \def\colorgray#1{\color[gray]{#1}}%
      \expandafter\def\csname LTw\endcsname{\color{white}}%
      \expandafter\def\csname LTb\endcsname{\color{black}}%
      \expandafter\def\csname LTa\endcsname{\color{black}}%
      \expandafter\def\csname LT0\endcsname{\color{black}}%
      \expandafter\def\csname LT1\endcsname{\color{black}}%
      \expandafter\def\csname LT2\endcsname{\color{black}}%
      \expandafter\def\csname LT3\endcsname{\color{black}}%
      \expandafter\def\csname LT4\endcsname{\color{black}}%
      \expandafter\def\csname LT5\endcsname{\color{black}}%
      \expandafter\def\csname LT6\endcsname{\color{black}}%
      \expandafter\def\csname LT7\endcsname{\color{black}}%
      \expandafter\def\csname LT8\endcsname{\color{black}}%
    \fi
  \fi
    \setlength{\unitlength}{0.0500bp}%
    \ifx\gptboxheight\undefined%
      \newlength{\gptboxheight}%
      \newlength{\gptboxwidth}%
      \newsavebox{\gptboxtext}%
    \fi%
    \setlength{\fboxrule}{0.5pt}%
    \setlength{\fboxsep}{1pt}%
\begin{picture}(5040.00,3150.00)%
    \gplgaddtomacro\gplbacktext{%
      \csname LTb\endcsname
      \put(573,638){\makebox(0,0)[r]{\strut{}$1.0$}}%
      \put(573,918){\makebox(0,0)[r]{\strut{}$2.0$}}%
      \put(573,1198){\makebox(0,0)[r]{\strut{}$3.0$}}%
      \put(573,1478){\makebox(0,0)[r]{\strut{}$4.0$}}%
      \put(573,1758){\makebox(0,0)[r]{\strut{}$5.0$}}%
      \put(573,2037){\makebox(0,0)[r]{\strut{}$6.0$}}%
      \put(573,2317){\makebox(0,0)[r]{\strut{}$7.0$}}%
      \put(573,2597){\makebox(0,0)[r]{\strut{}$8.0$}}%
      \put(705,158){\makebox(0,0){\strut{}$10^{-2}$}}%
      \put(1948,158){\makebox(0,0){\strut{}$10^{-1}$}}%
      \put(3191,158){\makebox(0,0){\strut{}$10^{0}$}}%
      \put(4434,158){\makebox(0,0){\strut{}$10^{1}$}}%
      \put(1008,3023){\makebox(0,0)[l]{\strut{}$Q_0=2\,\text{GeV}$}}%
      \put(3023,3023){\makebox(0,0)[l]{\strut{}$z=0.3$}}%
      \put(2318,1480){\makebox(0,0)[l]{\strut{}$\,\,M=0.20\,\text{GeV}$}}%
      \put(2318,1165){\makebox(0,0)[l]{\strut{}$m_D=0.30\,\text{GeV}$}}%
      \put(2318,850){\makebox(0,0)[l]{\strut{}$m_K=0.10\,\text{GeV}$}}%
    }%
    \gplgaddtomacro\gplfronttext{%
      \csname LTb\endcsname
      \put(57,1606){\rotatebox{-270}{\makebox(0,0){\strut{}$\tilde{\param{D}}^{(1,d_{\msbar})}(z,\Tsc{b}{};\mu_{Q_0},Q_0^2)$}}}%
      \put(2569,-172){\makebox(0,0){\strut{}$b_{\text{T}}(\text{GeV}^{-1})$}}%
      \csname LTb\endcsname
      \put(2016,2564){\makebox(0,0)[r]{\strut{}$a=2\,\text{GeV}$}}%
      \csname LTb\endcsname
      \put(2016,2212){\makebox(0,0)[r]{\strut{}$a=4\,\text{GeV}$}}%
    }%
    \gplbacktext
    \put(0,0){\includegraphics{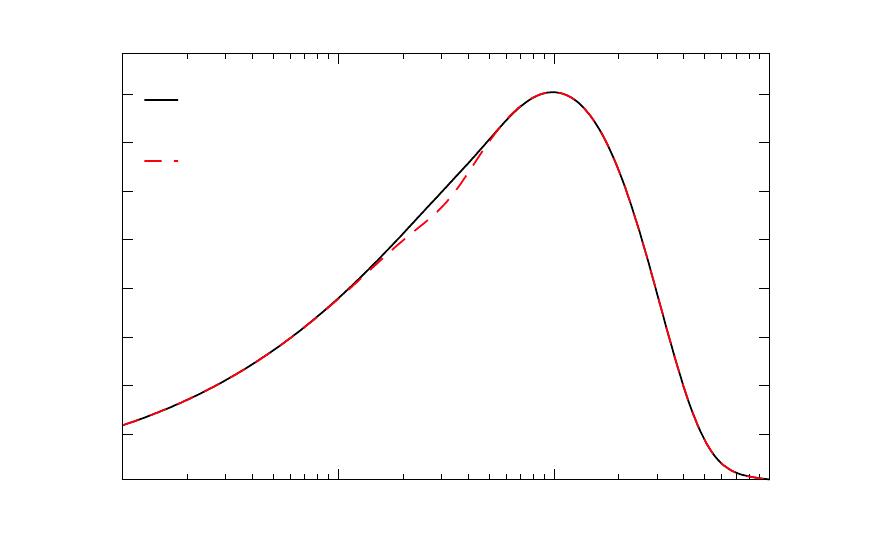}}%
    \gplfronttext
  \end{picture}%
\endgroup

%% file: WqT_step_B3_2_inline.tex
\begingroup
  \makeatletter
  \providecommand\color[2][]{%
    \GenericError{(gnuplot) \space\space\space\@spaces}{%
      Package color not loaded in conjunction with
      terminal option `colourtext'%
    }{See the gnuplot documentation for explanation.%
    }{Either use 'blacktext' in gnuplot or load the package
      color.sty in LaTeX.}%
    \renewcommand\color[2][]{}%
  }%
  \providecommand\includegraphics[2][]{%
    \GenericError{(gnuplot) \space\space\space\@spaces}{%
      Package graphicx or graphics not loaded%
    }{See the gnuplot documentation for explanation.%
    }{The gnuplot epslatex terminal needs graphicx.sty or graphics.sty.}%
    \renewcommand\includegraphics[2][]{}%
  }%
  \providecommand\rotatebox[2]{#2}%
  \@ifundefined{ifGPcolor}{%
    \newif\ifGPcolor
    \GPcolortrue
  }{}%
  \@ifundefined{ifGPblacktext}{%
    \newif\ifGPblacktext
    \GPblacktextfalse
  }{}%
  \let\gplgaddtomacro\g@addto@macro
  \gdef\gplbacktext{}%
  \gdef\gplfronttext{}%
  \makeatother
  \ifGPblacktext
    \def\colorrgb#1{}%
    \def\colorgray#1{}%
  \else
    \ifGPcolor
      \def\colorrgb#1{\color[rgb]{#1}}%
      \def\colorgray#1{\color[gray]{#1}}%
      \expandafter\def\csname LTw\endcsname{\color{white}}%
      \expandafter\def\csname LTb\endcsname{\color{black}}%
      \expandafter\def\csname LTa\endcsname{\color{black}}%
      \expandafter\def\csname LT0\endcsname{\color[rgb]{1,0,0}}%
      \expandafter\def\csname LT1\endcsname{\color[rgb]{0,1,0}}%
      \expandafter\def\csname LT2\endcsname{\color[rgb]{0,0,1}}%
      \expandafter\def\csname LT3\endcsname{\color[rgb]{1,0,1}}%
      \expandafter\def\csname LT4\endcsname{\color[rgb]{0,1,1}}%
      \expandafter\def\csname LT5\endcsname{\color[rgb]{1,1,0}}%
      \expandafter\def\csname LT6\endcsname{\color[rgb]{0,0,0}}%
      \expandafter\def\csname LT7\endcsname{\color[rgb]{1,0.3,0}}%
      \expandafter\def\csname LT8\endcsname{\color[rgb]{0.5,0.5,0.5}}%
    \else
      \def\colorrgb#1{\color{black}}%
      \def\colorgray#1{\color[gray]{#1}}%
      \expandafter\def\csname LTw\endcsname{\color{white}}%
      \expandafter\def\csname LTb\endcsname{\color{black}}%
      \expandafter\def\csname LTa\endcsname{\color{black}}%
      \expandafter\def\csname LT0\endcsname{\color{black}}%
      \expandafter\def\csname LT1\endcsname{\color{black}}%
      \expandafter\def\csname LT2\endcsname{\color{black}}%
      \expandafter\def\csname LT3\endcsname{\color{black}}%
      \expandafter\def\csname LT4\endcsname{\color{black}}%
      \expandafter\def\csname LT5\endcsname{\color{black}}%
      \expandafter\def\csname LT6\endcsname{\color{black}}%
      \expandafter\def\csname LT7\endcsname{\color{black}}%
      \expandafter\def\csname LT8\endcsname{\color{black}}%
    \fi
  \fi
    \setlength{\unitlength}{0.0500bp}%
    \ifx\gptboxheight\undefined%
      \newlength{\gptboxheight}%
      \newlength{\gptboxwidth}%
      \newsavebox{\gptboxtext}%
    \fi%
    \setlength{\fboxrule}{0.5pt}%
    \setlength{\fboxsep}{1pt}%
\begin{picture}(5040.00,3150.00)%
    \gplgaddtomacro\gplbacktext{%
      \csname LTb\endcsname
      \put(573,421){\makebox(0,0)[r]{\strut{}$-4.0$}}%
      \put(573,693){\makebox(0,0)[r]{\strut{}$-2.0$}}%
      \put(573,965){\makebox(0,0)[r]{\strut{}$0.0$}}%
      \put(573,1237){\makebox(0,0)[r]{\strut{}$2.0$}}%
      \put(573,1508){\makebox(0,0)[r]{\strut{}$4.0$}}%
      \put(573,1780){\makebox(0,0)[r]{\strut{}$6.0$}}%
      \put(573,2052){\makebox(0,0)[r]{\strut{}$8.0$}}%
      \put(573,2324){\makebox(0,0)[r]{\strut{}$10.0$}}%
      \put(573,2595){\makebox(0,0)[r]{\strut{}$12.0$}}%
      \put(705,158){\makebox(0,0){\strut{}$0.0$}}%
      \put(1637,158){\makebox(0,0){\strut{}$1.0$}}%
      \put(2570,158){\makebox(0,0){\strut{}$2.0$}}%
      \put(3502,158){\makebox(0,0){\strut{}$3.0$}}%
      \put(4434,158){\makebox(0,0){\strut{}$4.0$}}%
      \csname LTb\endcsname
      \put(1008,3023){\makebox(0,0)[l]{\strut{}$Q_0=2\,\text{GeV}$}}%
      \put(3023,3023){\makebox(0,0)[l]{\strut{}$z=0.3$}}%
      \put(1612,2362){\makebox(0,0)[l]{\strut{}$Q=Q_0$}}%
      \put(1764,1952){\makebox(0,0)[l]{\strut{}$Q=2\,Q_0$}}%
      \put(2066,1637){\makebox(0,0)[l]{\strut{}$Q=5\,Q_0$}}%
    }%
    \gplgaddtomacro\gplfronttext{%
      \csname LTb\endcsname
      \put(-75,1606){\rotatebox{-270}{\makebox(0,0){\strut{}$q_{\text{T}}\,W^{(1)}(Q,q_{\text{T}})/H^{(1)}$}}}%
      \put(2569,-172){\makebox(0,0){\strut{}$q_{\text{T}}(\text{GeV})$}}%
      \csname LTb\endcsname
      \put(4283,2212){\makebox(0,0)[r]{\strut{}$\tilde{\param{D}}^{(1,d_{\msbar})},\,\,\tilde{\param{K}}^{(1)}_{\hphantom{input}}$}}%
      \csname LTb\endcsname
      \put(4283,2564){\makebox(0,0)[r]{\strut{}$\inpt{\tilde{D}}^{(1,d_{\msbar})},\,\,\inpt{\tilde{K}}^{(1)}$}}%
    }%
    \gplbacktext
    \put(0,0){\includegraphics{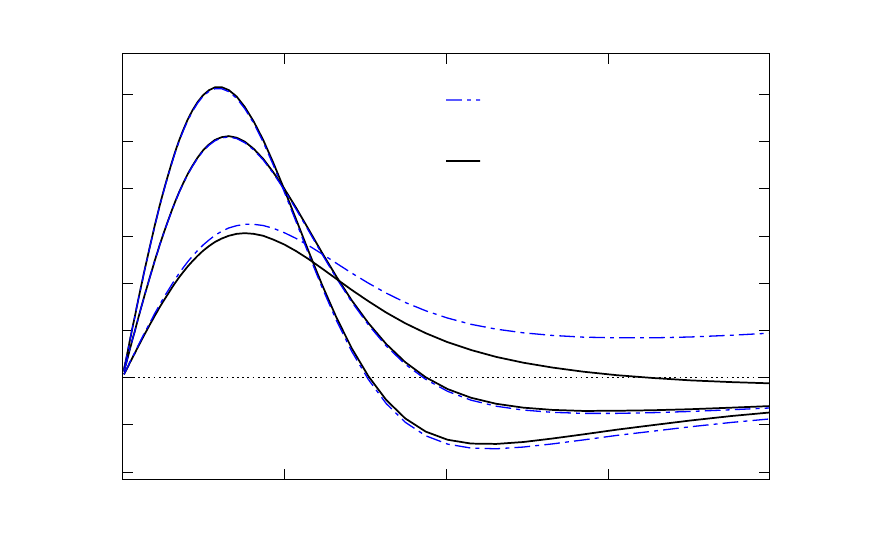}}%
    \gplfronttext
  \end{picture}%
\endgroup

%% file: WqT_bands_inline.tex
\begingroup
  \makeatletter
  \providecommand\color[2][]{%
    \GenericError{(gnuplot) \space\space\space\@spaces}{%
      Package color not loaded in conjunction with
      terminal option `colourtext'%
    }{See the gnuplot documentation for explanation.%
    }{Either use 'blacktext' in gnuplot or load the package
      color.sty in LaTeX.}%
    \renewcommand\color[2][]{}%
  }%
  \providecommand\includegraphics[2][]{%
    \GenericError{(gnuplot) \space\space\space\@spaces}{%
      Package graphicx or graphics not loaded%
    }{See the gnuplot documentation for explanation.%
    }{The gnuplot epslatex terminal needs graphicx.sty or graphics.sty.}%
    \renewcommand\includegraphics[2][]{}%
  }%
  \providecommand\rotatebox[2]{#2}%
  \@ifundefined{ifGPcolor}{%
    \newif\ifGPcolor
    \GPcolortrue
  }{}%
  \@ifundefined{ifGPblacktext}{%
    \newif\ifGPblacktext
    \GPblacktextfalse
  }{}%
  \let\gplgaddtomacro\g@addto@macro
  \gdef\gplbacktext{}%
  \gdef\gplfronttext{}%
  \makeatother
  \ifGPblacktext
    \def\colorrgb#1{}%
    \def\colorgray#1{}%
  \else
    \ifGPcolor
      \def\colorrgb#1{\color[rgb]{#1}}%
      \def\colorgray#1{\color[gray]{#1}}%
      \expandafter\def\csname LTw\endcsname{\color{white}}%
      \expandafter\def\csname LTb\endcsname{\color{black}}%
      \expandafter\def\csname LTa\endcsname{\color{black}}%
      \expandafter\def\csname LT0\endcsname{\color[rgb]{1,0,0}}%
      \expandafter\def\csname LT1\endcsname{\color[rgb]{0,1,0}}%
      \expandafter\def\csname LT2\endcsname{\color[rgb]{0,0,1}}%
      \expandafter\def\csname LT3\endcsname{\color[rgb]{1,0,1}}%
      \expandafter\def\csname LT4\endcsname{\color[rgb]{0,1,1}}%
      \expandafter\def\csname LT5\endcsname{\color[rgb]{1,1,0}}%
      \expandafter\def\csname LT6\endcsname{\color[rgb]{0,0,0}}%
      \expandafter\def\csname LT7\endcsname{\color[rgb]{1,0.3,0}}%
      \expandafter\def\csname LT8\endcsname{\color[rgb]{0.5,0.5,0.5}}%
    \else
      \def\colorrgb#1{\color{black}}%
      \def\colorgray#1{\color[gray]{#1}}%
      \expandafter\def\csname LTw\endcsname{\color{white}}%
      \expandafter\def\csname LTb\endcsname{\color{black}}%
      \expandafter\def\csname LTa\endcsname{\color{black}}%
      \expandafter\def\csname LT0\endcsname{\color{black}}%
      \expandafter\def\csname LT1\endcsname{\color{black}}%
      \expandafter\def\csname LT2\endcsname{\color{black}}%
      \expandafter\def\csname LT3\endcsname{\color{black}}%
      \expandafter\def\csname LT4\endcsname{\color{black}}%
      \expandafter\def\csname LT5\endcsname{\color{black}}%
      \expandafter\def\csname LT6\endcsname{\color{black}}%
      \expandafter\def\csname LT7\endcsname{\color{black}}%
      \expandafter\def\csname LT8\endcsname{\color{black}}%
    \fi
  \fi
    \setlength{\unitlength}{0.0500bp}%
    \ifx\gptboxheight\undefined%
      \newlength{\gptboxheight}%
      \newlength{\gptboxwidth}%
      \newsavebox{\gptboxtext}%
    \fi%
    \setlength{\fboxrule}{0.5pt}%
    \setlength{\fboxsep}{1pt}%
\begin{picture}(5040.00,6300.00)%
    \gplgaddtomacro\gplbacktext{%
      \csname LTb\endcsname
      \put(573,4222){\makebox(0,0)[r]{\strut{}$10^{-2}$}}%
      \put(573,4895){\makebox(0,0)[r]{\strut{}$10^{0}$}}%
      \put(573,5568){\makebox(0,0)[r]{\strut{}$10^{2}$}}%
      \put(1233,5301){\makebox(0,0)[l]{\strut{}$a=2\,\text{GeV}$}}%
      \put(860,3999){\makebox(0,0)[l]{\strut{}$M_{\hphantom{D}}=0.2^{+0.10}_{-0.10}\,\text{GeV}$}}%
      \put(860,3729){\makebox(0,0)[l]{\strut{}$m_D=0.3^{+0.15}_{-0.15}\,\text{GeV}$}}%
      \put(860,3459){\makebox(0,0)[l]{\strut{}$m_K=0.1^{+0.05}_{-0.05}\,\text{GeV}$}}%
      \put(1008,5890){\makebox(0,0)[l]{\strut{}$Q_0=2\,\text{GeV}$}}%
      \put(3023,5890){\makebox(0,0)[l]{\strut{}$z=0.3$}}%
      \put(2681,5448){\makebox(0,0)[l]{\strut{}$Q=2\,\text{GeV}\hphantom{000}$}}%
      \put(2681,5153){\makebox(0,0)[l]{\strut{}$\hphantom{Q=}10\,\text{GeV}\hphantom{00}$}}%
      \put(2681,4859){\makebox(0,0)[l]{\strut{}$\hphantom{Q=}100\,\text{GeV}\hphantom{0}$}}%
      \put(2681,4564){\makebox(0,0)[l]{\strut{}$\hphantom{Q=}1000\,\text{GeV}$}}%
    }%
    \gplgaddtomacro\gplfronttext{%
    }%
    \gplgaddtomacro\gplbacktext{%
      \csname LTb\endcsname
      \put(573,1765){\makebox(0,0)[r]{\strut{}$10^{-2}$}}%
      \put(573,2438){\makebox(0,0)[r]{\strut{}$10^{0}$}}%
      \put(573,3111){\makebox(0,0)[r]{\strut{}$10^{2}$}}%
      \put(705,536){\makebox(0,0){\strut{}$0.0$}}%
      \put(1327,536){\makebox(0,0){\strut{}$2.0$}}%
      \put(1948,536){\makebox(0,0){\strut{}$4.0$}}%
      \put(2570,536){\makebox(0,0){\strut{}$6.0$}}%
      \put(3191,536){\makebox(0,0){\strut{}$8.0$}}%
      \put(3813,536){\makebox(0,0){\strut{}$10.0$}}%
      \put(4434,536){\makebox(0,0){\strut{}$12.0$}}%
      \put(1233,2844){\makebox(0,0)[l]{\strut{}$Q_0<a<2\,Q_0$}}%
      \put(860,1468){\makebox(0,0)[l]{\strut{}$M_{\hphantom{D}}=0.2\,\text{GeV}$}}%
      \put(860,1223){\makebox(0,0)[l]{\strut{}$m_D=0.3\,\text{GeV}$}}%
      \put(860,977){\makebox(0,0)[l]{\strut{}$m_K=0.1\,\text{GeV}$}}%
    }%
    \gplgaddtomacro\gplfronttext{%
      \csname LTb\endcsname
      \put(26,3243){\rotatebox{-270}{\makebox(0,0){\strut{}$W^{(1)}(Q,q_{\text{T}})/H^{(1)}$}}}%
      \put(2569,206){\makebox(0,0){\strut{}$q_{\text{T}}(\text{GeV})$}}%
    }%
    \gplbacktext
    \put(0,0){\includegraphics{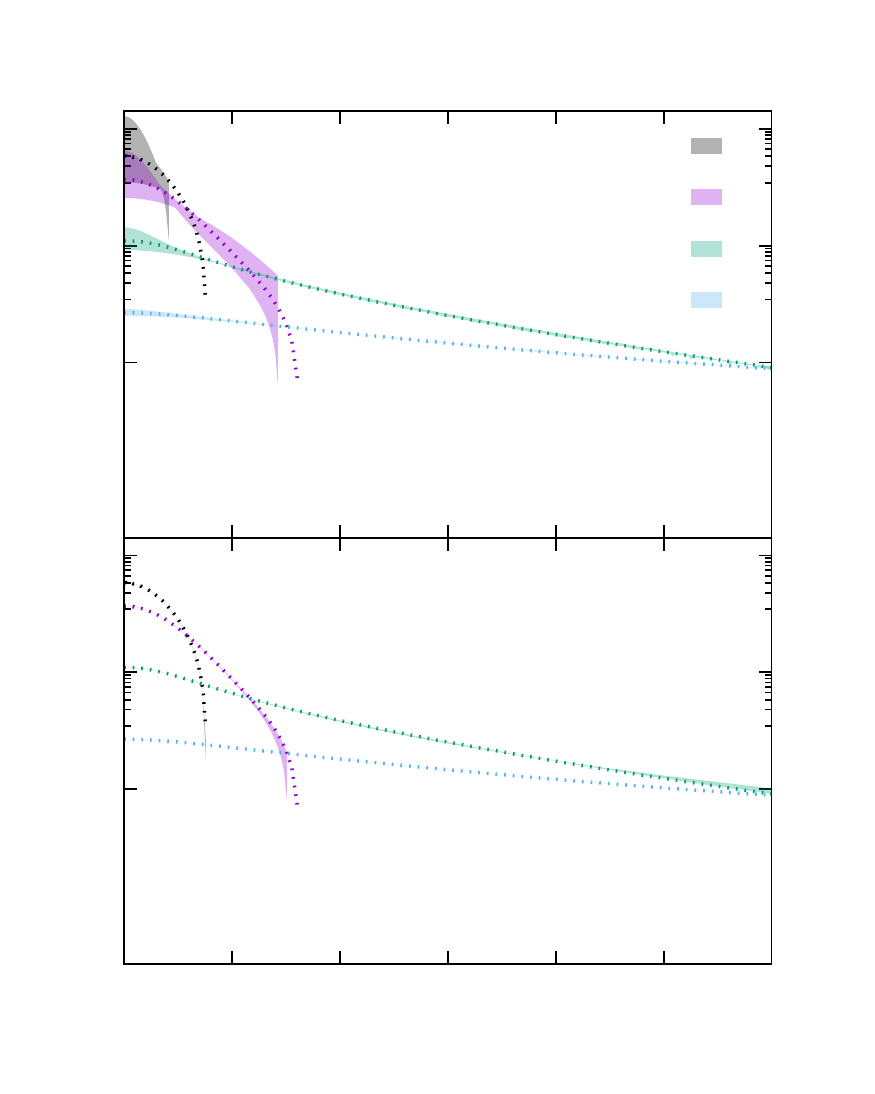}}%
    \gplfronttext
  \end{picture}%
\endgroup

%% file: WqT_bands_linear_inline.tex
\begingroup
  \makeatletter
  \providecommand\color[2][]{%
    \GenericError{(gnuplot) \space\space\space\@spaces}{%
      Package color not loaded in conjunction with
      terminal option `colourtext'%
    }{See the gnuplot documentation for explanation.%
    }{Either use 'blacktext' in gnuplot or load the package
      color.sty in LaTeX.}%
    \renewcommand\color[2][]{}%
  }%
  \providecommand\includegraphics[2][]{%
    \GenericError{(gnuplot) \space\space\space\@spaces}{%
      Package graphicx or graphics not loaded%
    }{See the gnuplot documentation for explanation.%
    }{The gnuplot epslatex terminal needs graphicx.sty or graphics.sty.}%
    \renewcommand\includegraphics[2][]{}%
  }%
  \providecommand\rotatebox[2]{#2}%
  \@ifundefined{ifGPcolor}{%
    \newif\ifGPcolor
    \GPcolortrue
  }{}%
  \@ifundefined{ifGPblacktext}{%
    \newif\ifGPblacktext
    \GPblacktextfalse
  }{}%
  \let\gplgaddtomacro\g@addto@macro
  \gdef\gplbacktext{}%
  \gdef\gplfronttext{}%
  \makeatother
  \ifGPblacktext
    \def\colorrgb#1{}%
    \def\colorgray#1{}%
  \else
    \ifGPcolor
      \def\colorrgb#1{\color[rgb]{#1}}%
      \def\colorgray#1{\color[gray]{#1}}%
      \expandafter\def\csname LTw\endcsname{\color{white}}%
      \expandafter\def\csname LTb\endcsname{\color{black}}%
      \expandafter\def\csname LTa\endcsname{\color{black}}%
      \expandafter\def\csname LT0\endcsname{\color[rgb]{1,0,0}}%
      \expandafter\def\csname LT1\endcsname{\color[rgb]{0,1,0}}%
      \expandafter\def\csname LT2\endcsname{\color[rgb]{0,0,1}}%
      \expandafter\def\csname LT3\endcsname{\color[rgb]{1,0,1}}%
      \expandafter\def\csname LT4\endcsname{\color[rgb]{0,1,1}}%
      \expandafter\def\csname LT5\endcsname{\color[rgb]{1,1,0}}%
      \expandafter\def\csname LT6\endcsname{\color[rgb]{0,0,0}}%
      \expandafter\def\csname LT7\endcsname{\color[rgb]{1,0.3,0}}%
      \expandafter\def\csname LT8\endcsname{\color[rgb]{0.5,0.5,0.5}}%
    \else
      \def\colorrgb#1{\color{black}}%
      \def\colorgray#1{\color[gray]{#1}}%
      \expandafter\def\csname LTw\endcsname{\color{white}}%
      \expandafter\def\csname LTb\endcsname{\color{black}}%
      \expandafter\def\csname LTa\endcsname{\color{black}}%
      \expandafter\def\csname LT0\endcsname{\color{black}}%
      \expandafter\def\csname LT1\endcsname{\color{black}}%
      \expandafter\def\csname LT2\endcsname{\color{black}}%
      \expandafter\def\csname LT3\endcsname{\color{black}}%
      \expandafter\def\csname LT4\endcsname{\color{black}}%
      \expandafter\def\csname LT5\endcsname{\color{black}}%
      \expandafter\def\csname LT6\endcsname{\color{black}}%
      \expandafter\def\csname LT7\endcsname{\color{black}}%
      \expandafter\def\csname LT8\endcsname{\color{black}}%
    \fi
  \fi
    \setlength{\unitlength}{0.0500bp}%
    \ifx\gptboxheight\undefined%
      \newlength{\gptboxheight}%
      \newlength{\gptboxwidth}%
      \newsavebox{\gptboxtext}%
    \fi%
    \setlength{\fboxrule}{0.5pt}%
    \setlength{\fboxsep}{1pt}%
\begin{picture}(5040.00,6300.00)%
    \gplgaddtomacro\gplbacktext{%
      \csname LTb\endcsname
      \put(573,3436){\makebox(0,0)[r]{\strut{}0.0}}%
      \put(573,3883){\makebox(0,0)[r]{\strut{}1.0}}%
      \put(573,4329){\makebox(0,0)[r]{\strut{}2.0}}%
      \put(573,4776){\makebox(0,0)[r]{\strut{}3.0}}%
      \put(573,5222){\makebox(0,0)[r]{\strut{}4.0}}%
      \put(705,2993){\makebox(0,0){\strut{}}}%
      \put(1637,2993){\makebox(0,0){\strut{}}}%
      \put(2570,2993){\makebox(0,0){\strut{}}}%
      \put(3502,2993){\makebox(0,0){\strut{}}}%
      \put(4434,2993){\makebox(0,0){\strut{}}}%
      \csname LTb\endcsname
      \put(2644,4687){\makebox(0,0)[l]{\strut{}$a_{\hphantom{D}}=2\,\text{GeV}$}}%
      \put(2570,4441){\makebox(0,0)[l]{\strut{}$M_{\hphantom{D}}=0.2^{+0.10}_{-0.10}\,\text{GeV}$}}%
      \put(2570,4195){\makebox(0,0)[l]{\strut{}$m_D=0.3^{+0.15}_{-0.15}\,\text{GeV}$}}%
      \put(2570,3950){\makebox(0,0)[l]{\strut{}$m_K=0.1^{+0.05}_{-0.05}\,\text{GeV}$}}%
      \put(1008,5890){\makebox(0,0)[l]{\strut{}$Q_0=2\,\text{GeV}$}}%
      \put(3023,5890){\makebox(0,0)[l]{\strut{}$z=0.3$}}%
    }%
    \gplgaddtomacro\gplfronttext{%
      \csname LTb\endcsname
      \put(3447,5496){\makebox(0,0)[r]{\strut{}$Q=2\,\text{GeV}\hphantom{000}$}}%
      \csname LTb\endcsname
      \put(3447,5276){\makebox(0,0)[r]{\strut{}$Q=10\,\text{GeV}\hphantom{00}$}}%
    }%
    \gplgaddtomacro\gplbacktext{%
      \csname LTb\endcsname
      \put(573,756){\makebox(0,0)[r]{\strut{}0.0}}%
      \put(573,1334){\makebox(0,0)[r]{\strut{}0.4}}%
      \put(573,1912){\makebox(0,0)[r]{\strut{}0.8}}%
      \put(573,2490){\makebox(0,0)[r]{\strut{}1.2}}%
      \put(573,3068){\makebox(0,0)[r]{\strut{}1.6}}%
      \put(705,536){\makebox(0,0){\strut{}0.0}}%
      \put(1637,536){\makebox(0,0){\strut{}1.0}}%
      \put(2570,536){\makebox(0,0){\strut{}2.0}}%
      \put(3502,536){\makebox(0,0){\strut{}3.0}}%
      \put(4434,536){\makebox(0,0){\strut{}4.0}}%
    }%
    \gplgaddtomacro\gplfronttext{%
      \csname LTb\endcsname
      \put(158,3243){\rotatebox{-270}{\makebox(0,0){\strut{}$W^{(1)}(Q,q_{\text{T}})/W^{(1)}(Q,q_{\text{T}}=0)$}}}%
      \put(2569,206){\makebox(0,0){\strut{}$\Tsc{q}{}$(\text{GeV})}}%
      \csname LTb\endcsname
      \put(3447,3039){\makebox(0,0)[r]{\strut{}$Q=100\,\text{GeV}\hphantom{0}$}}%
      \csname LTb\endcsname
      \put(3447,2819){\makebox(0,0)[r]{\strut{}$Q=1000\,\text{GeV}$}}%
    }%
    \gplbacktext
    \put(0,0){\includegraphics{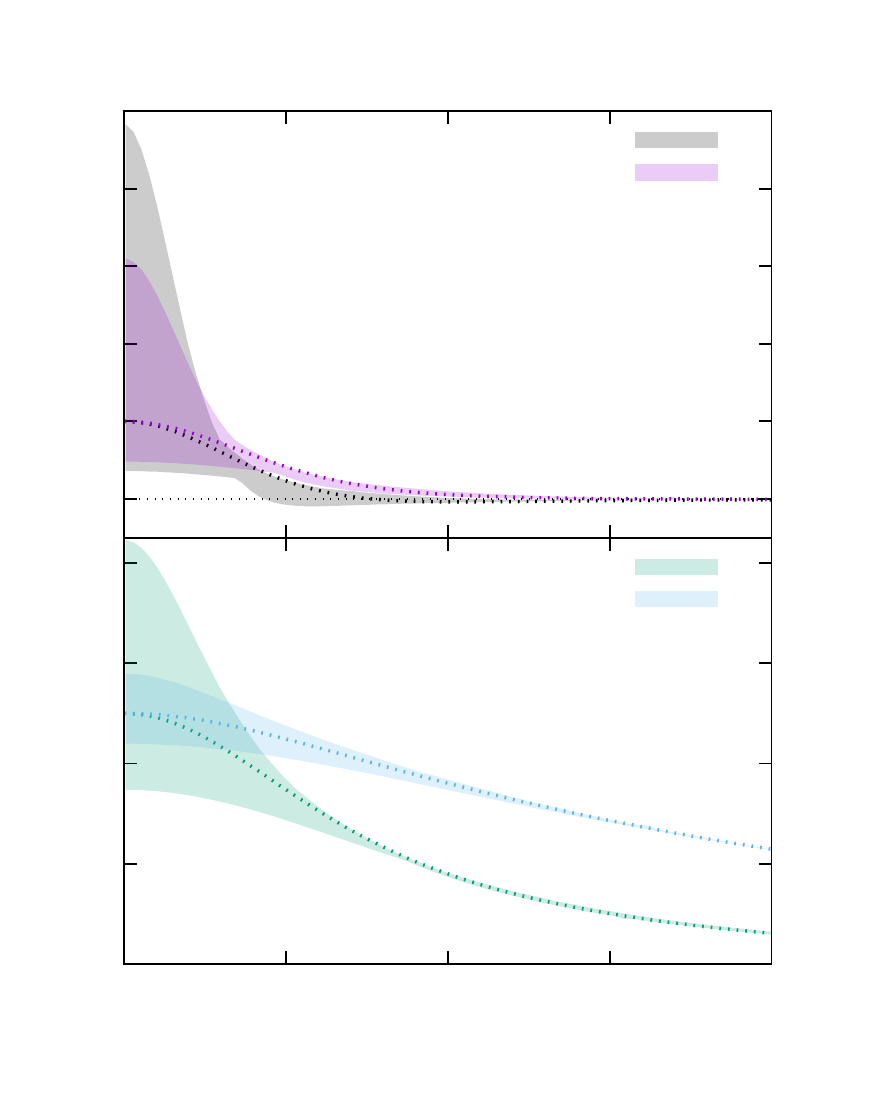}}%
    \gplfronttext
  \end{picture}%
\endgroup

%% file: gk_inline.tex
\begingroup
  \makeatletter
  \providecommand\color[2][]{%
    \GenericError{(gnuplot) \space\space\space\@spaces}{%
      Package color not loaded in conjunction with
      terminal option `colourtext'%
    }{See the gnuplot documentation for explanation.%
    }{Either use 'blacktext' in gnuplot or load the package
      color.sty in LaTeX.}%
    \renewcommand\color[2][]{}%
  }%
  \providecommand\includegraphics[2][]{%
    \GenericError{(gnuplot) \space\space\space\@spaces}{%
      Package graphicx or graphics not loaded%
    }{See the gnuplot documentation for explanation.%
    }{The gnuplot epslatex terminal needs graphicx.sty or graphics.sty.}%
    \renewcommand\includegraphics[2][]{}%
  }%
  \providecommand\rotatebox[2]{#2}%
  \@ifundefined{ifGPcolor}{%
    \newif\ifGPcolor
    \GPcolortrue
  }{}%
  \@ifundefined{ifGPblacktext}{%
    \newif\ifGPblacktext
    \GPblacktextfalse
  }{}%
  \let\gplgaddtomacro\g@addto@macro
  \gdef\gplbacktext{}%
  \gdef\gplfronttext{}%
  \makeatother
  \ifGPblacktext
    \def\colorrgb#1{}%
    \def\colorgray#1{}%
  \else
    \ifGPcolor
      \def\colorrgb#1{\color[rgb]{#1}}%
      \def\colorgray#1{\color[gray]{#1}}%
      \expandafter\def\csname LTw\endcsname{\color{white}}%
      \expandafter\def\csname LTb\endcsname{\color{black}}%
      \expandafter\def\csname LTa\endcsname{\color{black}}%
      \expandafter\def\csname LT0\endcsname{\color[rgb]{1,0,0}}%
      \expandafter\def\csname LT1\endcsname{\color[rgb]{0,1,0}}%
      \expandafter\def\csname LT2\endcsname{\color[rgb]{0,0,1}}%
      \expandafter\def\csname LT3\endcsname{\color[rgb]{1,0,1}}%
      \expandafter\def\csname LT4\endcsname{\color[rgb]{0,1,1}}%
      \expandafter\def\csname LT5\endcsname{\color[rgb]{1,1,0}}%
      \expandafter\def\csname LT6\endcsname{\color[rgb]{0,0,0}}%
      \expandafter\def\csname LT7\endcsname{\color[rgb]{1,0.3,0}}%
      \expandafter\def\csname LT8\endcsname{\color[rgb]{0.5,0.5,0.5}}%
    \else
      \def\colorrgb#1{\color{black}}%
      \def\colorgray#1{\color[gray]{#1}}%
      \expandafter\def\csname LTw\endcsname{\color{white}}%
      \expandafter\def\csname LTb\endcsname{\color{black}}%
      \expandafter\def\csname LTa\endcsname{\color{black}}%
      \expandafter\def\csname LT0\endcsname{\color{black}}%
      \expandafter\def\csname LT1\endcsname{\color{black}}%
      \expandafter\def\csname LT2\endcsname{\color{black}}%
      \expandafter\def\csname LT3\endcsname{\color{black}}%
      \expandafter\def\csname LT4\endcsname{\color{black}}%
      \expandafter\def\csname LT5\endcsname{\color{black}}%
      \expandafter\def\csname LT6\endcsname{\color{black}}%
      \expandafter\def\csname LT7\endcsname{\color{black}}%
      \expandafter\def\csname LT8\endcsname{\color{black}}%
    \fi
  \fi
    \setlength{\unitlength}{0.0500bp}%
    \ifx\gptboxheight\undefined%
      \newlength{\gptboxheight}%
      \newlength{\gptboxwidth}%
      \newsavebox{\gptboxtext}%
    \fi%
    \setlength{\fboxrule}{0.5pt}%
    \setlength{\fboxsep}{1pt}%
\begin{picture}(5040.00,6300.00)%
    \gplgaddtomacro\gplbacktext{%
      \csname LTb\endcsname
      \put(573,3704){\makebox(0,0)[r]{\strut{}$-1.0$}}%
      \put(573,4195){\makebox(0,0)[r]{\strut{}$-0.5$}}%
      \put(573,4687){\makebox(0,0)[r]{\strut{}$0.0$}}%
      \put(573,5178){\makebox(0,0)[r]{\strut{}$0.5$}}%
      \put(1568,2993){\makebox(0,0){\strut{}}}%
      \put(3001,2993){\makebox(0,0){\strut{}}}%
      \put(4434,2993){\makebox(0,0){\strut{}}}%
      \csname LTb\endcsname
      \put(1568,5350){\makebox(0,0)[l]{\strut{}$b_{\text{max}}=0.1\,\text{GeV}^{-1}$}}%
    }%
    \gplgaddtomacro\gplfronttext{%
      \csname LTb\endcsname
      \put(2681,4153){\makebox(0,0)[r]{\strut{}$\tilde{\param{K}}^{(1)}_{\hphantom{\text{K}}}(\Tsc{b}{};\mu_{Q_0})\,\,$}}%
      \csname LTb\endcsname
      \put(2681,3823){\makebox(0,0)[r]{\strut{}$\tilde{\param{K}}^{(1)}_{\hphantom{\text{K}}}(b_{*};\mu_{Q_0})\,\,$}}%
      \csname LTb\endcsname
      \put(2681,3493){\makebox(0,0)[r]{\strut{}$\,\,-\param{g}^{(1)}_{\text{K}}(\Tsc{b}{};\,b_{\text{max}})$}}%
    }%
    \gplgaddtomacro\gplbacktext{%
      \csname LTb\endcsname
      \put(573,1247){\makebox(0,0)[r]{\strut{}$-1.0$}}%
      \put(573,1738){\makebox(0,0)[r]{\strut{}$-0.5$}}%
      \put(573,2230){\makebox(0,0)[r]{\strut{}$0.0$}}%
      \put(573,2721){\makebox(0,0)[r]{\strut{}$0.5$}}%
      \put(1568,536){\makebox(0,0){\strut{}$10^{-1}$}}%
      \put(3001,536){\makebox(0,0){\strut{}$10^{0}$}}%
      \put(4434,536){\makebox(0,0){\strut{}$10^{1}$}}%
      \csname LTb\endcsname
      \put(1568,2893){\makebox(0,0)[l]{\strut{}$b_{\text{max}}=1.0\,\text{GeV}^{-1}$}}%
      \put(1008,5890){\makebox(0,0)[l]{\strut{}$Q_0=2\,\text{GeV}$}}%
    }%
    \gplgaddtomacro\gplfronttext{%
      \csname LTb\endcsname
      \put(2569,206){\makebox(0,0){\strut{}$\Tsc{b}{}(\text{GeV}^{-1})$}}%
    }%
    \gplbacktext
    \put(0,0){\includegraphics{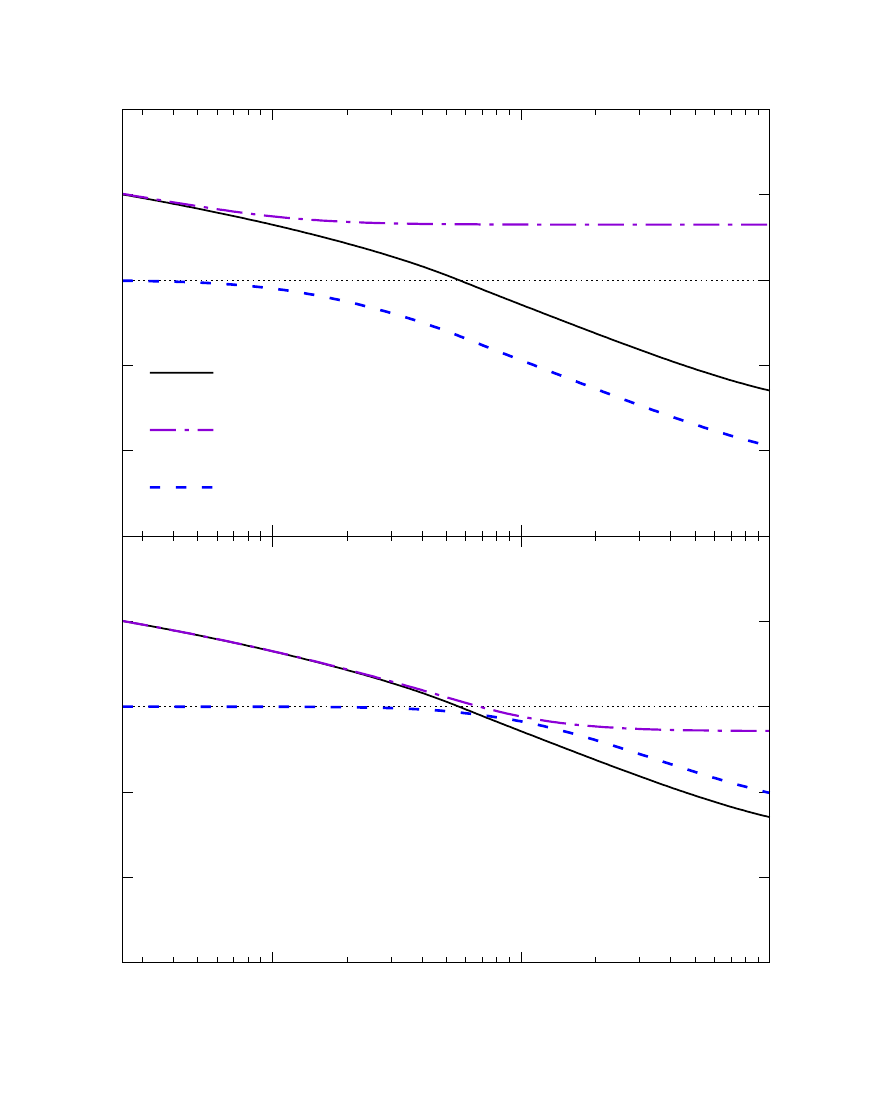}}%
    \gplfronttext
  \end{picture}%
\endgroup

%% file: gD_inline.tex
\begingroup
  \makeatletter
  \providecommand\color[2][]{%
    \GenericError{(gnuplot) \space\space\space\@spaces}{%
      Package color not loaded in conjunction with
      terminal option `colourtext'%
    }{See the gnuplot documentation for explanation.%
    }{Either use 'blacktext' in gnuplot or load the package
      color.sty in LaTeX.}%
    \renewcommand\color[2][]{}%
  }%
  \providecommand\includegraphics[2][]{%
    \GenericError{(gnuplot) \space\space\space\@spaces}{%
      Package graphicx or graphics not loaded%
    }{See the gnuplot documentation for explanation.%
    }{The gnuplot epslatex terminal needs graphicx.sty or graphics.sty.}%
    \renewcommand\includegraphics[2][]{}%
  }%
  \providecommand\rotatebox[2]{#2}%
  \@ifundefined{ifGPcolor}{%
    \newif\ifGPcolor
    \GPcolortrue
  }{}%
  \@ifundefined{ifGPblacktext}{%
    \newif\ifGPblacktext
    \GPblacktextfalse
  }{}%
  \let\gplgaddtomacro\g@addto@macro
  \gdef\gplbacktext{}%
  \gdef\gplfronttext{}%
  \makeatother
  \ifGPblacktext
    \def\colorrgb#1{}%
    \def\colorgray#1{}%
  \else
    \ifGPcolor
      \def\colorrgb#1{\color[rgb]{#1}}%
      \def\colorgray#1{\color[gray]{#1}}%
      \expandafter\def\csname LTw\endcsname{\color{white}}%
      \expandafter\def\csname LTb\endcsname{\color{black}}%
      \expandafter\def\csname LTa\endcsname{\color{black}}%
      \expandafter\def\csname LT0\endcsname{\color[rgb]{1,0,0}}%
      \expandafter\def\csname LT1\endcsname{\color[rgb]{0,1,0}}%
      \expandafter\def\csname LT2\endcsname{\color[rgb]{0,0,1}}%
      \expandafter\def\csname LT3\endcsname{\color[rgb]{1,0,1}}%
      \expandafter\def\csname LT4\endcsname{\color[rgb]{0,1,1}}%
      \expandafter\def\csname LT5\endcsname{\color[rgb]{1,1,0}}%
      \expandafter\def\csname LT6\endcsname{\color[rgb]{0,0,0}}%
      \expandafter\def\csname LT7\endcsname{\color[rgb]{1,0.3,0}}%
      \expandafter\def\csname LT8\endcsname{\color[rgb]{0.5,0.5,0.5}}%
    \else
      \def\colorrgb#1{\color{black}}%
      \def\colorgray#1{\color[gray]{#1}}%
      \expandafter\def\csname LTw\endcsname{\color{white}}%
      \expandafter\def\csname LTb\endcsname{\color{black}}%
      \expandafter\def\csname LTa\endcsname{\color{black}}%
      \expandafter\def\csname LT0\endcsname{\color{black}}%
      \expandafter\def\csname LT1\endcsname{\color{black}}%
      \expandafter\def\csname LT2\endcsname{\color{black}}%
      \expandafter\def\csname LT3\endcsname{\color{black}}%
      \expandafter\def\csname LT4\endcsname{\color{black}}%
      \expandafter\def\csname LT5\endcsname{\color{black}}%
      \expandafter\def\csname LT6\endcsname{\color{black}}%
      \expandafter\def\csname LT7\endcsname{\color{black}}%
      \expandafter\def\csname LT8\endcsname{\color{black}}%
    \fi
  \fi
    \setlength{\unitlength}{0.0500bp}%
    \ifx\gptboxheight\undefined%
      \newlength{\gptboxheight}%
      \newlength{\gptboxwidth}%
      \newsavebox{\gptboxtext}%
    \fi%
    \setlength{\fboxrule}{0.5pt}%
    \setlength{\fboxsep}{1pt}%
\begin{picture}(5040.00,6300.00)%
    \gplgaddtomacro\gplbacktext{%
      \csname LTb\endcsname
      \put(573,3591){\makebox(0,0)[r]{\strut{}-6.0}}%
      \put(573,3969){\makebox(0,0)[r]{\strut{}-4.0}}%
      \put(573,4347){\makebox(0,0)[r]{\strut{}-2.0}}%
      \put(573,4724){\makebox(0,0)[r]{\strut{}0.0}}%
      \put(573,5102){\makebox(0,0)[r]{\strut{}2.0}}%
      \put(573,5480){\makebox(0,0)[r]{\strut{}4.0}}%
      \put(1568,2993){\makebox(0,0){\strut{}}}%
      \put(3001,2993){\makebox(0,0){\strut{}}}%
      \put(4434,2993){\makebox(0,0){\strut{}}}%
      \csname LTb\endcsname
      \put(1568,5350){\makebox(0,0)[l]{\strut{}$b_{\text{max}}=0.1\,\text{GeV}^{-1}$}}%
    }%
    \gplgaddtomacro\gplfronttext{%
      \csname LTb\endcsname
      \put(3875,4153){\makebox(0,0)[r]{\strut{}$\ln\left(\tilde{\param{D}}^{(1,d_{\msbar})}(z,b_{\text{T}};\mu_{Q_0},Q_0^2)\right)$}}%
      \csname LTb\endcsname
      \put(3875,3823){\makebox(0,0)[r]{\strut{}$\ln\left(\tilde{\param{D}}^{(1,d_{\msbar})}(z,b_{*}\;;\mu_{Q_0},Q_0^2)\right)$}}%
      \csname LTb\endcsname
      \put(3875,3493){\makebox(0,0)[r]{\strut{}$-\param{g}^{(1,d_{\msbar})}(z,\Tsc{b}{};b_{\text{max}})\qquad$}}%
    }%
    \gplgaddtomacro\gplbacktext{%
      \csname LTb\endcsname
      \put(573,1134){\makebox(0,0)[r]{\strut{}-6.0}}%
      \put(573,1512){\makebox(0,0)[r]{\strut{}-4.0}}%
      \put(573,1890){\makebox(0,0)[r]{\strut{}-2.0}}%
      \put(573,2267){\makebox(0,0)[r]{\strut{}0.0}}%
      \put(573,2645){\makebox(0,0)[r]{\strut{}2.0}}%
      \put(573,3023){\makebox(0,0)[r]{\strut{}4.0}}%
      \put(1568,536){\makebox(0,0){\strut{}$10^{-1}$}}%
      \put(3001,536){\makebox(0,0){\strut{}$10^{0}$}}%
      \put(4434,536){\makebox(0,0){\strut{}$10^{1}$}}%
      \csname LTb\endcsname
      \put(1008,5890){\makebox(0,0)[l]{\strut{}$Q_0=2\,\text{GeV}$}}%
      \put(3023,5890){\makebox(0,0)[l]{\strut{}$z=0.3$}}%
      \put(1568,2893){\makebox(0,0)[l]{\strut{}$b_{\text{max}}=1.0\,\text{GeV}^{-1}$}}%
    }%
    \gplgaddtomacro\gplfronttext{%
      \csname LTb\endcsname
      \put(2569,206){\makebox(0,0){\strut{}$\Tsc{b}{}(\text{GeV}^{-1})$}}%
    }%
    \gplbacktext
    \put(0,0){\includegraphics{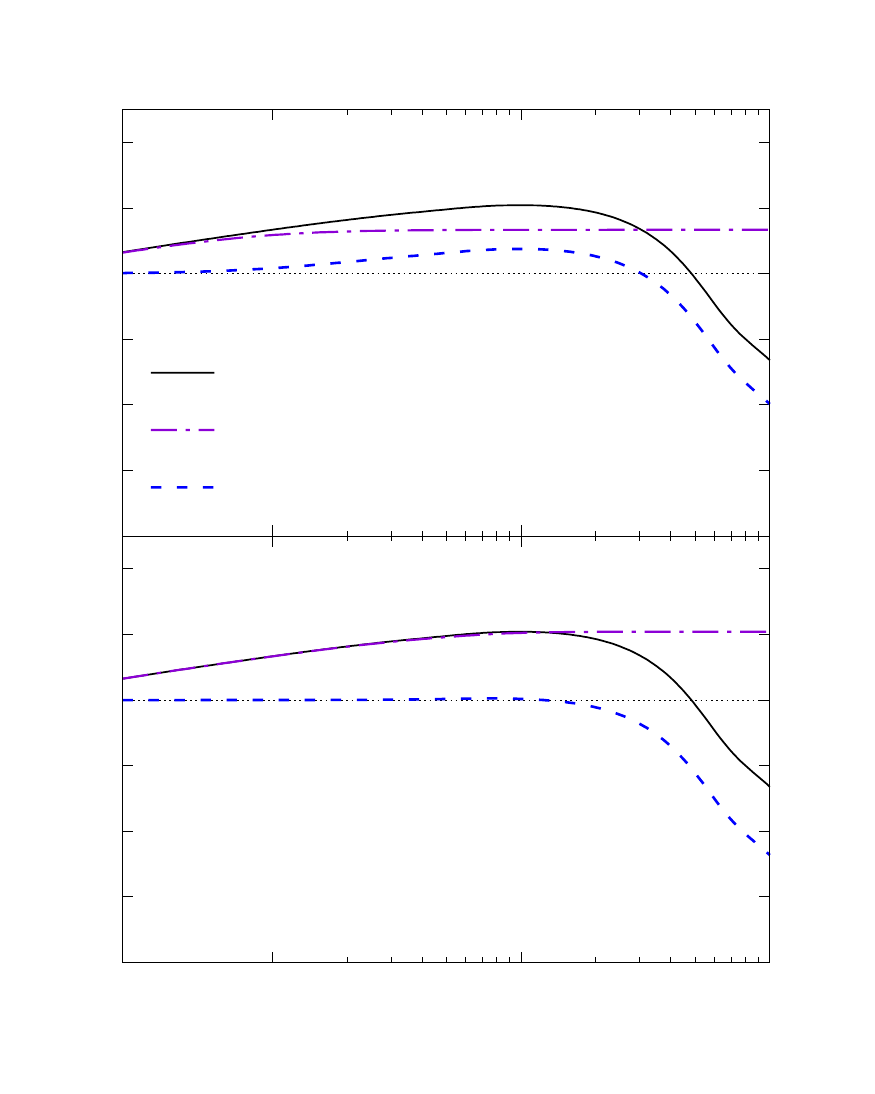}}%
    \gplfronttext
  \end{picture}%
\endgroup

%% file: WqT_ope_inline.tex
\begingroup
  \makeatletter
  \providecommand\color[2][]{%
    \GenericError{(gnuplot) \space\space\space\@spaces}{%
      Package color not loaded in conjunction with
      terminal option `colourtext'%
    }{See the gnuplot documentation for explanation.%
    }{Either use 'blacktext' in gnuplot or load the package
      color.sty in LaTeX.}%
    \renewcommand\color[2][]{}%
  }%
  \providecommand\includegraphics[2][]{%
    \GenericError{(gnuplot) \space\space\space\@spaces}{%
      Package graphicx or graphics not loaded%
    }{See the gnuplot documentation for explanation.%
    }{The gnuplot epslatex terminal needs graphicx.sty or graphics.sty.}%
    \renewcommand\includegraphics[2][]{}%
  }%
  \providecommand\rotatebox[2]{#2}%
  \@ifundefined{ifGPcolor}{%
    \newif\ifGPcolor
    \GPcolortrue
  }{}%
  \@ifundefined{ifGPblacktext}{%
    \newif\ifGPblacktext
    \GPblacktextfalse
  }{}%
  \let\gplgaddtomacro\g@addto@macro
  \gdef\gplbacktext{}%
  \gdef\gplfronttext{}%
  \makeatother
  \ifGPblacktext
    \def\colorrgb#1{}%
    \def\colorgray#1{}%
  \else
    \ifGPcolor
      \def\colorrgb#1{\color[rgb]{#1}}%
      \def\colorgray#1{\color[gray]{#1}}%
      \expandafter\def\csname LTw\endcsname{\color{white}}%
      \expandafter\def\csname LTb\endcsname{\color{black}}%
      \expandafter\def\csname LTa\endcsname{\color{black}}%
      \expandafter\def\csname LT0\endcsname{\color[rgb]{1,0,0}}%
      \expandafter\def\csname LT1\endcsname{\color[rgb]{0,1,0}}%
      \expandafter\def\csname LT2\endcsname{\color[rgb]{0,0,1}}%
      \expandafter\def\csname LT3\endcsname{\color[rgb]{1,0,1}}%
      \expandafter\def\csname LT4\endcsname{\color[rgb]{0,1,1}}%
      \expandafter\def\csname LT5\endcsname{\color[rgb]{1,1,0}}%
      \expandafter\def\csname LT6\endcsname{\color[rgb]{0,0,0}}%
      \expandafter\def\csname LT7\endcsname{\color[rgb]{1,0.3,0}}%
      \expandafter\def\csname LT8\endcsname{\color[rgb]{0.5,0.5,0.5}}%
    \else
      \def\colorrgb#1{\color{black}}%
      \def\colorgray#1{\color[gray]{#1}}%
      \expandafter\def\csname LTw\endcsname{\color{white}}%
      \expandafter\def\csname LTb\endcsname{\color{black}}%
      \expandafter\def\csname LTa\endcsname{\color{black}}%
      \expandafter\def\csname LT0\endcsname{\color{black}}%
      \expandafter\def\csname LT1\endcsname{\color{black}}%
      \expandafter\def\csname LT2\endcsname{\color{black}}%
      \expandafter\def\csname LT3\endcsname{\color{black}}%
      \expandafter\def\csname LT4\endcsname{\color{black}}%
      \expandafter\def\csname LT5\endcsname{\color{black}}%
      \expandafter\def\csname LT6\endcsname{\color{black}}%
      \expandafter\def\csname LT7\endcsname{\color{black}}%
      \expandafter\def\csname LT8\endcsname{\color{black}}%
    \fi
  \fi
    \setlength{\unitlength}{0.0500bp}%
    \ifx\gptboxheight\undefined%
      \newlength{\gptboxheight}%
      \newlength{\gptboxwidth}%
      \newsavebox{\gptboxtext}%
    \fi%
    \setlength{\fboxrule}{0.5pt}%
    \setlength{\fboxsep}{1pt}%
\begin{picture}(5040.00,6300.00)%
    \gplgaddtomacro\gplbacktext{%
      \csname LTb\endcsname
      \put(573,3857){\makebox(0,0)[r]{\strut{}$0.0$}}%
      \put(573,4451){\makebox(0,0)[r]{\strut{}$5.0$}}%
      \put(573,5045){\makebox(0,0)[r]{\strut{}$10.0$}}%
      \put(573,5639){\makebox(0,0)[r]{\strut{}$15.0$}}%
      \put(705,2993){\makebox(0,0){\strut{}}}%
      \put(1637,2993){\makebox(0,0){\strut{}}}%
      \put(2570,2993){\makebox(0,0){\strut{}}}%
      \put(3502,2993){\makebox(0,0){\strut{}}}%
      \put(4434,2993){\makebox(0,0){\strut{}}}%
      \csname LTb\endcsname
      \put(3241,5374){\makebox(0,0)[l]{\strut{}$b_{\text{max}}(\text{GeV}^{-1})$}}%
      \put(891,3950){\makebox(0,0)[l]{\strut{}$Q=2\,\text{GeV}$}}%
      \put(1008,5890){\makebox(0,0)[l]{\strut{}$Q_0=2\,\text{GeV}$}}%
      \put(3023,5890){\makebox(0,0)[l]{\strut{}$z=0.3$}}%
    }%
    \gplgaddtomacro\gplfronttext{%
      \csname LTb\endcsname
      \put(3949,3946){\makebox(0,0)[r]{\strut{}\eref{before_app_bstar} \hphantom{$\quad 0.10$}}}%
      \csname LTb\endcsname
      \put(3949,4298){\makebox(0,0)[r]{\strut{}\eref{before_app_bstar_css} $\quad 0.10$}}%
      \csname LTb\endcsname
      \put(3949,4650){\makebox(0,0)[r]{\strut{}\eref{before_app_bstar_css} $\quad 0.50$}}%
      \csname LTb\endcsname
      \put(3949,5002){\makebox(0,0)[r]{\strut{}\eref{before_app_bstar_css} $\quad 1.00$}}%
    }%
    \gplgaddtomacro\gplbacktext{%
      \csname LTb\endcsname
      \put(573,1400){\makebox(0,0)[r]{\strut{}$0.0$}}%
      \put(573,1994){\makebox(0,0)[r]{\strut{}$5.0$}}%
      \put(573,2588){\makebox(0,0)[r]{\strut{}$10.0$}}%
      \put(573,3182){\makebox(0,0)[r]{\strut{}$15.0$}}%
      \put(705,536){\makebox(0,0){\strut{}$0$}}%
      \put(1637,536){\makebox(0,0){\strut{}$1$}}%
      \put(2570,536){\makebox(0,0){\strut{}$2$}}%
      \put(3502,536){\makebox(0,0){\strut{}$3$}}%
      \put(4434,536){\makebox(0,0){\strut{}$4$}}%
      \csname LTb\endcsname
      \put(891,1493){\makebox(0,0)[l]{\strut{}$Q=10\,\text{GeV}$}}%
      \put(2420,2402){\makebox(0,0)[l]{\strut{}$\,\,M=0.20\,\text{GeV}$}}%
      \put(2420,2156){\makebox(0,0)[l]{\strut{}$m_D=0.30\,\text{GeV}$}}%
      \put(2420,1910){\makebox(0,0)[l]{\strut{}$m_K=0.10\,\text{GeV}$}}%
    }%
    \gplgaddtomacro\gplfronttext{%
      \csname LTb\endcsname
      \put(-75,3243){\rotatebox{-270}{\makebox(0,0){\strut{}$q_{\text{T}}\,W^{(1)}(Q,q_{\text{T}})/H^{(1)}$}}}%
      \put(2569,206){\makebox(0,0){\strut{}$\Tsc{q}{}(\text{GeV})$}}%
    }%
    \gplbacktext
    \put(0,0){\includegraphics{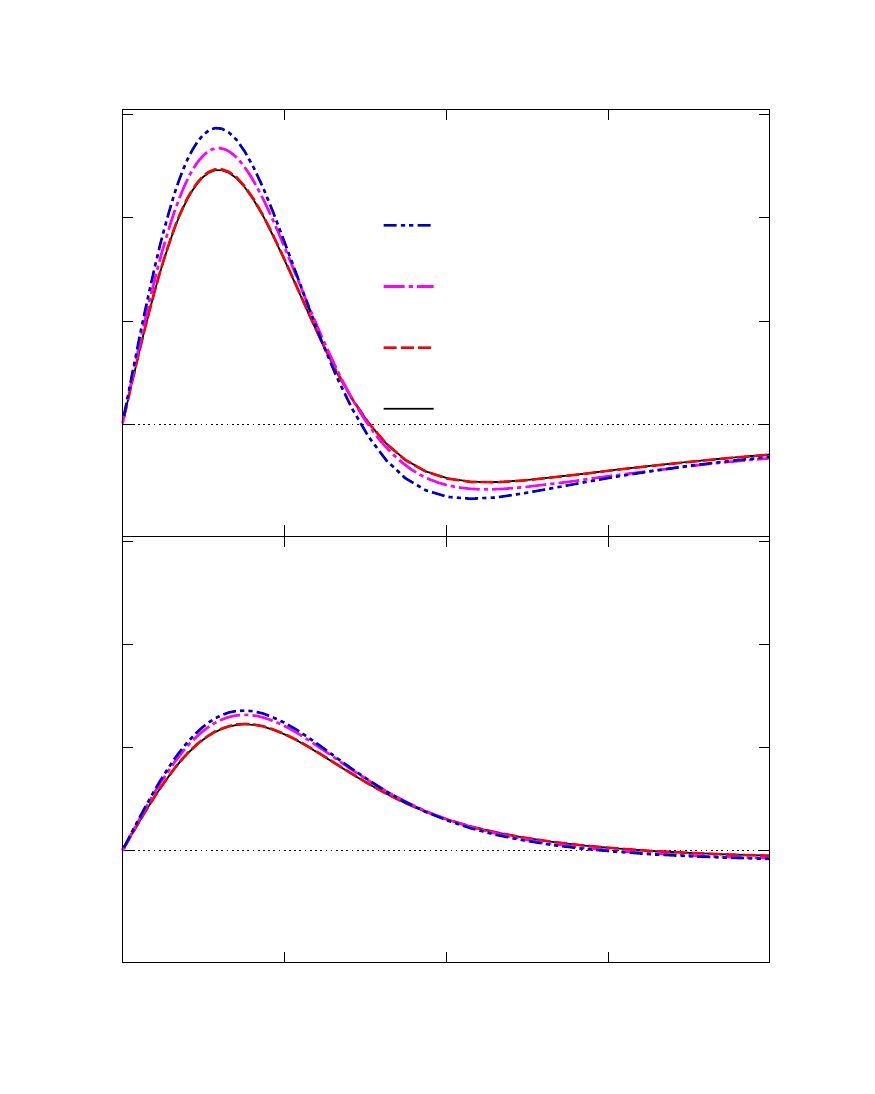}}%
    \gplfronttext
  \end{picture}%
\endgroup